\documentclass{article}
\usepackage{authblk}

\usepackage{geometry}
\usepackage{comment}
\usepackage{graphics,enumitem,epsfig,textcomp,tikz}
\usepackage{amsfonts,amsmath,amssymb,euscript,color,mathrsfs,bm}
\usepackage{dsfont}
\usepackage{float} 
\usepackage{hyperref}
\usepackage{cleveref}
\usepackage{cases}

\newtheorem{theorem}{Theorem}[section]
\newtheorem{proposition}[theorem]{Proposition}

\newtheorem{lemma}[theorem]{Lemma}
\newtheorem{remark}[theorem]{Remark}
\def\begproof{\noindent{\bf Proof: }}
\def\endproof{\par\rightline{\vrule height5pt width5pt depth0pt}\medskip}

\def\R{\mathbb{R}}
\def\Norm#1{\left\| #1 \right\|}

\def\C{\mathbb{C}}

\newcommand{\Cspace}{\bm{C}}
\newcommand{\Pspace}{P}
\newcommand{\Cfunc}[1]{\mathbb{#1}}
\newcommand{\Pfunc}[1]{#1}
\newcommand{\ptest}{\varphi}
\newcommand{\ptrial}{\varphi}
\newcommand{\Ctest}{\Psi}
\newcommand{\Ctrial}{\Psi}
\newcommand{\ctrial}{\psi}
\newcommand{\dd}[2]{\frac{\partial#1}{\partial #2}}
\renewcommand{\u}{\bm{u}}
\newcommand{\udot}{\dot{\u}}
\newcommand{\dx}{\mathrm{d}x}
\newcommand{\figtabsep}{2pt}

\title{Robust and scalable nonlinear solvers for finite element discretizations of biological transportation networks}
\author[1]{Jan Haskovec}
\author[1,2]{Peter Markowich}
\author[3]{Simone Portaro}
\author[1]{Stefano Zampini}
\affil[1]{Mathematical and Computer Sciences and Engineering Division, King Abdullah University of Science and Technology, Thuwal 23955-6900, Kingdom of Saudi Arabia}
\affil[2]{Faculty of Mathematics; University of Vienna, Oskar-Morgenstern-Platz 1, 1090 Vienna}
\affil[3]{Dipartimento di Ingegneria e Scienze dell'Informazione e Matematica (DISIM), Università degli Studi dell'Aquila}

\begin{document}

\maketitle 
\begin{abstract}
We develop robust and scalable, fully implicit nonlinear finite element solvers for the simulations of biological transportation networks driven by the gradient flow minimization of a nonconvex energy cost functional.
Our approach employs a discontinuous space for the conductivity tensor that allows us to guarantee the preservation of its positive semidefiniteness throughout the entire minimization procedure 
arising from the time integration of the gradient flow dynamics using a backward Euler scheme.
Extensive tests in two and three dimensions demonstrate the robustness and performance of the solver, highlight the sensitivity of the emergent network structures to mesh resolution and topology, and validate the resilience of the linear preconditioner to the ill-conditioning of the model. The implementation achieves near-optimal parallel scaling on large-scale, high-performance computing platforms. To the best of our knowledge, the network formation system has never been simulated in three dimensions before. Consequently, our three-dimensional results are the first of their kind.
\end{abstract}

\section{Introduction}\label{sec:intro}
    A transportation network can be understood as a spatial configuration that facilitates the movement or transfer of various resources. In biological systems, network structures such as leaf venation patterns, vascular systems, and neural connections have garnered considerable attention in recent research \cite{astuto2023self, bernot2009optimal, bohn2002constitutive, katifori2010damage}. A key area of study focuses on optimizing transport properties—whether electrical, fluid-based, or material flows—by balancing competing factors such as cost, transport efficiency, and resilience to failure \cite{barthelemy2011spatial}. Unlike engineered networks, biological transportation systems evolve without centralized control \cite{tero2010rules}. Instead, they are shaped by numerous cycles of evolutionary adaptation, making them prime examples of emergent structures governed by self-regulating processes.

    This paper investigates a class of self-regulating processes governed by the minimization of a biological energy cost functional given by
    \begin{align} \label{eq:energy}
        E[\C] := \int_{\Omega}  c^2 \,\nabla p \cdot ( \C + r \mathbb{I} ) \,\nabla p + \frac{\nu}{\gamma} \left( \Norm{\C}^{2} + \varepsilon \right)^{\frac{\gamma}{2}}  \dx,
    \end{align}
    where $\C(x)$ is a symmetric, positive semidefinite (SPSD) conductivity tensor defined on an open, bounded domain $\Omega \subset \R^d$ (with $d=2, 3$ for physical applications). The scalar field $p(x)$ represents the material pressure in a porous medium, and it is assumed to obey a Darcy law
    \begin{align} \label{eq:poisson}
    - \nabla \cdot \left( (\C + r \mathbb{I} ) \,\nabla p \right) = S, \quad \mathrm{in} \; \Omega,
    \end{align}
    where $S(x)$ represents a prescribed spatial distribution of sources and sinks of material.
    The parameter $c$ quantifies the system's propensity to align with the pressure gradient, $r > 0$ describes the isotropic background permeability of the medium, $\nu > 0$ denotes the metabolic constant, $\gamma > 0$ is the metabolic exponent, and $\varepsilon > 0$ is a regularization parameter. In biological contexts, such as plant leaf venation, the metabolic exponent typically falls within $1/2\leq \gamma \leq 1$, as shown in \cite{hu2013optimization,hu2013adaptation}. In contrast, for blood vessels, metabolic costs are proportional to the vessel's cross-sectional area, implying $\gamma= 1/2$, see \cite{murray1926physiological}. The notation $\Norm{\cdot}$ refers to the Frobenius norm, although other norms may be considered depending on the specific context or application of interest. We emphasize that setting $\varepsilon = 0$ recovers the metabolic term originally introduced in \cite{hu2013optimization, hu2013adaptation}.

    Evolutionary selection is modeled as the minimization of the energy functional \eqref{eq:energy} subject to the constraint \eqref{eq:poisson} with respect to $\C$.
    To be able to study the formation and, eventually, adaptation of the network patterns,
    we consider the formal $L^2$-gradient flow of the constrained energy functional \eqref{eq:energy}--\eqref{eq:poisson},
    \begin{align}\label{eq:flow}
        \frac{\partial \C}{\partial t} = c^2 \nabla p \otimes \nabla p - \nu \left(\Norm{\C }^2 + \varepsilon \right)^\frac{\gamma-2}{2} \C, \quad \mathrm{in} \; \Omega \times[0,\infty),
    \end{align}
    where $t\ge0$ is a time-like variable associated with the gradient flow dynamics; see \cite{marko_pilli}. We assume that the source term $S$ remains constant in time; time-varying sources can be accommodated by adding an extra first-order term to the energy functional.
    
    For $\gamma \ge 1$, the system \eqref{eq:poisson}-\eqref{eq:flow} admits a unique global weak solution in an appropriate Sobolev space, given reasonable assumptions on the data; see \cite[Proposition 6]{marko_pilli}. The proof deploys the convexity of the energy functional for $\gamma \ge 1$, allowing the application of Rockafellar's theorem. Currently, no known results establish well-posedness in Sobolev spaces for $\gamma < 1$. However, under stronger regularity assumptions on the data, a local well-posedness result in H\"{o}lder spaces has been recently obtained \cite{astuto2024self}. For further details on the derivation and comprehensive analysis of the gradient flow structure in this model, we refer the reader to \cite{marko_pilli, haskovec2018ode, portaro2022emergence, portaro2024measure, portaro2024mathematical}.

    A key novelty of this work is the treatment of the energy cost functional without the commonly assumed linear diffusion term for the conductivity matrix, \(\Delta \C\), which has been extensively studied in the literature \cite{astuto2023self, marko_pilli, haskovec2018ode, marko_perthame_2}. This diffusion term is typically introduced to account for random fluctuations in the medium, often modeled as Brownian motion. Remarkably, even in the absence of linear diffusion, complex and rich network structures emerge in our simulations. Their formation is driven by the positive feedback loop between the tensor-product term \(\nabla p \otimes \nabla p\) in \eqref{eq:flow} and the flux \((\C + r\mathbb{I})\nabla p\) induced by the Poisson equation \eqref{eq:poisson}, which is counter-balanced by the metabolic term $\nu (\Norm{\C }^2 + \varepsilon )^\frac{\gamma-2}{2} \C$.
    This underscores the central role of the conductivity tensor in shaping emergent network patterns in biological systems. In contrast, for the vectorial model studied in \cite{marko_perthame_2}, where the conductivity matrix is assumed to be of the form \(\C = \mathbf{m}\otimes\mathbf{m}\), linear diffusion was found to be essential for the initial propagation of material layers and, consequently, for the formation of network structures. Finally, for a recent derivation of a simplified scalar model related to our approach, with connections to \(p\)-Laplacian equations, we refer to \cite{haskovec2025gradient}.

    The energy functional \eqref{eq:energy}--\eqref{eq:poisson} has been obtained as a refinement limit of the discrete network model introduced in \cite{hu2013adaptation, hu2019optimization}. A formal derivation of \eqref{eq:energy}--\eqref{eq:poisson} as a $\Gamma$-limit of discrete energy functionals posed on a sequence of discrete graphs has been carried out in \cite{haskovec2018ode} in the special setting of diagonal permeability tensors, and in \cite{marko_pilli} extended to general symmetric, positive semidefinite conductivity tensors. Rigorous derivation with respect to the strong $L^2$-topology has been carried out in \cite{haskovec2019rigorous}. Finally, in \cite{alves2025rigorous}, a graphon limit has been derived rigorously, where \eqref{eq:poisson} is replaced by an integral equation.

    The structure of optimal discrete transportation networks for \(\gamma < 1\) is analyzed in \cite[Theorem~2.1]{burger2019mesoscopic}, where it is shown that the graph obtained by discarding edges with zero conductivity is acyclic (loop-free), that is, any pair of nodes is connected by a unique path of edges. Conversely, in \cite[Theorem~1]{HV24} it is proved that every acyclic graph generates a local minimizer of the discrete energy for \(\gamma < 1\). Moreover, for \(\gamma = 1\), the set of minimizers consists of acyclic graphs. Assuming that sequences of discrete minimizers converge, in an appropriate sense, to minimizers of \eqref{eq:energy}--\eqref{eq:poisson} in the continuum limit, one should not expect the limiting structures to exhibit a finite resolution scale. As a consequence, successive mesh refinements in the discretization of \eqref{eq:poisson}--\eqref{eq:flow} naturally lead to increasingly fine structures, with characteristic length scales determined by the mesh size, and different meshes may produce distinct network configurations. These theoretical considerations are consistent with the numerical results presented in \Cref{sec:experiments}, where mesh refinement and mesh topology are observed to have a significant impact on the resulting network structures. A robust numerical solver therefore plays a key role in enabling a systematic investigation of these effects.

    Various finite element–based methods have been proposed in the literature for the numerical solution of the vectorial model \cite{marko_perthame_2}, including Crank--Nicolson schemes \cite{marko_albi}, semi-implicit first-order time discretizations \cite{marko_perthame,marko_perthame_schlo}, and, more recently, energy-dissipation–preserving methods \cite{hong2021energy,yang2024discontinuous}. Semi-implicit solvers for the differential-algebraic system arising from the constrained gradient-flow dynamics \eqref{eq:flow}, incorporating an additional diffusion term for the conductivity matrix, have been recently proposed in \cite{astuto2023finite, astuto2023asymmetry}. In particular, finite difference discretizations in space combined with semi-implicit time-stepping based on a symmetric alternating direction scheme are proposed in \cite{astuto2023asymmetry}, while a finite element discretization coupled with a semi-implicit backward Euler scheme is considered in \cite{astuto2023finite}. Both studies report severe ill-conditioning for \(\gamma < 1\) as \(r \to 0\). In this regime, the energy functional is nonconvex, and numerical errors—further amplified by the ill-conditioning of the Poisson problem as \(r \to 0\)—accumulate over time, leading to the appearance of spurious secondary branches in the computed network structures. Moreover, numerical results obtained with continuous bilinear finite element discretizations for the conductivity matrix exhibit oscillatory behavior of the energy in the long-time regime and fail to preserve energy decay \cite{astuto2023finite}.
    
    The primary objective of this work is to develop a novel fully implicit nonlinear solution strategy, based on finite element discretizations, that preserves energy decay, remains robust to the severe ill-conditioning of the problem, and guarantees the preservation of symmetries—when present—in the final network structure.
    A key novelty of the proposed approach is the treatment of the conductivity matrix without artificial diffusion, which enables its discretization using discontinuous finite element spaces and thereby facilitates the preservation of positive semidefiniteness. This is complemented by a robust and scalable linear preconditioning strategy that effectively addresses the ill-conditioning of the Poisson problem. Together, these ingredients enable the efficient solution of the nonlinear systems arising from fully implicit time-stepping schemes. The entire algorithm is implemented in a distributed-memory framework and demonstrated on large-scale simulations, including three-dimensional network structures. While the numerical building blocks employed are well established, their combination and adaptation to this strongly nonconvex and ill-conditioned network formation model—together with the discontinuous treatment of the conductivity tensor—constitute the main methodological contribution of this work.

    For the first time, we present three-dimensional numerical results for the system \eqref{eq:poisson}--\eqref{eq:flow}. Previous numerical studies \cite{marko_perthame_schlo, albi_burger, astuto2022comparison} were restricted to lower-dimensional settings due to the prohibitive computational cost of three-dimensional simulations. This advancement enables the study of biological transportation network formation in realistic three-dimensional settings at high spatial resolution.

    The paper is structured as follows: \Cref{sec:model} introduces the model and the main assumptions. \Cref{sec:numerics} details the finite element discretization, the choice of the time-advancing scheme, and the design of the nonlinear solution strategy. Numerical results are presented and discussed in \Cref{sec:experiments}. Conclusions are drawn in \Cref{sec:conclusions}.

\section{The Model}\label{sec:model}

The primary objective of this paper is to design robust solvers and to conduct several numerical experiments; as a first step, we introduce a parameter reduction through a suitable scaling of the energy functional \eqref{eq:energy}. Specifically, we consider the following rescaling
\begin{align}
    \label{eq:energy_scaled}
    E_{c, \nu, \gamma}[\C] &= c^2 \int_{\Omega} \nabla p \cdot (\C + r \mathbb{I}) \nabla p + \frac{\nu}{c^2 \gamma} \left( \Norm{\C}^{2} + \varepsilon \right)^{\frac{\gamma}{2}}  dx \notag \\
    &:= c^2 E_{\tilde{\nu}, \gamma}[\C],
\end{align}
where $\tilde{\nu}=\frac{\nu}{c^2}$. Furthermore, under this scaling, the natural time variable for the evolution of the system is $\tilde{t}=c^2 t$. From now on, we will drop the tilde superscript, using $E$, $t$, and $\nu$ for notational simplicity, and focus on the numerical solution of the following differential-algebraic system:
    \begin{equation}
    \label{eq:system}
    \left\{
    \begin{aligned}
        &\frac{\partial \C}{\partial t} = \nabla p \otimes \nabla p - \nu \left(\Norm{\C }^2 + \varepsilon \right)^\frac{\gamma-2}{2} \C &&\quad \mathrm{in} \; \Omega \times[0,\infty),\\
        &- \nabla \cdot \left( \left (\C + r \mathbb{I} \right) \nabla p \right) = S &&\quad \mathrm{in} \; \Omega \times[0,\infty),\\
        &(\mathbb{C} + r \mathbb{I} ) \nabla p \cdot \bm{n} = 0 &&\quad \text{in} \; \partial \Omega \times[0,\infty),\\
        &\C(x,t=0) = \C_0(x) \ge 0 &&\quad \mathrm{in} \; \Omega,
    \end{aligned}
    \right.
    \end{equation}
    where the non-negativity of the initial condition is understood in the sense of matrices, and $\bm{n}$ denotes the outward-pointing unit normal vector to the boundary $\partial \Omega$. To ensure the solvability of the Poisson equation subject to the homogeneous Neumann boundary conditions, we impose the global conservation of mass,
    \begin{align}\label{eq:S}
        \int_{\Omega} S(x) ~\mathrm{d}x = 0.
    \end{align}
    
    The model under consideration is derived as the $L^2$-gradient flow of \eqref{eq:energy_scaled}, subject to the elliptic constraint \eqref{eq:poisson} with no-flux boundary conditions. Its derivation relies on well-established techniques as presented in \cite{ambrosio2008gradient, mielke2023introduction, santambrogio2017euclidean}, and the explicit computation in the biological network setting is presented in \cite{haskovec2018ode, portaro2022emergence}. We note that we employ the gradient flow approach not only as a procedure to find minima of the energy functional. Instead, we are interested in studying network formation as a dynamical process. Starting from a generic (and `simple') initial datum, e.g., $\C_0(x) \equiv \mathbb{I}$ in $\Omega$, we follow the gradient flow dynamics to observe the self-organization of the network, creation, and refinement of branches, until a steady state is reached.

    Well-posedness of the solutions of \eqref{eq:system} with $\gamma\geq 1$ is obtained by a slight modification of \cite[Proposition 7]{marko_pilli}. In particular, we have the following result.
    \begin{proposition}\label{thm:wellp}
    Let $\gamma\geq 1$, $r>0$, $S\in L^2(\Omega)$ verifying \eqref{eq:S}, and $\C_0\in [L^2(\Omega)]^{d\times d}$,
    symmetric and positive semidefinite almost everywhere in $\Omega$, and such that $E_{\tilde{\nu}, \gamma}[\C_0] <+\infty$. Then the problem \eqref{eq:system} admits a unique weak solution $\C\in H^1((0,\infty); L^2(\Omega))$, symmetric and positive semidefinite for almost all $t\geq 0$ and $x\in\Omega$. Moreover, we have the energy dissipation inequality
    \begin{align} \label{eq:energyIneq}
        E_{\tilde{\nu}, \gamma}[\C(t)] + \int_0^t \int_\Omega \left| \frac{\partial\C}{\partial t}(s,x)\right|^2  \mathrm{d} x\,\mathrm{d} s \leq E_{\tilde{\nu}, \gamma}[\C_0] \qquad \forall t \geq 0.
    \end{align}
    \end{proposition}

    The proof follows from the gradient flow structure of \eqref{eq:system}, combined with the convexity and lower semicontinuity of the energy functional \eqref{eq:energy_scaled} with $\gamma\geq 1$, by an application of the Rockafellar theorem.
    We note that for $\gamma<1$ the problem no longer enjoys convexity properties, and no well-posedness result is currently known. Nevertheless, assuming the global existence of sufficiently regular solutions, one can establish the following result on the preservation of positive semidefiniteness of the solution, which holds for all $\gamma>0$.

    \begin{lemma}
    Let $\C_0(x) \geq 0$ for some $x\in\Omega$. Then, the positive semidefiniteness of $\C$ is preserved along the flow, i.e. $\C(x,t) \geq 0$, $\forall t \geq 0 $
    \end{lemma}    
    \begproof
    Fix an arbitrary vector $\xi \in \R^d$ and define $z(x,t):= \xi \cdot \C(x,t)\ \xi$. Multiplying \eqref{eq:flow} by $\xi$ on both sides yields 
    \begin{align*}
        \frac{\partial z}{\partial t} = |\nabla p \cdot \xi|^2 - {\nu} \left(\Norm{\C }^2 + \varepsilon \right)^\frac{\gamma-2}{2} z \ge - {\nu} \left(\Norm{\C }^2 + \varepsilon \right)^\frac{\gamma-2}{2} z,
    \end{align*}
    with initial condition $z(x,0)=z_0(x):= \xi \cdot \C_0(x) \ \xi \ge 0$. By applying a Grönwall-type argument, we deduce that
    \begin{align*}
        z \ge z_0 \exp \left({-{\nu} \int_0^t \left(\Norm{\C(x,s) }^2 + \varepsilon \right)^\frac{\gamma-2}{2}} ds \right) \ge 0.
    \end{align*}
    Hence, $\C(x,t)$ remains positive semidefinite for all $t \ge 0$.
    \endproof
    
\section{Numerical scheme}\label{sec:numerics}

In this section, we describe the numerical discretization and solution strategy adopted for the system \eqref{eq:system}. While the main algorithmic ingredients—finite element discretizations, implicit time integration, Newton’s method for nonlinear problems, and Schur complement–based preconditioning—are well established in the literature, their application to the present model requires particular care and, in part, novel adaptations. In particular, although finite element methods have been used for related problems, continuous finite element discretizations for the conductivity tensor are not suitable in the present setting, motivating a different discretization strategy. The remaining components of the algorithm are specifically designed to address the strong nonlinearity and nonconvexity of the energy functional, as well as the severe ill-conditioning of the Poisson problem as the regularization parameter \(r\) approaches zero. The novelty, therefore, lies not in the numerical techniques themselves, but in how they are combined and tailored to respect the model's structural properties and remain effective in the strongly nonconvex, ill-conditioned regimes considered.

We first introduce the spatial and temporal discretizations employed for the system \eqref{eq:system} in \Cref{sec:semidiscrete,sec:time_disc}. We then derive the Jacobian associated with the fully implicit formulation, establish the well-posedness of the resulting linearized problems, and exploit the block structure of the linear systems to design a scalable and robust Schur complement–based preconditioning strategy in \Cref{sec:linear}. 

\subsection{Semi-discrete formulation}\label{sec:semidiscrete}

We base the numerical scheme on the Finite Element Method (FEM, \cite{ciarletFEM}). We thus proceed in a standard way and introduce the variational formulation of \eqref{eq:flow}, working with the Sobolev spaces
\[
\Cspace := \bigg\{ \Cfunc{B} \in [L^2(\Omega)]^{d\times d} : \Cfunc{B} = \Cfunc{B}^T\bigg\}, \quad \Pspace := \bigg\{ \Pfunc{q} \in H^1(\Omega) : \int_\Omega \Pfunc{q} ~\dx = 0 \bigg\}.
\]
Multiplying by test functions $(\Cfunc{B},\Pfunc{q})$, integrating over the domain, and by using integration by parts (taking into account the natural boundary conditions on $\C$), we obtain the following variational problem. Let $T > 0$, $S \in L^2(\Omega)$, find $\C(x,t) \in H^1([0,T]; \Cspace)$, $p(x,t)\in \Pspace$ such that
\begin{align*}
&\int_\Omega(\C + r\mathbb{I})\nabla p \cdot \nabla q ~ \dx = \int_\Omega S q ~ \dx, &\forall q \in \Pspace,\\
&\int_\Omega \left( \dd{\C}{t} : \Cfunc{B} - (\nabla p \otimes \nabla p) : \Cfunc{B} + {\nu} (\Norm{\C}^2 + \varepsilon)^{(\gamma - 2)/2} \C :\Cfunc{B} \right) \dx = 0, &\forall \Cfunc{B} \in \Cspace,
\end{align*}
which consists of $d (d+1)/2$ equations for the components of the symmetric conductivity matrix and one equation for the pressure. We note that $(\C  + r\mathbb{I})\nabla p$ is the usual matrix-vector product, while products of the form $\C: \Cfunc{B}$ are entry-wise, i.e., $\C : \mathbb{B} := \mathrm{tr}(\C^T \mathbb{B})$.

Introducing a tessellation $\Omega_h\subset \Omega$ with a given mesh size $h$ and the finite-dimensional subspaces
\[
\Cspace_h = \bigg{\{} \Cfunc{C}_h \in \Cspace ~|~ \Cfunc{C}_h = \sum^{N_c}_{j=1} a_j \Ctrial_j \bigg{\}}, \quad 
\Pspace_h = \bigg{\{} \Pfunc{p}_h \in \Pspace ~|~ \Pfunc{p}_h = \sum^{N_p}_{j=1} a_j \ptrial_j \bigg{\}},
\]
we arrive at a semi-discrete formulation: Given $S \in L^2(\Omega_h)$, find $\Pfunc{p}_h \in \Pspace_h$, $\C_h \in H^1([0,T]; \Cspace_h)$, such that
\begin{align*}
&\int_\Omega(\C_h + r\mathbb{I})\nabla p_h \cdot \nabla q_h ~ \dx = \int_\Omega S q_h ~ \dx, &\forall\, q_h \in \Pspace_h,\\
&\int_\Omega \left( \dd{\C_h}{t} : \Cfunc{B}_h - (\nabla p_h \otimes \nabla p_h) : \Cfunc{B}_h + {\nu} (\Norm{\C_h}^2 + \varepsilon)^{(\gamma - 2)/2} \C_h :\Cfunc{B}_h \right) \dx = 0, &\forall\, \Cfunc{B}_h \in \Cspace_h,
\end{align*}
which leads to a set of Differential-Algebraic equations (DAE)
\begin{equation}\label{eq:semidiscrete}
\begin{aligned} %
&\int_\Omega(\C_h + r\mathbb{I})\nabla p_h \cdot \nabla \ptest_i ~ \dx = \int_\Omega S \ptest_i ~ \dx, \, &\forall i \in \{1,\dots,N_p\}, \\
&\int_\Omega \left( \dd{\C_h}{t} : \Ctest_i - (\nabla p_h \otimes \nabla p_h) : \Ctest_i + {\nu} (\Norm{\C_h}^2 + \varepsilon)^{(\gamma - 2)/2} \C_h : \Ctest_i \right) \dx = 0, \, &\forall i \in \{1,\dots,N_c\},
\end{aligned}
\end{equation}
where $\ptest_i$ (resp. $\Ctest_i$) are test functions for $\Pspace_h$ (resp. $\Cspace_h$).

Motivated by the need to preserve the structural properties of the conductivity matrix, in particular positive semidefiniteness, we discretize the conductivity using piecewise constant finite element spaces and the pressure using continuous piecewise linear finite elements, namely
\begin{equation}\label{eq:fem_spaces}
\begin{aligned}
\Cspace_h &:= \big\{ \C_h \in \Cspace \;|\; {\C_h}_{|e} \text{ is constant for all } e \in \Omega_h \big\},\\
\Pspace_h &:= \big\{ p_h \in \Pspace \;|\; p_h \in C^0(\Omega_h) \text{ and } {p_h}_{|e} \text{ is linear for all } e \in \Omega_h \big\},
\end{aligned}
\end{equation}
where \(e\) denotes an element of the tessellated domain \(\Omega_h\). Note that the term \emph{linear} in the definition of \(\Pspace_h\) refers to bi-linear or tri-linear elements when \(K\) is a tensor-product cell.

We then proceed by stating the system of equations explicitly in two dimensions; similar arguments hold for the three-dimensional case and are omitted. Specifically, we discretize the symmetric conductivity matrix with a three-component FEM space
\[
\C_h := \begin{bmatrix}\C^{0}_h & \C^{1}_h \\ \C^{1}_h & \C^{2}_h\end{bmatrix},
\]
using the same scalar basis for each component $\ctrial$. The conductivity equations of \eqref{eq:semidiscrete} can thus be rewritten 
\begin{equation*}
\begin{aligned} %
&\int_\Omega \dd{\C^0_h}{t} ~ \ctrial_i - \dd{p_h}{x} \dd{p_h}{x} ~ \ctrial_i + {\nu} (\Norm{\C_h}^2 + \varepsilon)^{(\gamma - 2)/2} ~\C^0_h ~ \ctrial_i \, \dx = 0,\\
&\int_\Omega \dd{\C^1_h}{t} ~ \ctrial_i - \dd{p_h}{x} \dd{p_h}{y} ~ \ctrial_i + {\nu} (\Norm{\C_h}^2 + \varepsilon)^{(\gamma - 2)/2} ~\C^1_h ~ \ctrial_i \, \dx = 0,\\
&\int_\Omega \dd{\C^2_h}{t} ~ \ctrial_i - \dd{p_h}{y}\dd{p_h}{y} ~ \ctrial_i + {\nu} (\Norm{\C_h}^2 + \varepsilon)^{(\gamma - 2)/2} ~ \C^2_h ~ \ctrial_i \, \dx = 0.
\end{aligned}
\end{equation*}
For symmetry reasons (see \Cref{sec:linear} for details), we further rescale the entire system of equations, including the constraint, as
\begin{equation}\label{eq:semidiscrete_csym}
\begin{aligned} %
&-\int_\Omega(\C_h + r\mathbb{I})\nabla p_h \cdot \nabla q_h ~ \dx = -\int_\Omega S q_h ~ \dx,\\
&\frac{1}{2}\int_\Omega  \dd{\C^0_h}{t} ~  \ctrial_i - \dd{p_h}{x} \dd{p_h}{x} ~ \ctrial_i + {\nu} (\Norm{\C_h}^2 + \varepsilon)^{(\gamma - 2)/2} ~\C^0_h ~ \ctrial_i \, \dx = 0,\\
&\phantom{\frac{1}{2}}\int_\Omega  \dd{\C^1_h}{t} ~ \ctrial_i - \dd{p_h}{x} \dd{p_h}{y} ~ \ctrial_i + {\nu} (\Norm{\C_h}^2 + \varepsilon)^{(\gamma - 2)/2} ~\C^1_h ~ \ctrial_i \, \dx = 0,\\
&\frac{1}{2}\int_\Omega \dd{\C^2_h}{t} ~ \ctrial_i - \dd{p_h}{y}\dd{p_h}{y} ~ \ctrial_i + {\nu} (\Norm{\C_h}^2 + \varepsilon)^{(\gamma - 2)/2} ~ \C^2_h ~ \ctrial_i \, \dx = 0.
\end{aligned}
\end{equation}

\subsection{Time discretization}\label{sec:time_disc}

In order to discuss the selection of a suitable time integration scheme, we first observe that the semi-explicit index-1 DAE in \eqref{eq:semidiscrete} forms a compact set of nonlinear equations of the type \cite{ascher1998computer}
\begin{equation}\label{eq:ifunction}
F(\udot,\u,t) = 0,
\end{equation}
where we have used $\u$ to denote the vector of degrees of freedom, ordered by fields, associated with the finite element functions $\C_h$ and $p_h$; for example, in two dimensions,
\[
\u := [\C^{0}_h ,\C^{1}_h , \C^{2}_h , p_h ],
\]
where, with abuse of notation, we used the finite element function to denote the vector of associated degrees of freedom.
Specific to the choice of the time integrator is the approximation of the time derivative vector $\udot$; in particular, using the backward Euler (BE) method as the time integrator results in a first-order time-advancing scheme by solving the set of nonlinear equations
\begin{equation}\label{eq:be}    
F\left(\frac{\u_{n+1}-\u_n}{\delta t_n},\u_{n+1},t_n + \delta t_n\right) = 0,
\end{equation}
where $t_n$ is the current time, $\delta t_n$ is the current time step,
$\u_n$ 
denotes the known solution at the $n$-th time step, and $\u_{n+1}$ is the vector of unknowns for the solution at the $n+1$-th step.

In our case, the energy functional \eqref{eq:energy_scaled} is convex if and only if $\gamma \ge 1$; in such cases, the backward Euler scheme will guarantee convergence to a global minimum and will maintain the energy decay property (note that $\lambda$-convexity is enough if $\delta t$ is sufficiently small). However, if the convexity of the energy is dropped, then the existence and uniqueness of the solution $\u_{n+1}$ is questionable, as is its energy decay. It is well established that the backward Euler scheme is unconditionally stable, meaning it converges to a stationary point regardless of the chosen time step. Nevertheless, it does not necessarily preserve the structural properties of the system. For a detailed discussion on the convergence of BE to equilibrium, we refer to \cite{dello2024local, merlet2009convergence}. 
Conversely, the explicit Euler method ensures dissipation but often requires very restrictive conditions on the time step in relation to the spatial discretization. We further note here that when the energy functional satisfies suitable smoothness, coercivity, and convexity conditions, discrete gradient methods provide unconditional stability and ensure a monotone decrease of the energy throughout the iterations (see, e.g., \cite{grimm2017discrete}).

For higher-order time integrations of DAEs, a popular choice is the so-called BDF method based on backward differential formulas for the approximation of the time-derivative vector \cite{gear1967numerical}

\begin{equation}\label{eq:bdf}    
F\left(\sum^s_{k=0}\alpha_k\u_{n+1-k},\u_{n+1},t_n + \delta t_n\right) = 0,
\end{equation}
where the coefficients $\alpha_k$ are chosen using Lagrangian interpolation polynomials to achieve a global $s$-order convergence for the method. For $k=1$, we recover the backward Euler scheme \eqref{eq:be}.
Another popular second-order method is the Crank--Nicolson method, which uses a trapezoidal rule to evaluate the right-hand side of explicit ordinary differential equations
\begin{equation}\label{eq:cn}    
\u_{n+1}=\u_n + \frac{\delta t_n}{2}[f(\u_n) + f(\u_{n+1})],
\end{equation}
where the equation is expressed in the explicit form $\udot = f(\u,t)$.

It is straightforward to prove that the backward Euler scheme preserves the positive semidefiniteness of the conductivity matrix at the discrete level. In fact, we have the following.

\begin{lemma}\label{lem:be_posdef}
Let \(\C^e_n\) denote the restriction of the discrete conductivity matrix \(C_h(x,t_n)\) to a given element \(e\) at time \(t_n\). Then, under backward Euler time-stepping, if \(\C^e_n\) is  SPSD, the updated matrix \(\C^e_{n+1}\) is also SPSD.
\end{lemma}
\begproof
We first rewrite the equations for $\C$ \eqref{eq:flow} as
\begin{equation*}
\frac{\partial \C}{\partial t} = \mathbb{G}(p) - m(\C)\,\C,
\end{equation*}
where
\begin{equation}\label{eq:lem_func}
\mathbb{G}(p) := \nabla p \otimes \nabla p , \quad m(\C) := 
{\nu} \left(\Norm{\C }^2 + \varepsilon \right)^\frac{\gamma-2}{2},
\end{equation}
and note that $\mathbb{G}(p)$ is a SPSD matrix and $m(\C)$ is a positive scalar. Using the fact that $\Cspace_h$ given in \Cref{eq:fem_spaces} consists of element-by-element discontinuous functions, the BE equations \eqref{eq:be} arising from \eqref{eq:semidiscrete} can be rewritten as
\begin{equation}\label{eq:be_eqs}
\int_e  \left( 1 +  m(\C^e_{n+1})\,\delta t \right)\,
\C^e_{n+1}:\Ctest_i  \,\dx = \int_e  \left( \C^e_n +  \mathbb{G}(p_{n+1})\,\delta t \right) :\Ctest_i  \,\dx,
\end{equation}
Thus, if $\C^e_n$ is SPSD, $\C^e_{n+1}$ will also be SPSD since $1 +  m(\C^e_{n+1})\delta t > 0 $, $\Ctest_i$ is a matrix test function with positive and constant entries, and the right-hand side consists of the weighted sum, with positive weights, of SPSD matrices sampled at the quadrature nodes used by the FEM method.
\endproof

\begin{remark}
Since the discretization space for the conductivity does not enforce positivity explicitly, the equations in \eqref{eq:be_eqs} must be solved with sufficient accuracy to preserve the positive semidefiniteness of the discrete conductivity matrix at the numerical level.
While higher-order discretizations could, in principle, be used, guaranteeing positivity may require specialized polynomial bases with non-negative basis functions, such as Bernstein polynomials \cite{kirby2011fast}.
Moreover, the solutions generated by the network formation process exhibit strongly localized and branched structures with sharp spatial variations and limited regularity. In this regime, higher-order polynomial approximations are not expected to provide significant accuracy gains. For these reasons, we restrict our attention to low-order discretizations, which offer a robust and efficient choice for the problems considered.
 \end{remark}

\begin{remark}
We stress that the discontinuity of the discretization space is crucial to guarantee the positive semidefiniteness of $\C_h$. The use of a $H^1(\Omega)$ conforming space, as needed in the presence of the diffusion term for $\C$ in the energy functional, will require the finite element spaces to guarantee the preservation of the maximum principle, a condition very hard to realize in practice; see, e.g., \cite{farago2006discrete} and the references therein.
\end{remark}

The positive semidefiniteness of $\C_h$ is generally not preserved by the BDF or CN methods. Using the notation given in \Cref{eq:lem_func}, the equations arising from \eqref{eq:bdf} for BDF of order 2 (similar arguments hold for higher-order formulas) are 
\begin{equation*}
\int_e \left( 1+ m(\C^e_{n+1})\,\delta t \right)\,\C^e_{n+1}:\Ctest_i  \,\dx = \int_e \left(\frac{4}{3} \C^e_n  - \frac{1}{3} \C^e_{n-1} + \mathbb{G}(p_{n+1})\,\delta t\right):\Ctest_i \, \dx,
\end{equation*}
where, without loss of generality, we have assumed constant time steps. In this case, we cannot guarantee the positive semidefiniteness of $\C^e_{n+1}$ since the $\C^e_{n-1}$ term contributes a negative shift by an SPSD matrix. Similar arguments hold for the CN update
\begin{equation*}
\int_e \left(1+\frac{\delta t}{2} m(\C^e_{n+1})\right) \C^e_{n+1}:\Ctest_i \,\dx = \int_e \left( \C^e_n + \frac{\delta t}{2}\big[\mathbb{G}(p_{n}) + \mathbb{G}(p_{n+1}) - m(\C^e_n)\,\C^e_n\big] \right) :\Ctest_i \,\dx,
\end{equation*}
where now the SPSD term $m(\C^e_n)\,\C^e_n$ contributes a negative shift to the update.

\subsection{Linear solver}\label{sec:linear}
Once the time integrator is set, at each time step of the simulation, we solve the associated nonlinear equations
using the inexact Newton's method \cite{nocedal1999numerical}. 
Assuming that the time discretization used to define $\udot$ is linear in the unknown, the Jacobian of the system can be obtained by considering the generalized Jacobian of \eqref{eq:ifunction}
\begin{equation}\label{eq:ijacobian}
J = \sigma\,\dd{F}{\udot} + \dd{F}{\u}
\end{equation}
where $\sigma$ is a scalar associated with the current time step choice \cite{ascher1998computer}. For example, for BE, $\sigma = 1/\delta t_n$.

We are interested in very fine discretizations and therefore rely on Krylov methods to solve the Jacobian linear systems. Since the convergence of such methods critically depends on the choice of an effective preconditioner, it is essential to establish the well-posedness of the linearized problems. To this end, we explicitly derive the Jacobians for the two-dimensional case using the symmetrized residual equations \eqref{eq:semidiscrete_csym} and prove that the resulting linear systems are well posed. Analogous arguments apply in three dimensions, and we omit the corresponding details for brevity.

The Jacobian of the time-dependent part of \eqref{eq:ijacobian} is independent of the linearization point, and it is given by
\[
\dd{F}{\udot} := 
\begin{bmatrix}
\frac{1}{2}M & 0 & 0 & 0\\
0 & M & 0 & 0\\
0 & 0 & \frac{1}{2}M & 0\\
0 & 0 & 0 & 0\\
\end{bmatrix},
\]
where $M$ is the mass matrix associated with the basis of the conductivity components
\[
M_{ij} = \int_\Omega \ctrial_i \,\ctrial_j ~\dx.
\]
The Jacobian of the time-independent part of \eqref{eq:ijacobian} is instead a block matrix with structure
\[
\dd{F}{\u} := 
\begin{bmatrix}
A^{00} & 2A^{01} & A^{02} & B^{0}\\
2A^{10} & 4A^{11} & 2A^{12} & B^{1}\\
A^{20} & 2A^{21} & A^{22} & B^{2}\\
{B^0}^T & {B^1}^T & {B^2}^T & -D\\
\end{bmatrix},
\]
where the matrix entries of the individual blocks ($i,j$ denote row and column indices) are given by
\begin{align*}
&A^{mn}_{ij} = \nu \int_\Omega (\alpha \,\delta_{mn} + \beta ~\C^{m}_h \C^{n}_h) \ctrial_i \ctrial_j \dx,\quad m,n =0,1,2 \\ 
&B^{0}_{ij} = -\int_\Omega \ctrial_i \dd{p_h}{x} \dd{\ptrial_j}{x}\dx,\\ 
&B^{1}_{ij} = -\int_\Omega \ctrial_i \left(\dd{p_h}{y} \dd{\ptrial_j}{x} + \dd{p_h}{x} \dd{\ptrial_j}{y}\right)\dx,\\
&B^{2}_{ij} = -\int_\Omega \ctrial_i \dd{p_h}{y} \dd{\ptrial_j}{y}\dx,\\
&D_{ij} = \int_\Omega  (\C_h + r \mathbb{I}) \nabla \ptrial_i \cdot \nabla \ptrial_j \dx,
\end{align*}
where $\delta_{mn}$ is the usual Dirac's delta coefficient, and the scalar coefficients $\alpha$ and $\beta$ are defined as
\begin{equation}\label{eq:jac_coeffs}
\alpha := (\Norm{\C_h}^2 + \varepsilon)^{(\gamma - 2)/2},\quad  \beta :=\frac{\gamma - 2}{2} (\Norm{\C_h}^2 + \varepsilon)^{(\gamma - 4)/2}.
\end{equation}

The following Lemma is crucial for establishing the well-posedness and invertibility of the linearized discrete problem (see Theorem~3.4 in \cite{benzi2005numerical}). A priori error estimates could be obtained by adapting Theorem~5.5.1 in \cite{boffi2013mixed}; however, as this paper focuses on numerical methodology, we do not pursue a detailed theoretical error analysis here and leave it for future work.

\begin{lemma}\label{lem:J00_posdef}
With our choice of discretization for the conductivity matrix, the matrix 
\[
A=
\begin{bmatrix}
A^{00} & 2A^{01} & A^{02}\\
2A^{10} & 4A^{11} & 2A^{12}\\
A^{20} & 2A^{21} & A^{22}
\end{bmatrix},
\]
is symmetric positive definite.
\end{lemma}

\begproof
Due to our choice of discretization space for $\C_h$ we have  $\ctrial_i\ctrial_j=\delta_{ij}$, and
$A$ is thus a block diagonal matrix with a block for each element $e$.
Therefore, it suffices to show that each $3\times3$ element block, in the sequel denoted by $A^{e}$, is symmetric and positive definite.

The symmetry of the element matrix is clear. For the positivity, since $\C_h$, and thus $\alpha$ and $\beta$ given in \Cref{eq:jac_coeffs}, are constant on each element, we can define
\[
\alpha_e :=\nu\,\alpha|e|,\qquad
\beta_e :=\nu\,\beta|e|,
\]
and obtain the following compact representation
\[
A^{e}=\alpha_e M+\beta_e\,c_V^Tc_V,
\]
where $M=\mathrm{diag}(1,4,1)$ and the rank-one update is given in terms of the vector
\[
c_V=[\C_h^0,\,2\C_h^1,\,\C_h^2].
\]
Let $M^{1/2} = \mathrm{diag}(1,2,1)$  and set $\tilde c^T := M^{-1/2}c^T_V$. Then
\[
A^{e}
=M^{1/2}\bigl(\alpha_e I+\beta_e\,\tilde c^T\tilde c\bigr)M^{1/2},
\]
so $A^{e}> 0$ if and only if $H:=\alpha_e I+\beta_e\tilde c^T\tilde c > 0$.
Because $\tilde c^T\tilde c$ has rank one, the eigenvalues of $H$ are
$\alpha_e$ (with multiplicity $2$) and $\alpha_e+\beta_e\|\tilde c\|^2$ \cite{HornJohnson2013}, where $\|\tilde c\|$ is the norm of the vector $\tilde c$.

Since $\alpha_e>0$, it remains to prove that $\alpha_e+\beta_e\|\tilde c\|^2>0$.
If $\gamma\ge2$, then this trivially holds since $\beta_e\ge0$ and $\alpha_e+\beta_e\|\tilde c\|^2\ge\alpha_e>0$.
On the other hand, when $0<\gamma<2$, we have $\beta_e < 0$. However, since
\begin{equation*} 
\|\tilde c\|^2=c_V M^{-1}c^T_V=(\C_h^0)^2+(\C_h^1)^2+(\C_h^2)^2\le \|\C_h\|^2,
\end{equation*}
we have
\[
\frac{\alpha_e+\beta_e\|\tilde c\|^2}{\nu\,|e|}
=(\|\C_h\|^2+\varepsilon)^{\frac{\gamma-4}{2}}(\|\C_h\|^2+\varepsilon+\frac{\gamma-2}{2}\|\tilde c\|^2)
\ge (\|\C_h\|^2+\varepsilon)^{\frac{\gamma-4}{2}}(\varepsilon+\frac{\gamma}{2}\|\C_h\|^2)>0.
\]
\endproof

Grouping the conductivity blocks, we can write the symmetric indefinite Jacobian \eqref{eq:ijacobian} as
\[
J = \begin{bmatrix}
J_{00} & J_{01} \\
J^T_{01} & -D
\end{bmatrix}.
\]
The use of a discontinuous discretization space for the entries of \(\C_h\) yields a trivially invertible \(J_{00}\) block, since, by Lemma~\ref{lem:J00_posdef}, this block is positive definite and block diagonal, with independent \(3\times3\) sub-blocks associated with each element. We can thus consider an exact Schur complement factorization of $J$ as
\begin{equation}\label{eq:schurPre}
\begin{bmatrix}
J_{00} & J_{01} \\
J^T_{01} & -D
\end{bmatrix} =
\begin{bmatrix} 
J_{00} & 0\\
J^T_{01} & I
\end{bmatrix} 
\begin{bmatrix} 
J^{-1}_{00}& 0\\
0 & -G
\end{bmatrix} 
\begin{bmatrix} 
J_{00} & J_{01}\\
0 & I 
\end{bmatrix},
\quad G := D + J^T_{01} J^{-1}_{00} J_{01},
\end{equation}
that requires solving the inexpensive block-diagonal problem $J_{00}$ twice, and the Schur complement system $G$ only once.
The Schur complement \(G\) is a symmetric positive semidefinite matrix representing a perturbed scalar diffusion problem, since \(J_{00}\) is symmetric positive definite. The positive semidefiniteness of \(G\) follows from the fact that the constant function belongs to the kernel of the column spaces of both \(D\) and \(J_{01}\).

In practice, the exact inverse of the Schur complement \(G\) is replaced by the action of a suitable preconditioner, and the factorization in \eqref{eq:schurPre} is employed as a preconditioner within an outer Krylov method for indefinite systems.  Since the Schur complement matrix can be constructed at relatively low cost, an algebraic multigrid (AMG) method is a natural and effective choice for its preconditioning, providing robust performance with an overall computational cost that scales linearly with the matrix size \cite{xu2017algebraic}. Non-overlapping domain decomposition methods, such as the Balancing Domain Decomposition by Constraints (BDDC) preconditioner \cite{zampiniBDDC}, also represent a viable alternative, particularly when the conductivity matrix exhibits strong heterogeneity across the domain.

\section{Numerical experiments}\label{sec:experiments}

The code used to conduct the numerical experiments described in this section is public\footnote{Available at \url{https://gitlab.com/petsc/petsc/-/blob/main/src/ts/tutorials/ex30.c}}. It is based on the Portable and Extensible Toolkit for Scientific Computing (PETSc), a software library specifically designed to solve large-scale nonlinear systems of equations arising from partial differential equations in a distributed memory fashion using the Message Passing Interface (MPI) \cite{petsc-user-ref,petsc-web-page,petsc-efficient}. Numerical simulations are performed on the Shaheen III supercomputer at KAUST, an HPE Cray EX system with 4608 AMD-powered compute nodes, tightly connected via a low-latency/high-speed Slingshot network. Each compute node is a dual-socket AMD EPYC 9654 of the Genoa family, with 192 cores per node and 384GB of DDR5 memory. Instructions for running the code and reproducing the results shown in this section are available at \cite{zampini_2025_15162173}.

The management of unstructured grids and the implementation of the finite element discretization are based on the {\tt DMPLEX} infrastructure \cite{knepley2013achieving,lange2015flexible}, which supports both simplicial and tensor-product cells, facilitating the development of dimension-independent computational kernels. Time integration is handled through the {\tt TS} module \cite{abhyankar2018petsc}, enabling seamless switching between time integrators for experimentation. Adaptive time-stepping is performed using digital filtering–based techniques \cite{soderlind2003digital}, combined with local truncation error estimation via extrapolation.

Unless otherwise stated, the nonlinear systems arising from the time-stepping schemes are solved to a tight absolute tolerance of \(10^{-14}\). The associated Jacobian linear systems are solved using right-preconditioned GMRES \cite{saad1986gmres} with the Schur complement–based preconditioner defined in \eqref{eq:schurPre}, where the inverse of the Schur complement is approximated by a smoothed aggregation AMG preconditioner \cite{adams2003parallel}. Linear solver tolerances are dynamically adjusted using the Eisenstat–Walker strategy \cite{ew96}, which avoids oversolving in the early Newton iterations and progressively tightens the accuracy as convergence is approached. A standard cubic backtracking line search on the residual function is employed to ensure global convergence.

\begin{figure}[htbp]
\centering
\begingroup
\setlength{\tabcolsep}{\figtabsep}
\begin{tabular}{c c}
\includegraphics[width=0.3\textwidth,trim={60 40 60 40},clip]{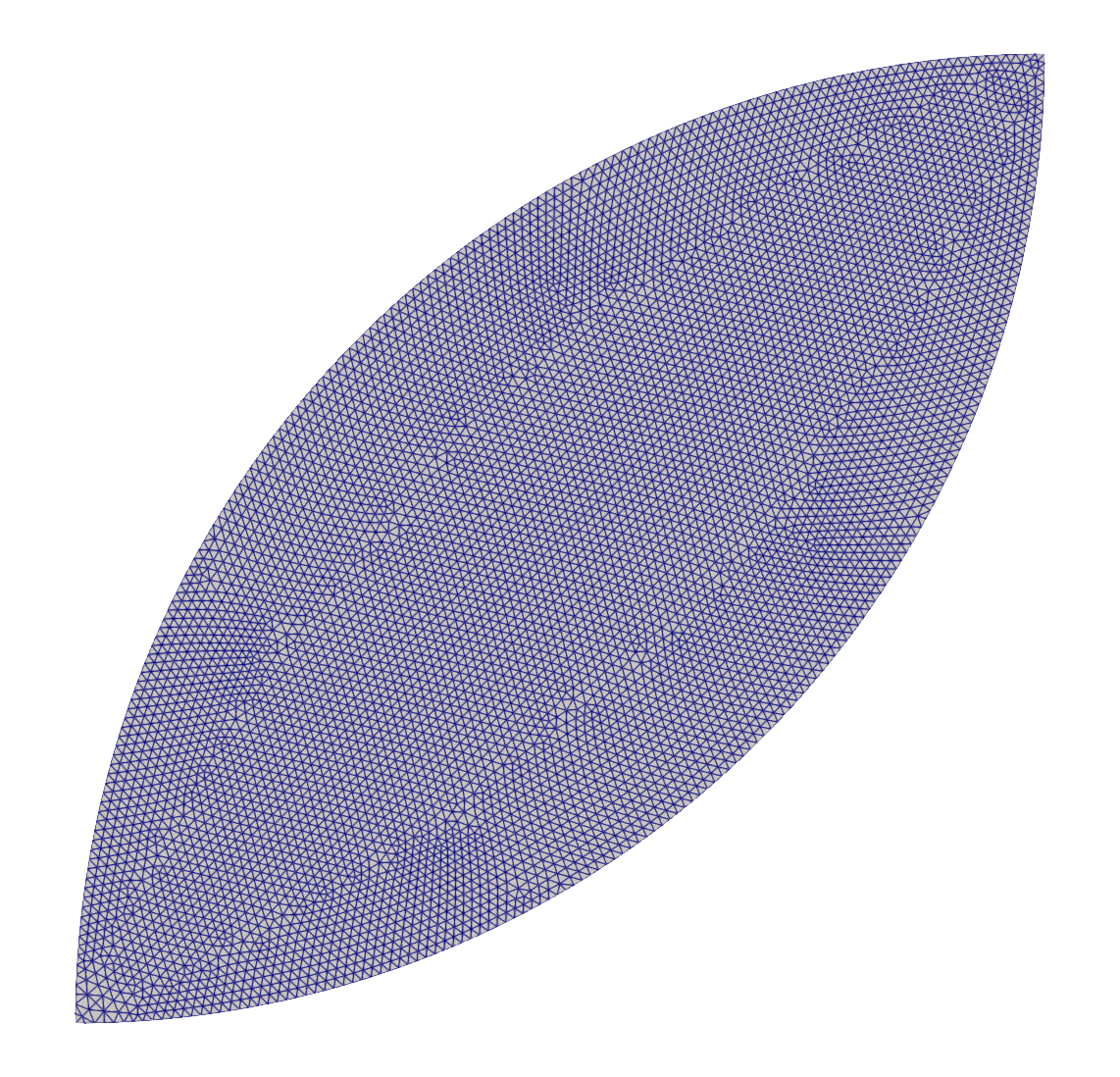} &
\includegraphics[width=0.3\textwidth,trim={60 40 60 40},clip]{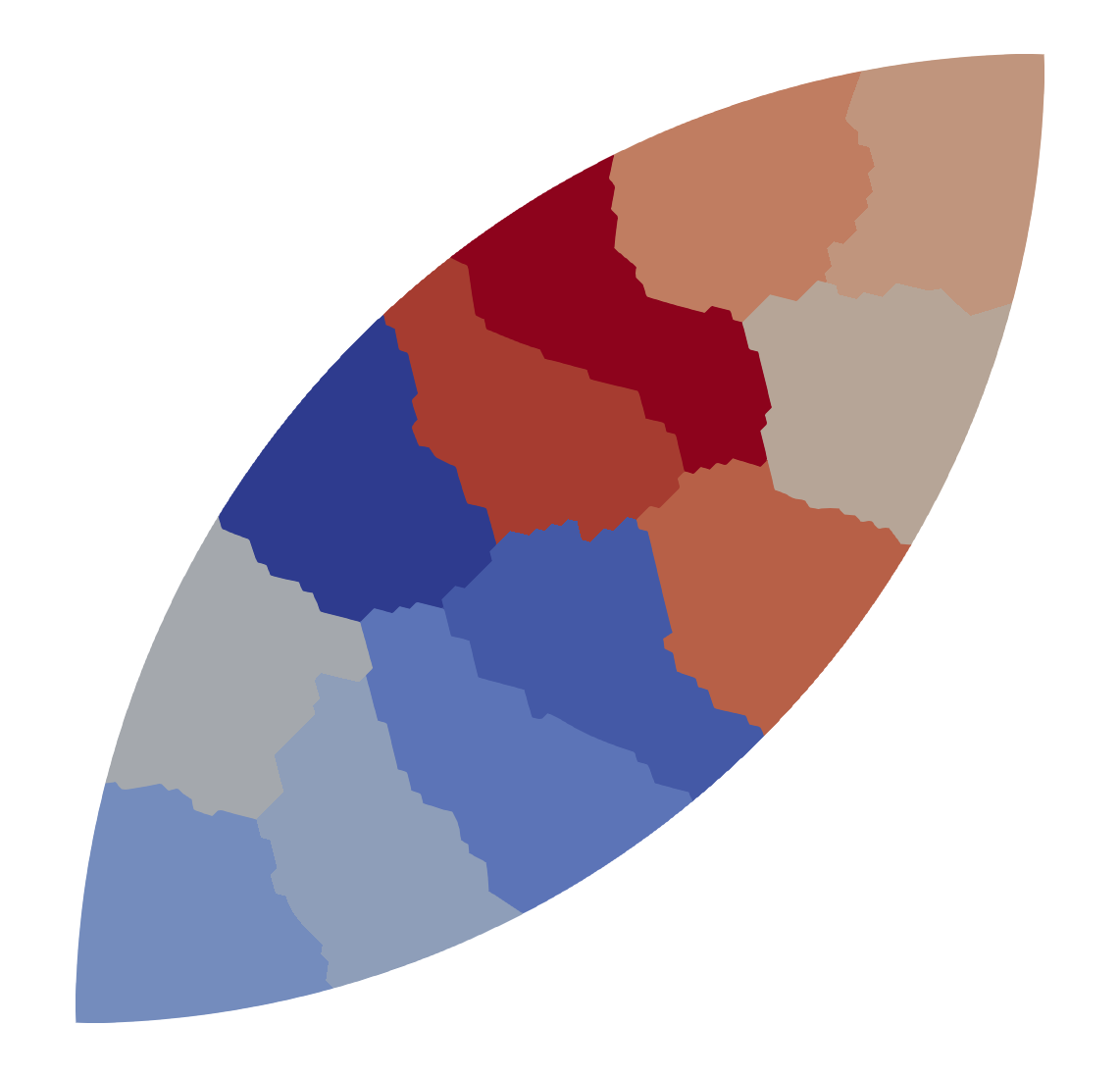}
\end{tabular}
\endgroup
\caption{Left: leaf mesh,  right: example decomposition with 12 subdomains.}
\label{fig:leaf_mesh}
\end{figure}

\subsection{Parallel scalability}\label{sec:scalability}

\begin{figure}[htbp]
\centering
\includegraphics[width=0.8\textwidth]{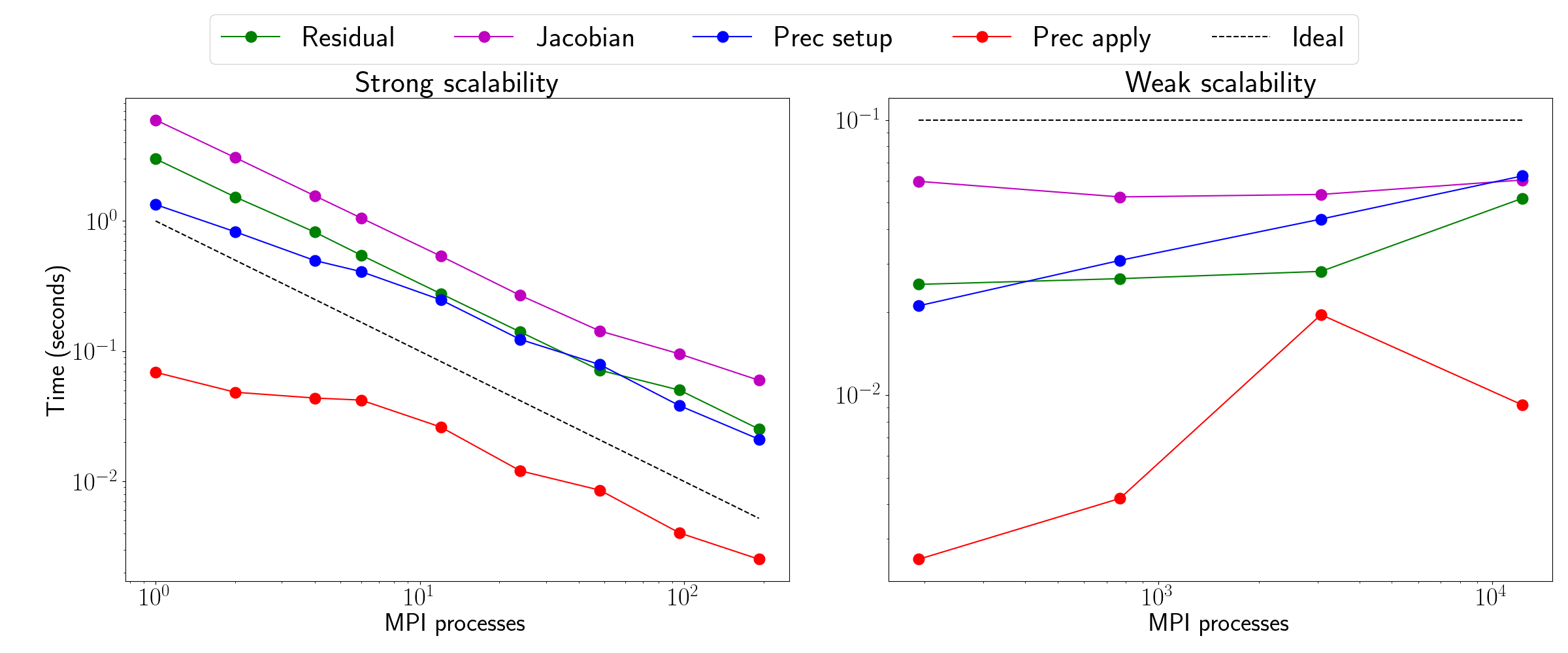}
\caption{Parallel scalability. Wall-clock timings for strong (left panel) and weak (right panel) scaling of the solver's components (see legends for color coding) compared against the ideal scaling (black dashed lines). See \Cref{sec:scalability} for additional details.}
\label{fig:scalability}
\end{figure}

We first report on the parallel scalability of the main components of the solver, namely the evaluation of the nonlinear residual \(F\) \eqref{eq:ifunction} and the Jacobian \(J\) \eqref{eq:ijacobian}, as well as the setup and application of the Schur complement–based preconditioner \eqref{eq:schurPre}. For these tests, we run the backward Euler solver for five time steps and report average wall-clock timings in \Cref{fig:scalability}. 

The left panel shows the results of a strong scaling experiment, in which the mesh is fixed and obtained by three uniform refinements of the leaf mesh\footnote{Available at \url{https://gitlab.com/petsc/datafiles/-/blob/main/meshes/leaf.h5}} shown in \Cref{fig:leaf_mesh}, generated using the ngsPETSc package \cite{betteridge2024ngspetsc}. The number of MPI processes is increased from 1 to 192, resulting in a total of 2.9 million degrees of freedom. The right panel reports the results of a weak scaling experiment, where both the mesh resolution and the number of MPI processes are increased proportionally. In this case, the number of degrees of freedom ranges from 2.9 million to 186.5 million, while the number of processes is increased from 192 to 12\,288, keeping the workload per process approximately constant at about 15\,000 degrees of freedom.

All solver components exhibit near-ideal scaling over the range of process counts considered, in both the strong and weak scaling regimes. The remaining factors that could potentially affect the scalability of the overall algorithm are the number of linear and nonlinear iterations required by the Newton solver. In the remainder of this section, we demonstrate, through extensive numerical experiments, that these iteration counts are essentially independent of both the mesh resolution and the number of processes and therefore do not affect the observed scaling behavior.

\subsection{Network formation}\label{sec:sequence_formation}

In this section, we report on the network formation process using the backward Euler integrator. For this experiment, we fix the model parameters to $\varepsilon=10^{-5}$,  $r=10^{-4}$, $\gamma=0.75$, and $\nu=0.03$. The domain is the unit square $\Omega=[0,1]^2$, discretized with a structured quadrilateral mesh with 1024 cells for each dimension; the total number of dofs is  4.2 million. The initial conductivity $\C_0(x)$ is chosen to be the identity matrix. The simulations are run up to the final time $T=200$, and we use the Gaussian source
\begin{equation}\label{eq:source}
S(x) = S_0(x) - \int_\Omega S_0(x) ~\dx, \quad S_0(x) = e^{-500\Norm{x-x_0}^2},
\end{equation}
with $x_0=(0.25,0.25)$.

\begin{figure}[htbp]
\centering
\begingroup
\setlength{\tabcolsep}{\figtabsep}
\begin{tabular}{c c c}
\includegraphics[width=0.32\textwidth,trim={10 0 80 35},clip]{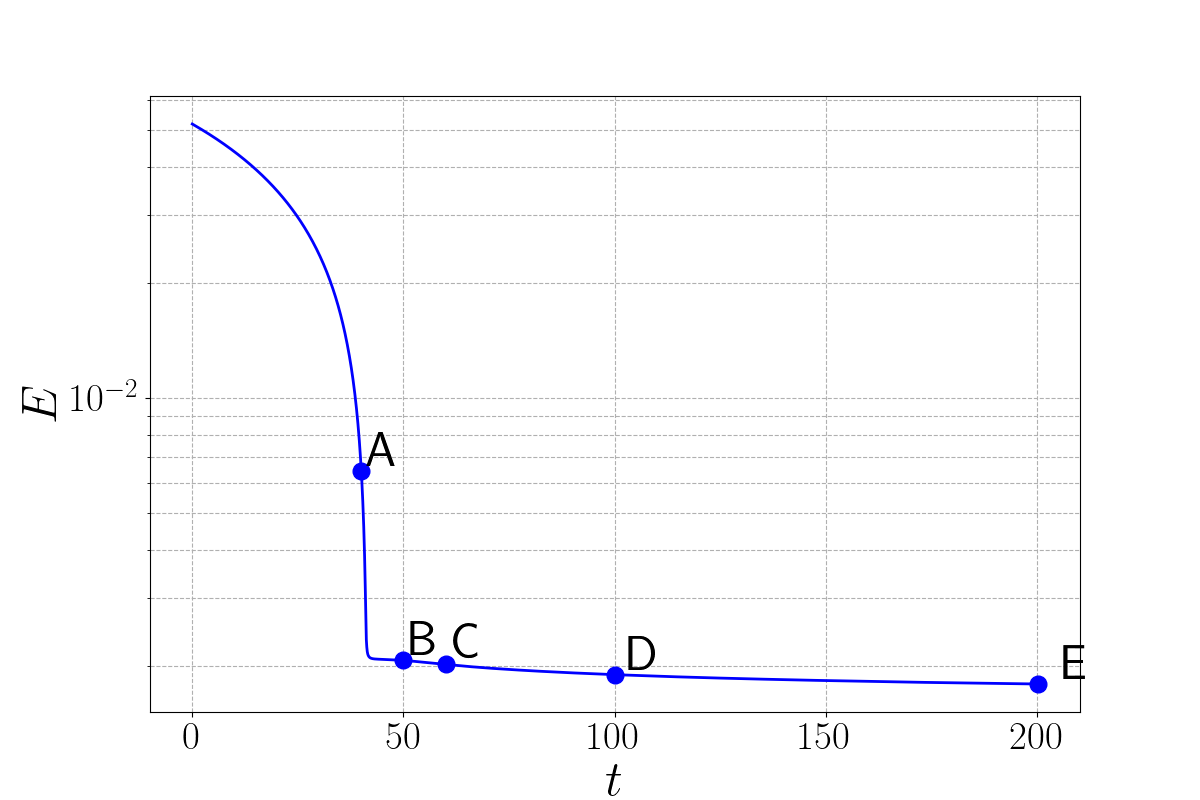} &
\includegraphics[width=0.32\textwidth,trim={10 0 80 35},clip]{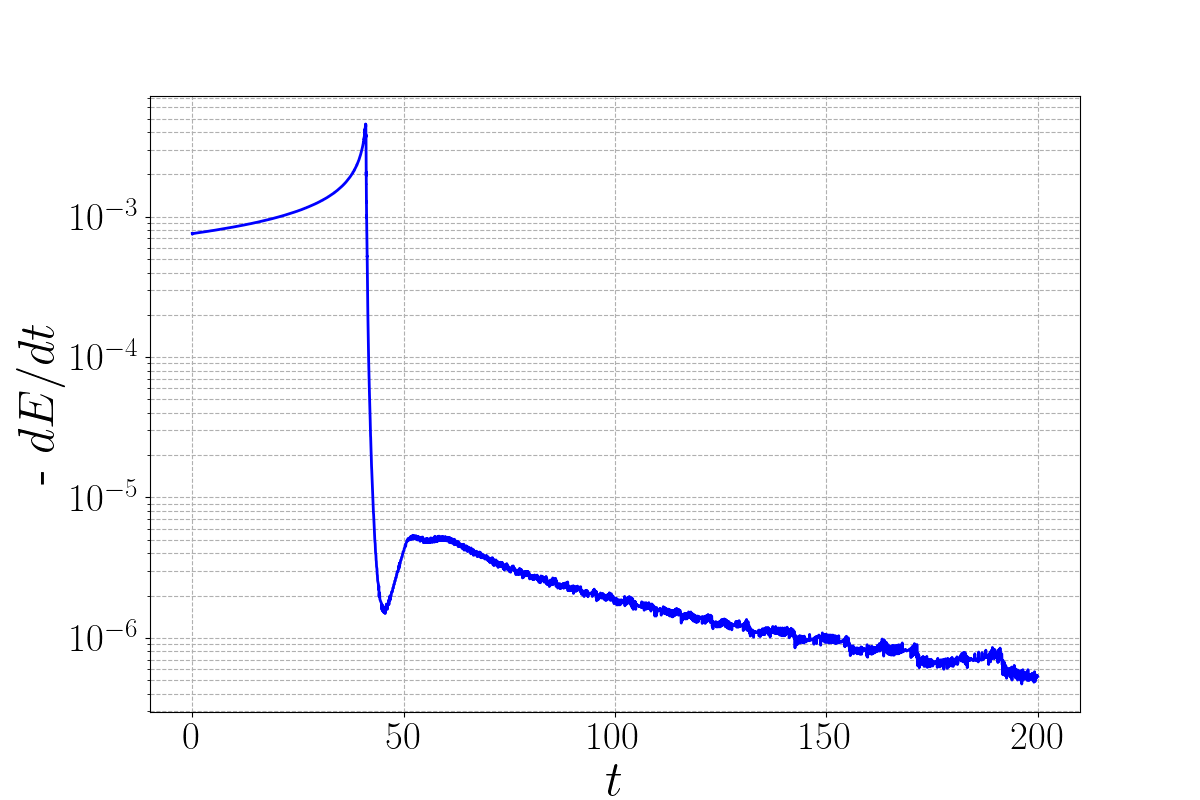} &
\includegraphics[width=0.32\textwidth,trim={10 0 80 35},clip]{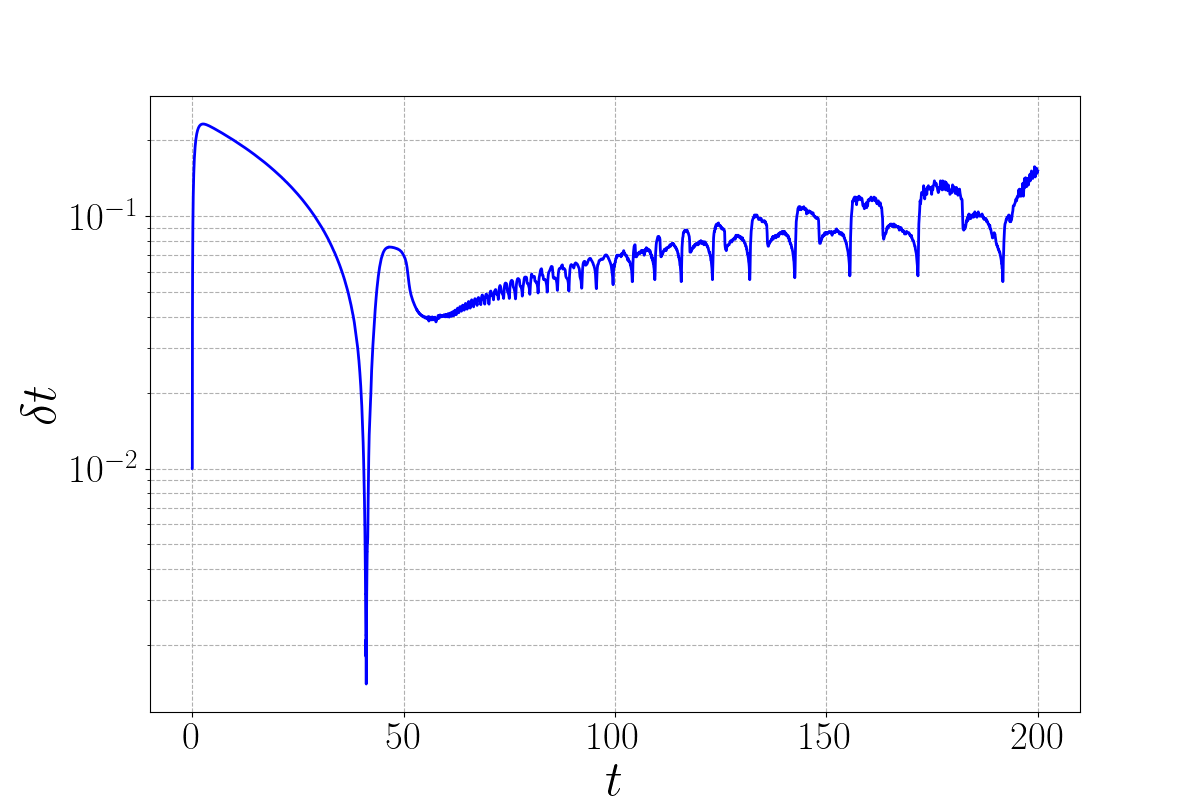}
\end{tabular}
\endgroup
\caption{Network formation. Energy ($E$, left panel), its negative time derivative ($-dE/dt$, center panel), and the time step ($\delta t$, right panel) in logarithmic scale as a function of simulated time. See \Cref{sec:sequence_formation} for additional details. }
\label{fig:box_sequence_logs}
\end{figure}

\begin{figure}[htbp]
\centering
\begingroup
\setlength{\tabcolsep}{\figtabsep}
\begin{tabular}{c c c c c}
\bf{A} & \bf{B} & \bf{C} & \bf{D} & \bf{E}\\
\includegraphics[width=0.19\textwidth,trim={70 30 30 5},clip]{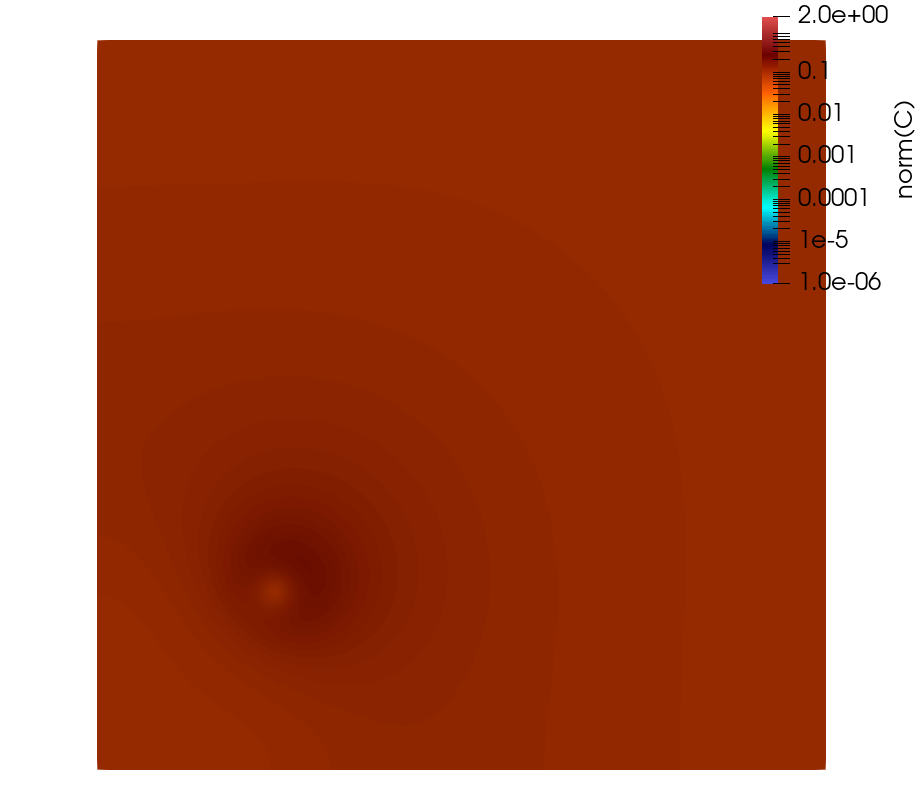} & 
\includegraphics[width=0.19\textwidth,trim={70 30 30 5},clip]{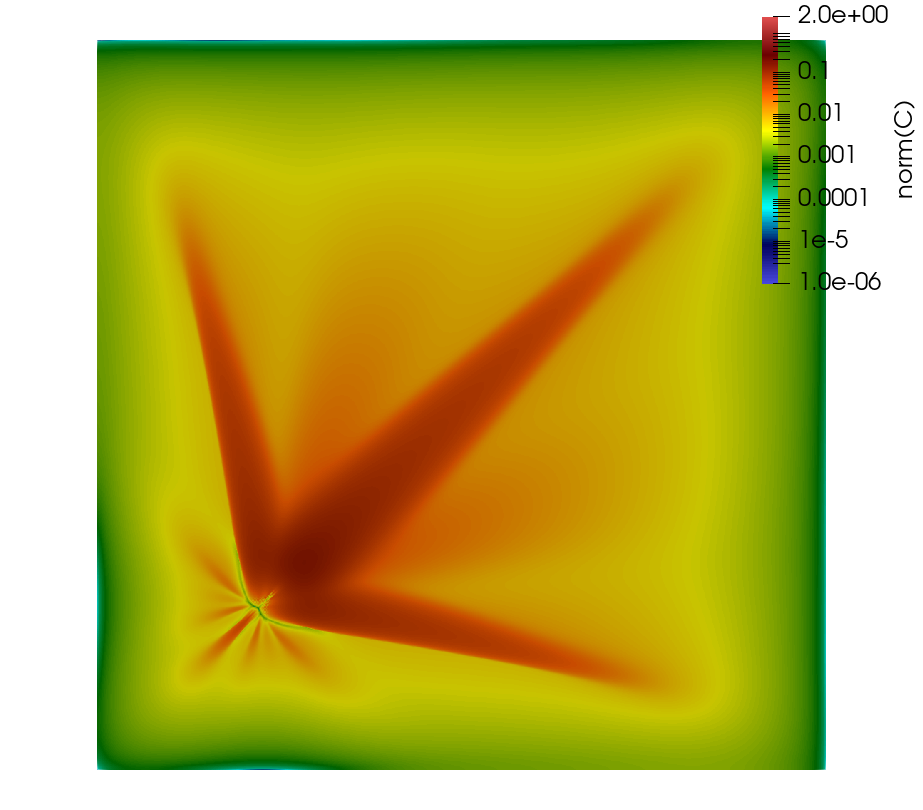} & 
\includegraphics[width=0.19\textwidth,trim={70 30 30 5},clip]{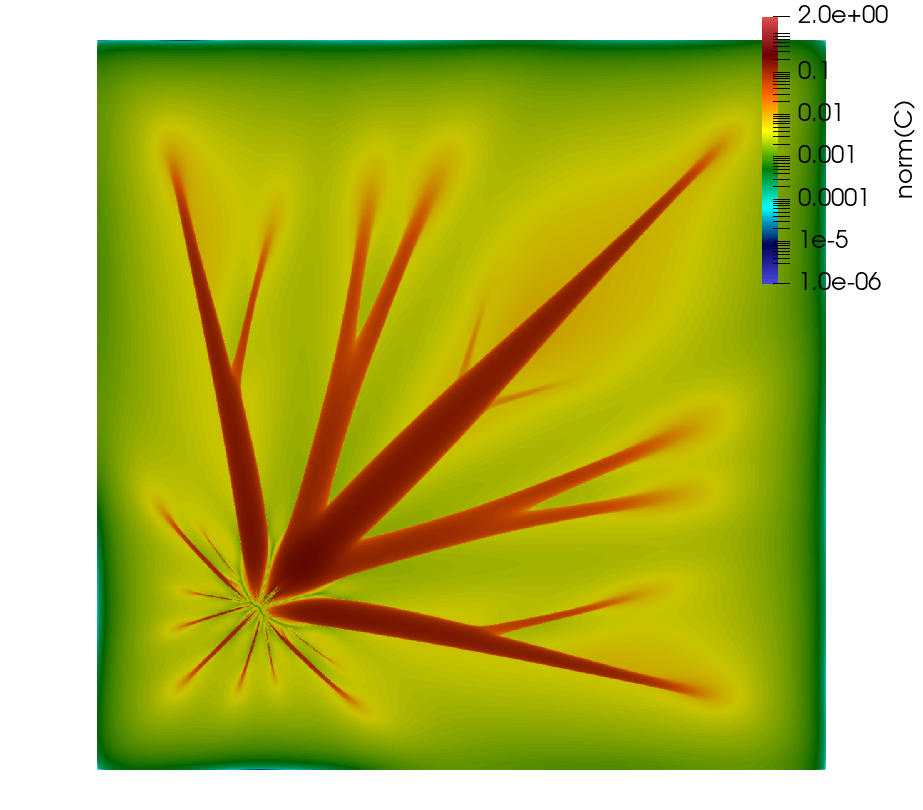} & 
\includegraphics[width=0.19\textwidth,trim={70 30 30 5},clip]{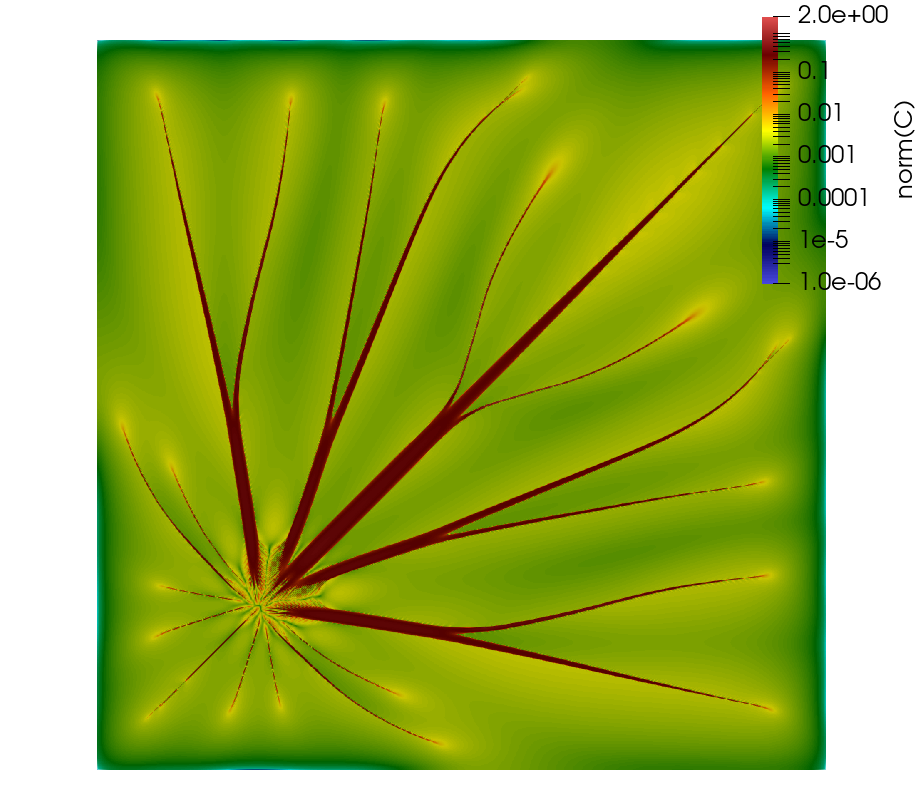} & 
\includegraphics[width=0.19\textwidth,trim={70 30 30 5},clip]{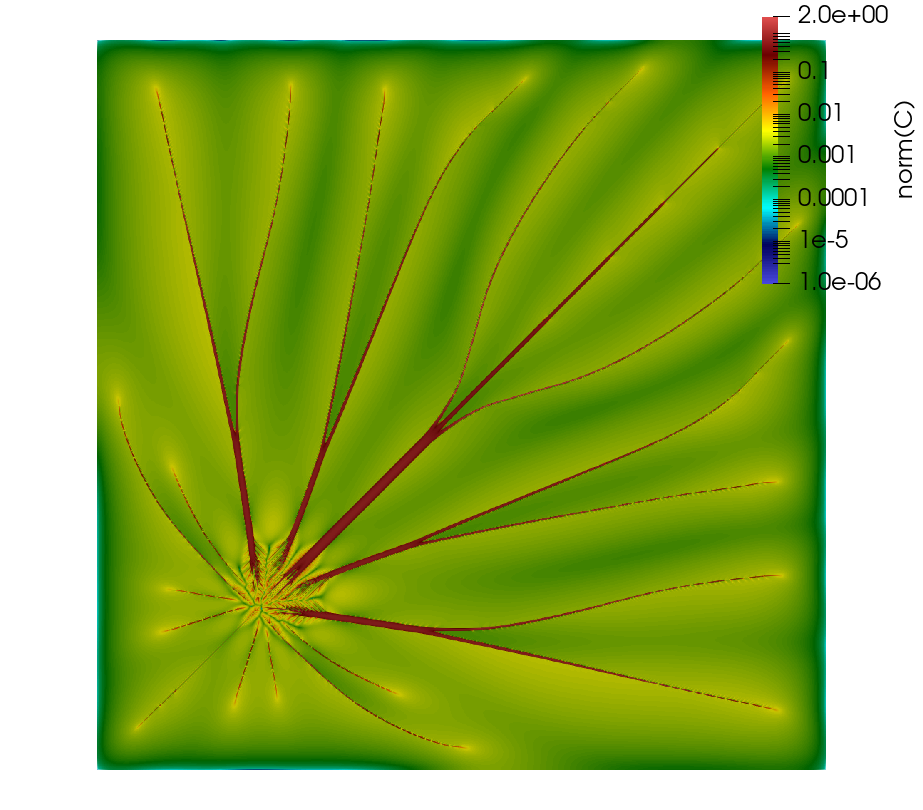} 
\end{tabular}
\endgroup
\caption{Network formation. $\Norm{\C_h}$ in logarithmic scale at selected time instances (see labels A--E in the left panel of \Cref{fig:box_sequence_logs}). See \Cref{sec:sequence_formation} for additional details. }
\label{fig:box_sequence_imgs}
\end{figure}

\Cref{fig:box_sequence_logs} shows the energy of the system ($E$, left panel), the negative time derivative of the energy ($-\frac{dE}{dt}$, central panel), and the time step ($\delta t$, right) taken by the solver in a semi-logarithmic scale as a function of the simulated time.
In \Cref{fig:box_sequence_imgs}, we present snapshots of $\Norm{\C_h}$ at selected time instances, as indicated in the text boxes (labeled from A to E) in \Cref{fig:box_sequence_logs}. These plots are displayed on a logarithmic scale to fully appreciate the network ramifications. 

The system's energy is monotonically decreasing; we observe a rapid decay in the early phases of the simulation,  as the conductivity field evolves from an initially homogeneous state to a spatially heterogeneous configuration (see panels A and B in \Cref{fig:box_sequence_imgs}), allowing for relatively large time steps to be taken. During the early stage, a smooth pattern in $\C_h$ is formed due to the positive tensor-product term $\nabla p_h\otimes\nabla p_h$ in \eqref{eq:flow} and the energy variations are associated with the $\nabla p_h \cdot ( \C_h + r \mathbb{I} )\ \nabla p_h$ term. After this initial phase, the network begins to form (see panels B and C in \Cref{fig:box_sequence_imgs}) due to the interplay of the positive tensor product term and the negative term associated with the metabolic cost $- \nu (\Norm{\C_h }^2 + \varepsilon)^\frac{\gamma-2}{2} \C_h$; energy variations are smaller, and the time step automatically selected by the solver is also decreased. In the final part of the simulation, the energy continues to decrease toward its steady state at a slower pace, with minimal variations in the branching structure (see panels D and E). Consequently, relatively larger time steps can be taken. The network patterns at all time instances are symmetric across the domain bisector, confirming the robustness of the fully implicit solver. The positive semidefiniteness of $\C_h$ is preserved during the time-stepping process (see \Cref{sec:ode_robust}).

\subsection{Robustness with respect to the mesh resolution}\label{sec:be_robust}

\begin{figure}[htb]
\centering
\begingroup
\setlength{\tabcolsep}{\figtabsep}
\begin{tabular}{c c c}
\includegraphics[width=0.32\textwidth,trim={10 0 80 35},clip]{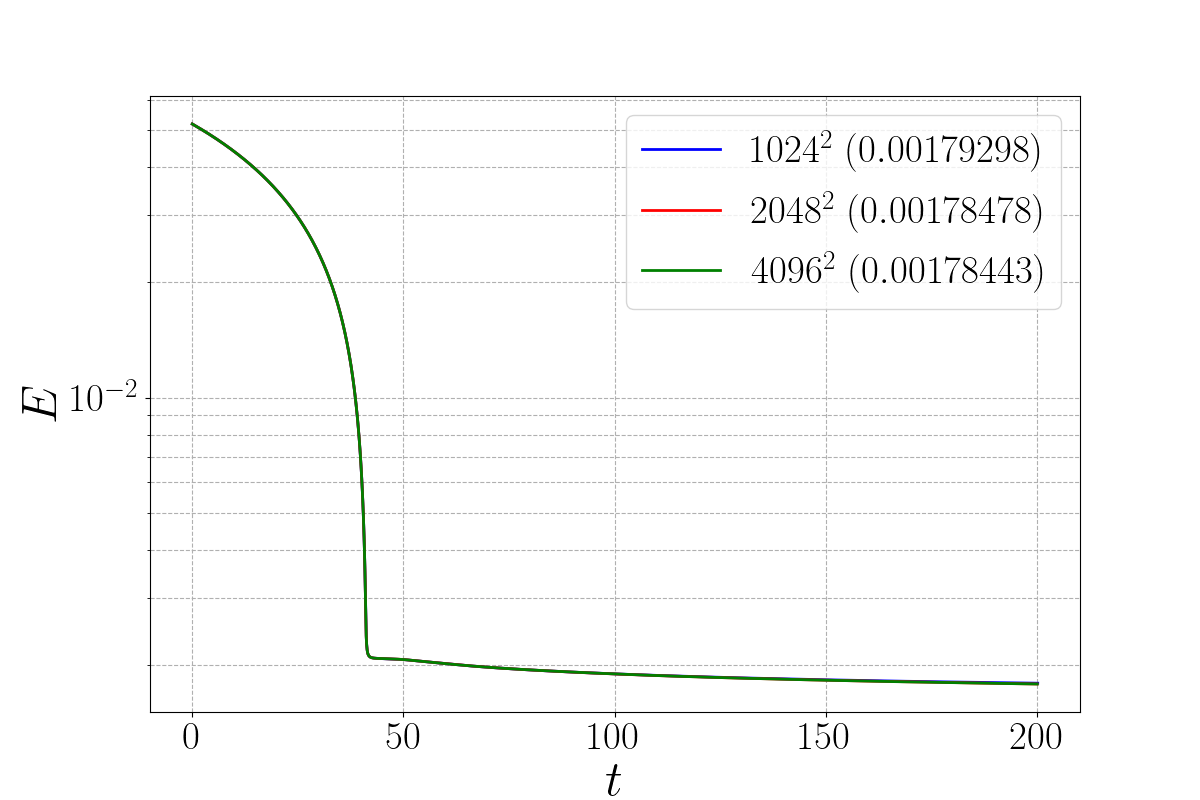} &
\includegraphics[width=0.32\textwidth,trim={10 0 80 35},clip]{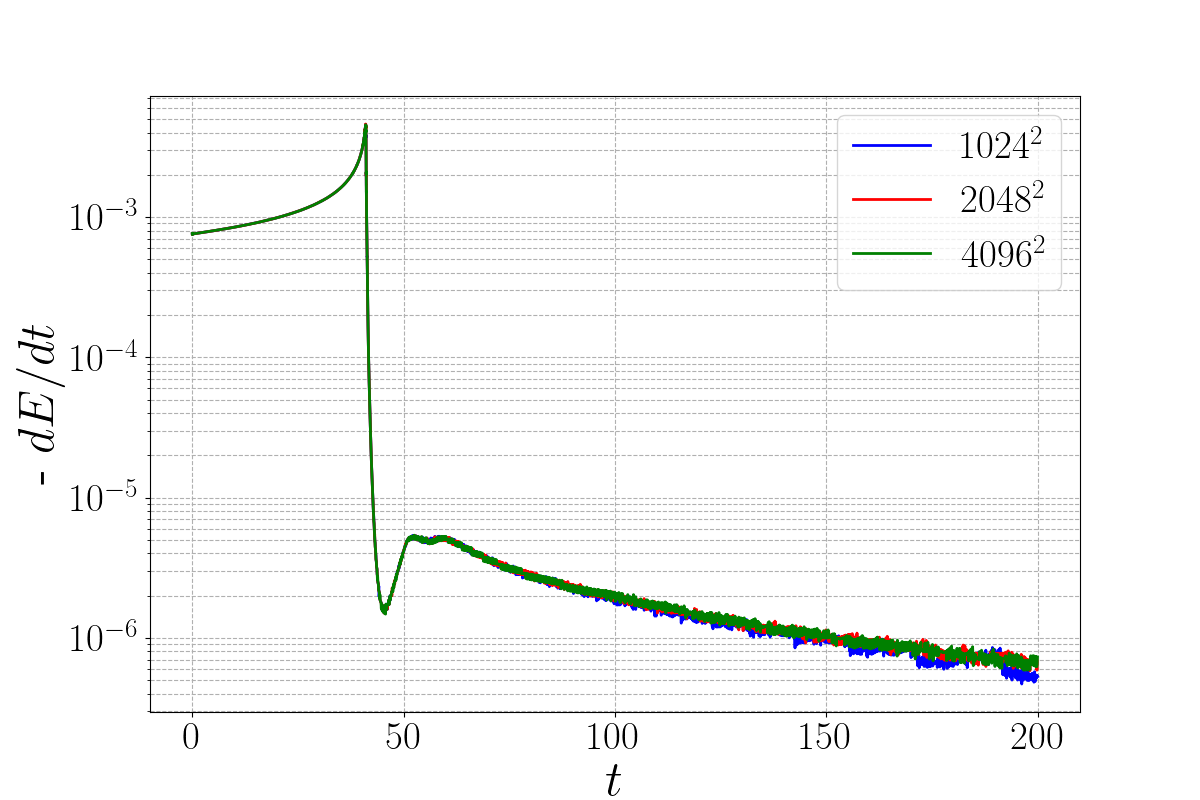} &
\includegraphics[width=0.32\textwidth,trim={10 0 80 35},clip]{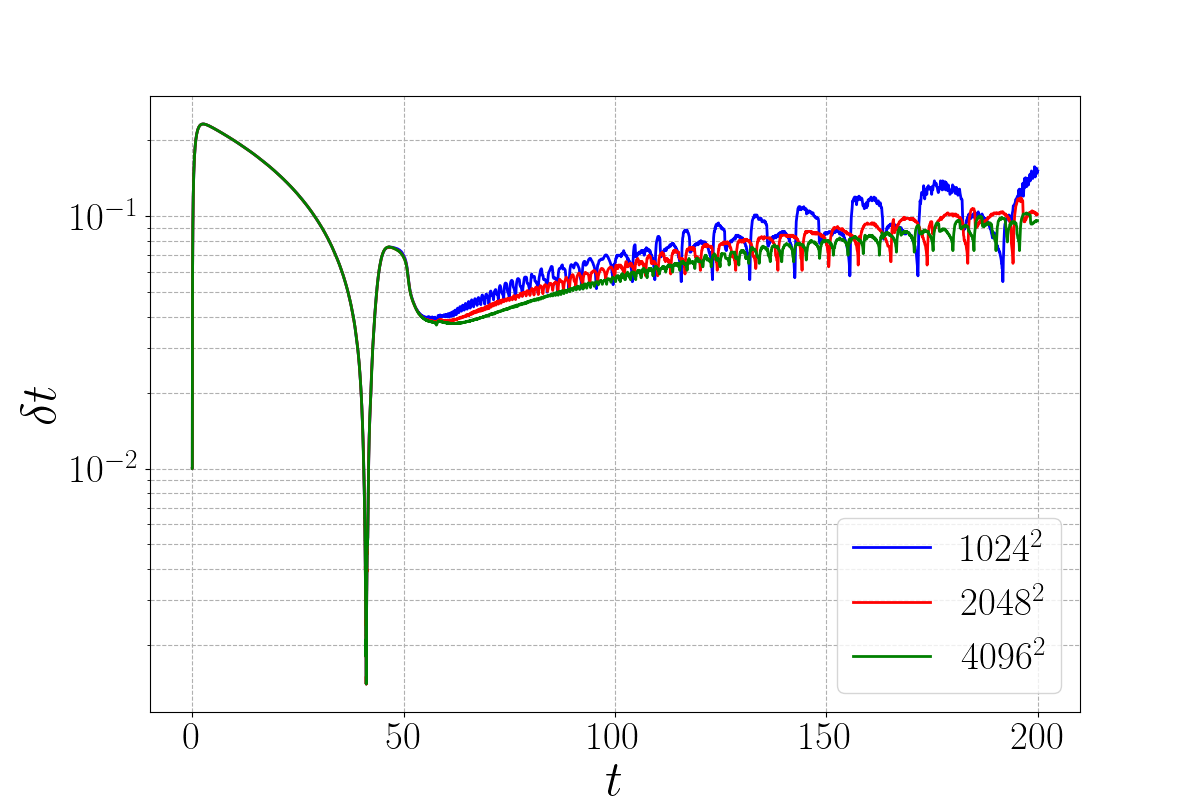}
\end{tabular}
\endgroup
\caption{Robustness with respect to the mesh resolution. Energy ($E$, left panel), its negative time derivative ($-dE/dt$, center), and the time step ($\delta t$, right) in logarithmic scale as a function of the simulated time for $1024^2$ (blue), $2048^2$(red), and $4096^2$ (green) mesh. See \Cref{sec:be_robust} for additional details. }
\label{fig:box_ref_logs}
\end{figure}

In the next set of experiments, we study the robustness of the backward Euler solver with respect to the mesh resolution. For these experiments, we use the same model parameters as in \Cref{sec:sequence_formation} and consider three levels of refinements of a structured quadrilateral mesh; the number of cells is $1024^2$, $2048^2$, and $4096^2$, and the corresponding number of dofs is 4.2, 16.8, and 67.1 million, respectively. We consider a weak scaling regime, where the size of the local meshes is kept approximately constant, and use an increasing number of MPI processes: 384, 1536, and 6144.

\Cref{fig:box_ref_logs} shows the energy of the system ($E$, left panel), the negative time derivative of the energy ($-dE/dt$, central panel), and the time step ($\delta t$, right) taken by the solver in a semi-logarithmic scale as a function of the simulated time. The final energy value is reported in parentheses in the legend of the left panel. The energy variation and the final value are similar across all three cases, as are the selected time steps.

\begin{figure}[htbp]
\centering
\begingroup
\setlength{\tabcolsep}{\figtabsep}
\begin{tikzpicture}
\node[anchor=south west, inner sep=0] (img) at (0,0) {
  \begin{tabular}{c c c}
$1024^2$ & $2048^2$ & $4096^2$\\
\includegraphics[width=0.26\textwidth,trim={70 15 30 5},clip]{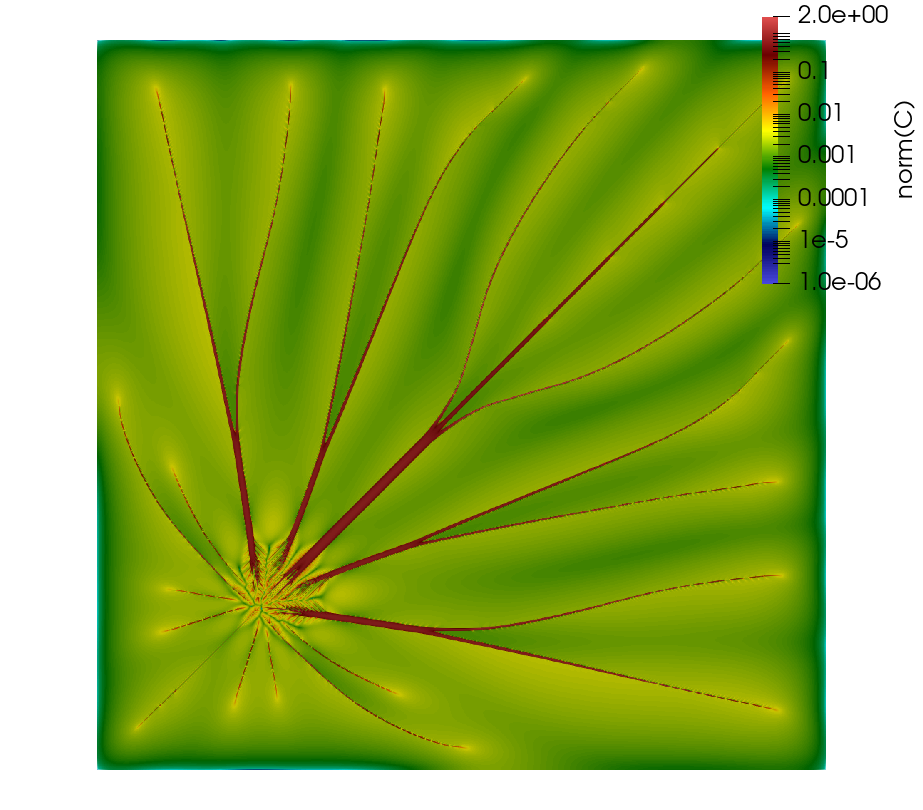} &
\includegraphics[width=0.26\textwidth,trim={70 15 30 5},clip]{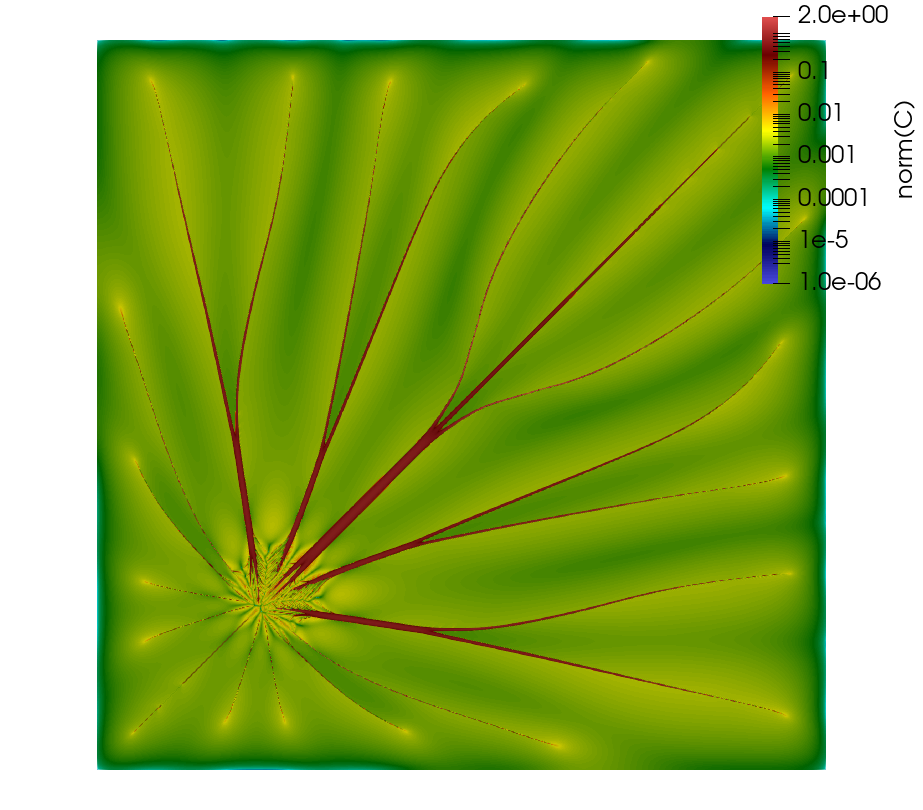} &
\includegraphics[width=0.26\textwidth,trim={70 15 30 5},clip]{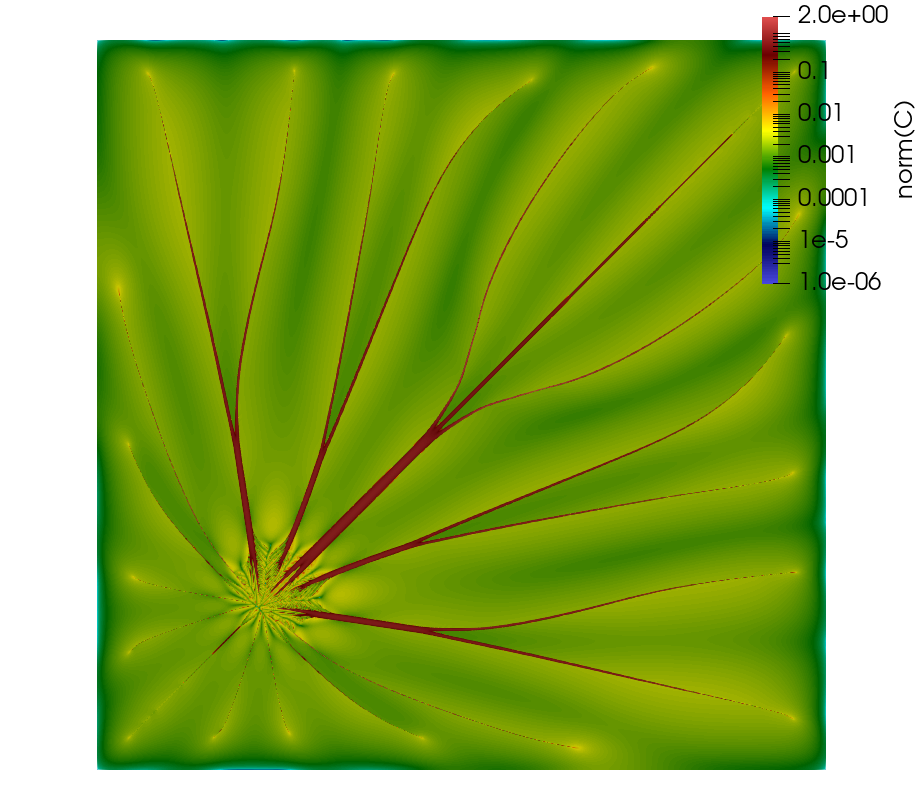} \\
\includegraphics[width=0.26\textwidth,trim={70 15 30 5},clip]{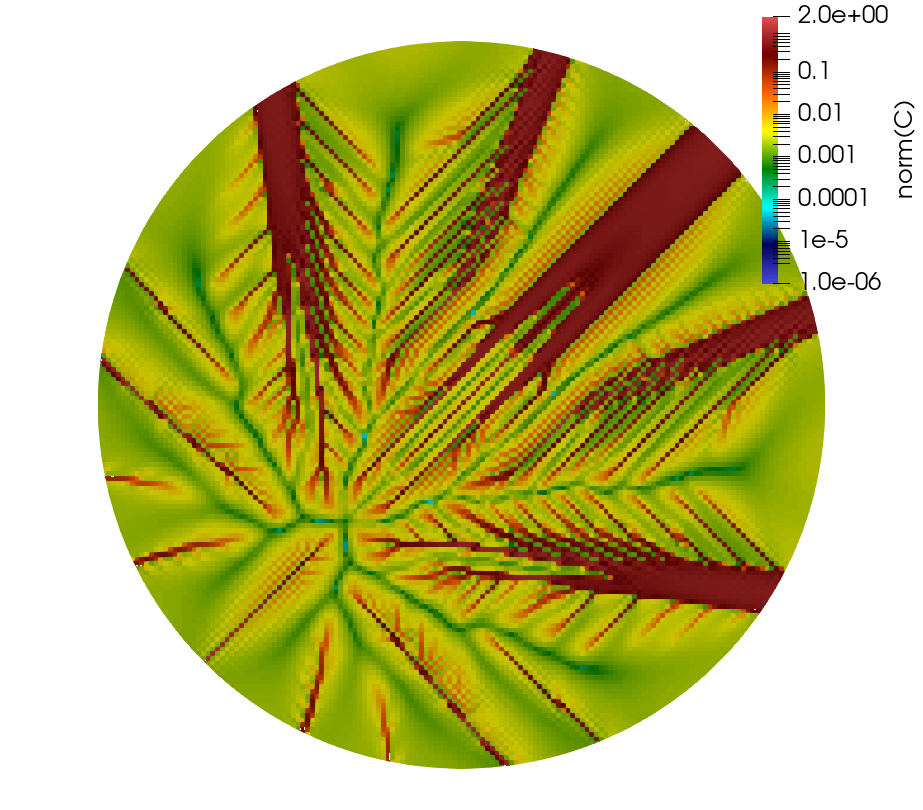} &
\includegraphics[width=0.26\textwidth,trim={70 15 30 5},clip]{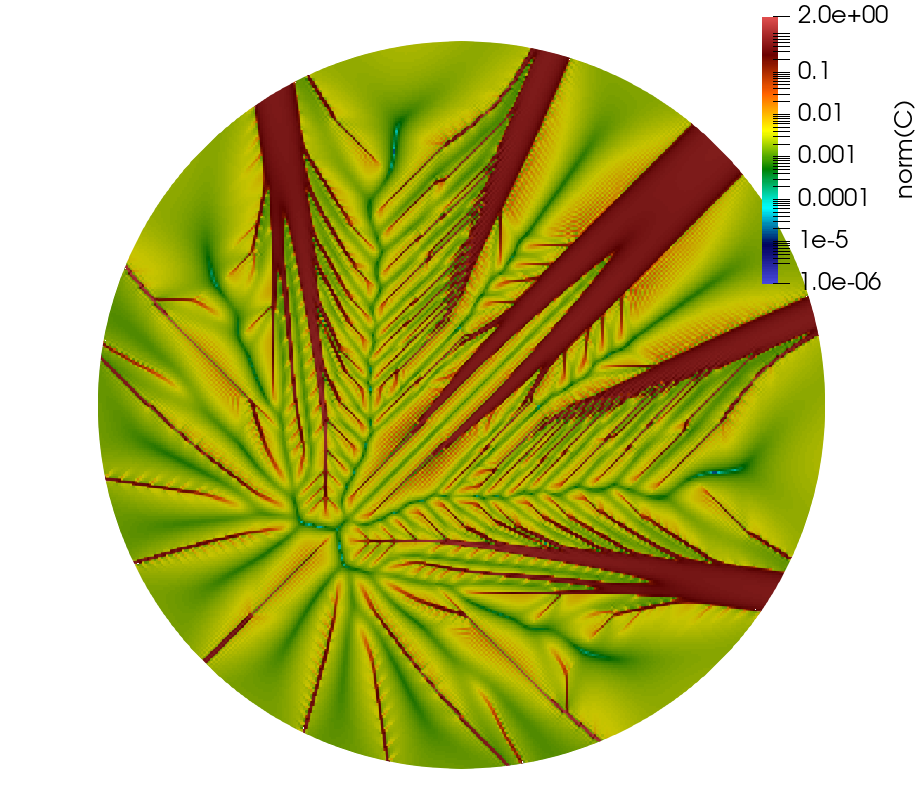} &
\includegraphics[width=0.26\textwidth,trim={70 15 30 5},clip]{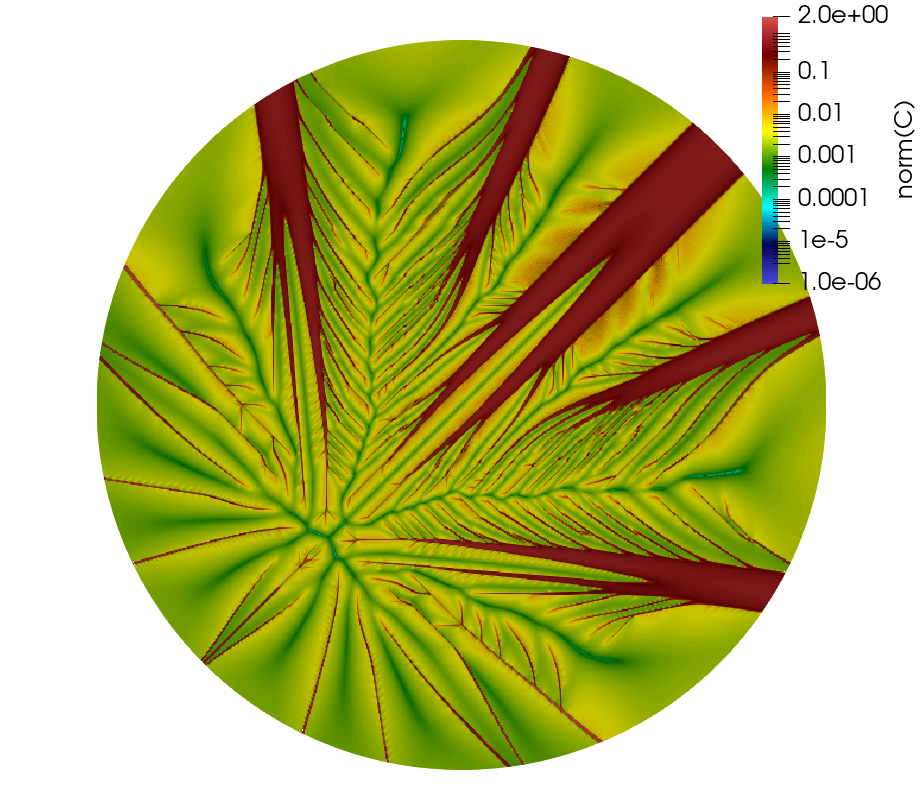}
\end{tabular}
};
\begin{scope}[x=1cm,y=1cm]
    \draw[line width=0.6pt] (1.06,5.0) circle (0.27);
    \draw[line width=0.6pt] (5.11,5.0) circle (0.27);
    \draw[line width=0.6pt] (9.16,5.0) circle (0.27);
    \draw[line width=0.6pt] (0.79,5.05) -- (0.21,2.2);
    \draw[line width=0.6pt] (1.315,5.1) -- (3.21,3.2);
    \draw[line width=0.6pt] (4.84,5.05) -- (4.27,2.2);
    \draw[line width=0.6pt] (5.365,5.1) -- (7.26,3.2);
    \draw[line width=0.6pt] (8.89,5.05) -- (8.35,2.2);
    \draw[line width=0.6pt] (9.415,5.1) -- (11.34,3.2);

  \end{scope}
\end{tikzpicture}
\endgroup
\caption{Robustness with respect to the mesh resolution. Top row: final state of $\Norm{\C_h}$ for $1024^2$ (left), $2048^2$ (center), and $4096^2$ (right) mesh. Bottom row: close-ups at the source location. See \Cref{sec:be_robust} for additional details.}
\label{fig:box_ref_imgs}
\end{figure}

The final states of $\Norm{\C_h}$ in logarithmic scale are shown in \Cref{fig:box_ref_imgs}. The network is well resolved away from the source location at all three resolutions; the robustness of the fully nonlinear solver is evident, as the network patterns are symmetric along the domain bisector. Zooming on the source location (see \Cref{fig:box_ref_imgs}, bottom row), we can observe a checkerboard pattern for $\Norm{\C_h}$ for the coarser mesh that naturally arises due to our choice of a discontinuous space for the conductivity matrix. Such a checkerboard pattern resolves as the mesh size decreases and finer structures form, leading to new checkerboard patterns at finer resolutions. We expect that for $\gamma\leq 1$ the analytical steady states, being refinement limits of sparse (acyclic) graphs, do not possess any definite scale. Therefore, any mesh refinement leads to the appearance of new, finer structures in the solution of the discretized problem, as can be observed in \Cref{fig:box_ref_imgs}. From this perspective, the numerical results reflect the choice of mesh used to solve the discretized problem. Different meshes lead to different network structures, as demonstrated in \Cref{sec:meshdependence}.

\begin{figure}[htbp]
\centering
\begingroup
\setlength{\tabcolsep}{\figtabsep}
\begin{tabular}{c c c}
\includegraphics[width=0.32\textwidth,trim={15 0 80 0},clip]{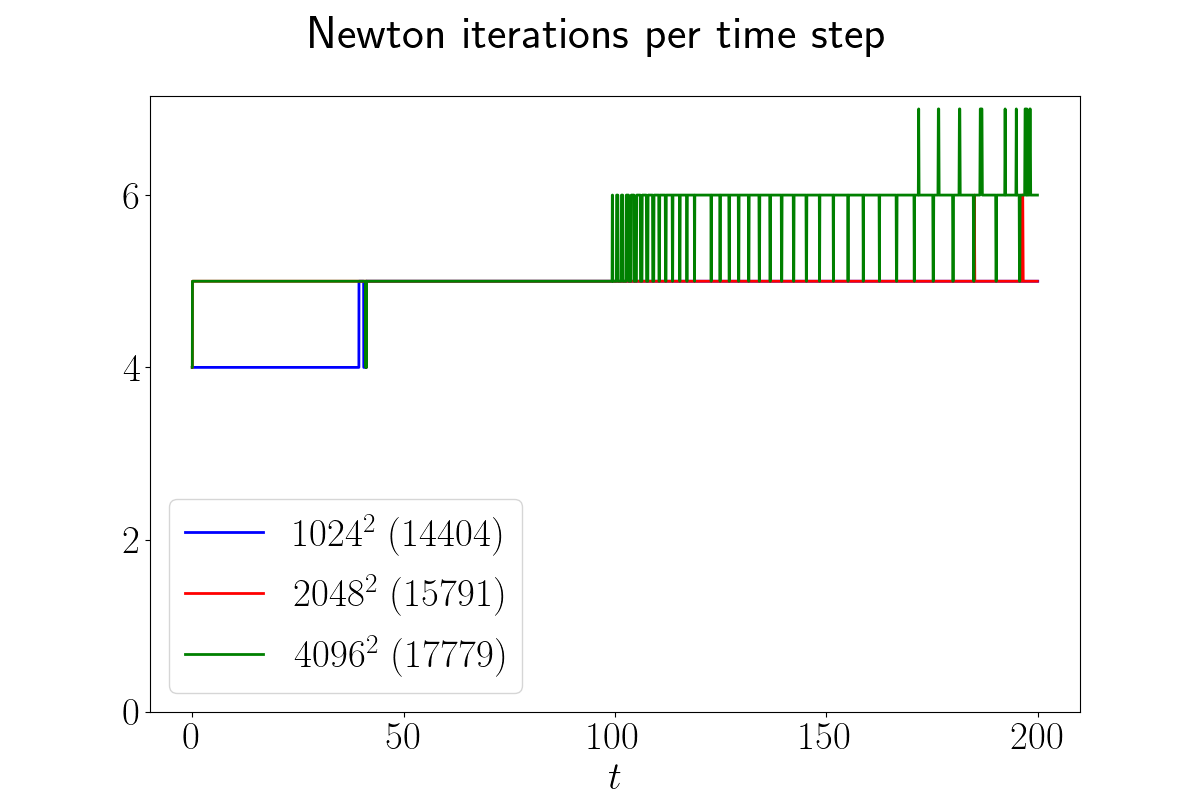} &
\includegraphics[width=0.32\textwidth,trim={15 0 80 0},clip]{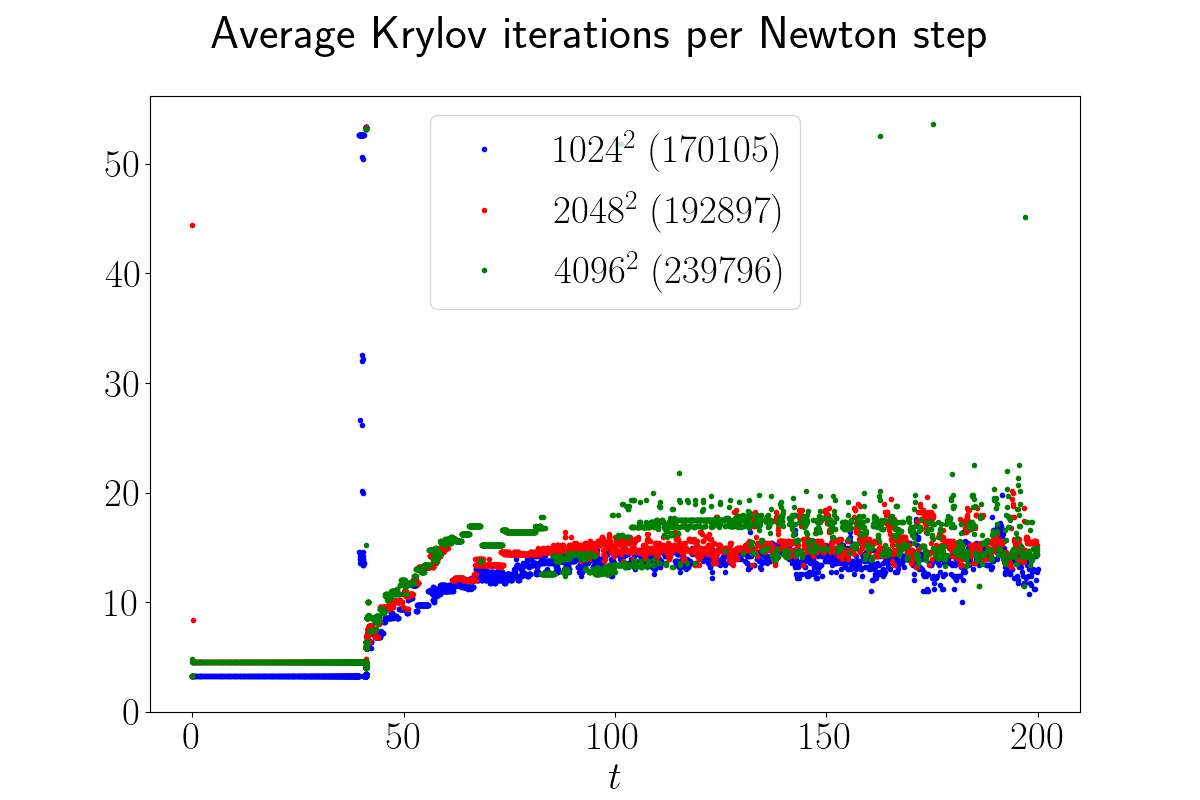} &
\includegraphics[width=0.32\textwidth,trim={15 0 80 0},clip]{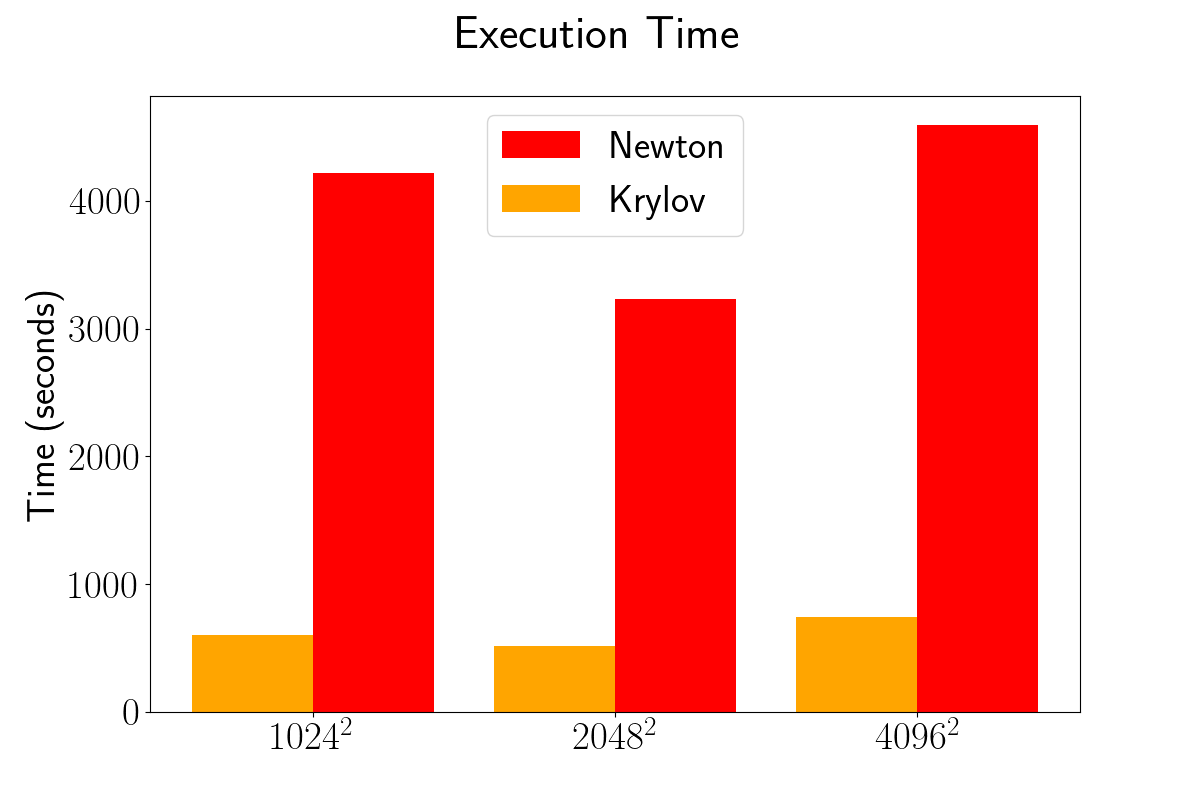}
\end{tabular}
\endgroup
\caption{Robustness with respect to the mesh resolution. Newton iterations (left panel), average number of linear iterations per Newton step (central panel), and wall-clock timings (right panel) for the BE solver on $1024^2$ (blue), $2048^2$(red), and $4096^2$ (green) meshes. The legend in the left panel reports the total number of Newton iterations (in parentheses), while the legend in the central panel reports the total number of linear iterations (in parentheses) for the different resolutions. See \Cref{sec:be_robust} for additional details.}
\label{fig:box_ref_its}
\end{figure}

In \Cref{fig:box_ref_its} we report the number of nonlinear (left panel) and linear (central panel) iterations of the three simulations, together with the time taken by the entire Newton solver and for solving the linearized equations (right). The nonlinearity of the problem is not severe, and the total number of iterations is almost independent of the resolution, with the number of nonlinear iterations per time step ranging from 4 to 7. The robustness of the Schur complement-based preconditioner is confirmed by the results shown in the central panel, where the average number of linear iterations is independent of the time step and almost always below 30 with a few outliers;  linear systems are easy to solve in the early stages of time advancing, while more iterations are needed as the network formation process progresses. The efficiency of the preconditioning step is also confirmed by the wall-clock timings shown in the right panel. These timings scale with the workload per process and are independent of the number of processes, with the linear solver contributing less than 20\% of the total nonlinear solver time.

\subsection{Time integrators comparison}\label{sec:ode_robust}

\begin{figure}[htbp]
\centering
\begingroup
\setlength{\tabcolsep}{\figtabsep}
\begin{tabular}{c c c}
\includegraphics[width=0.32\textwidth,trim={10 0 80 35},clip]{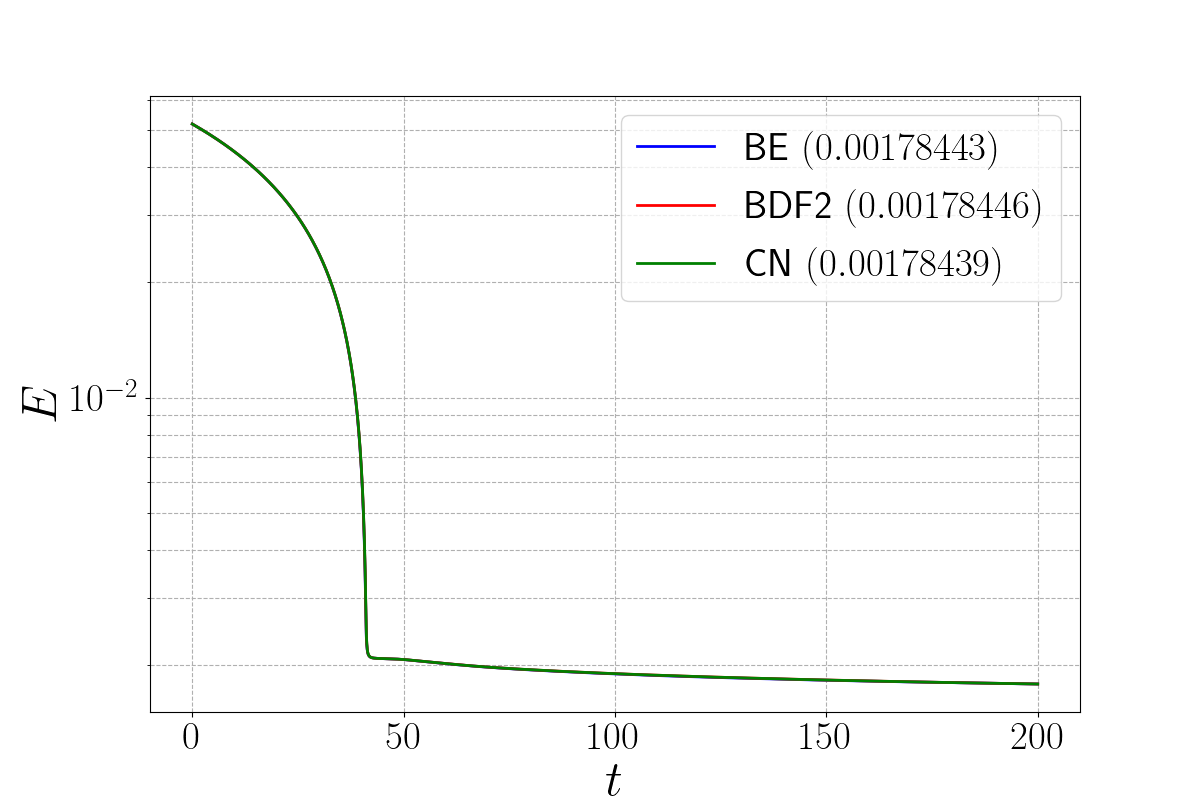} &
\includegraphics[width=0.32\textwidth,trim={10 0 80 35},clip]{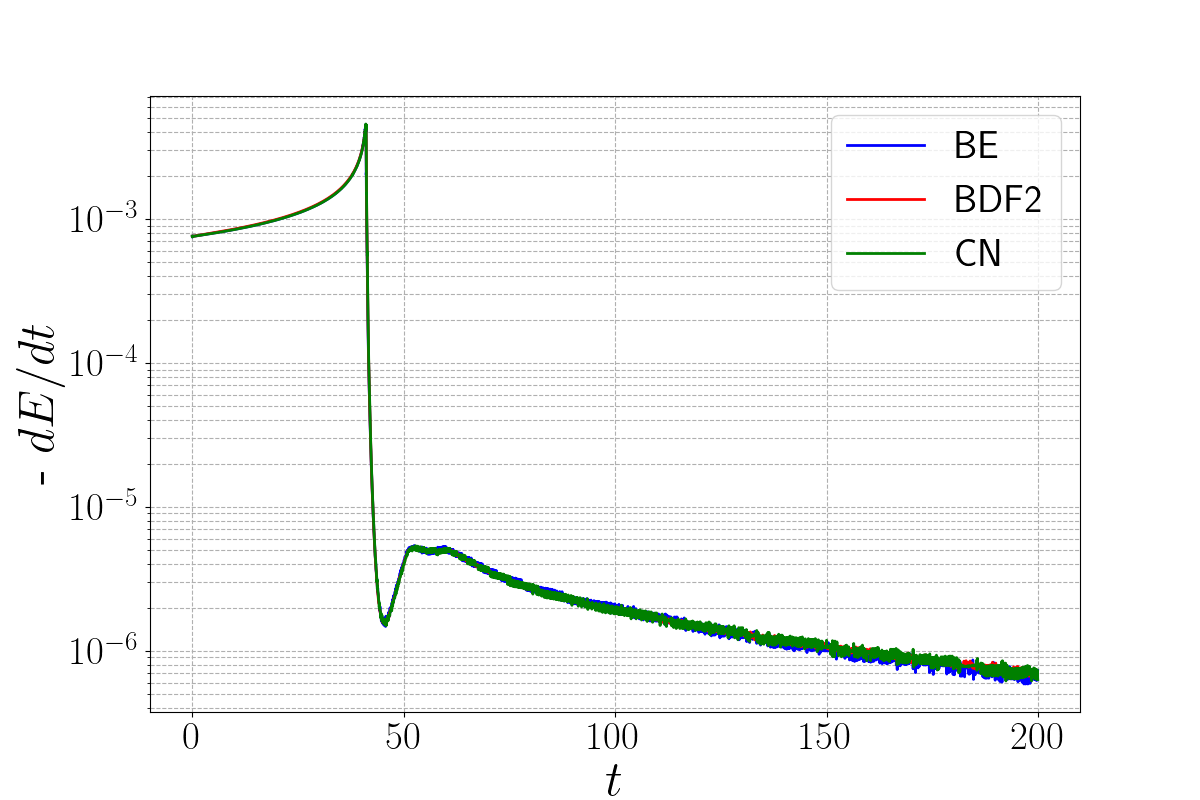} &
\includegraphics[width=0.32\textwidth,trim={10 0 80 35},clip]{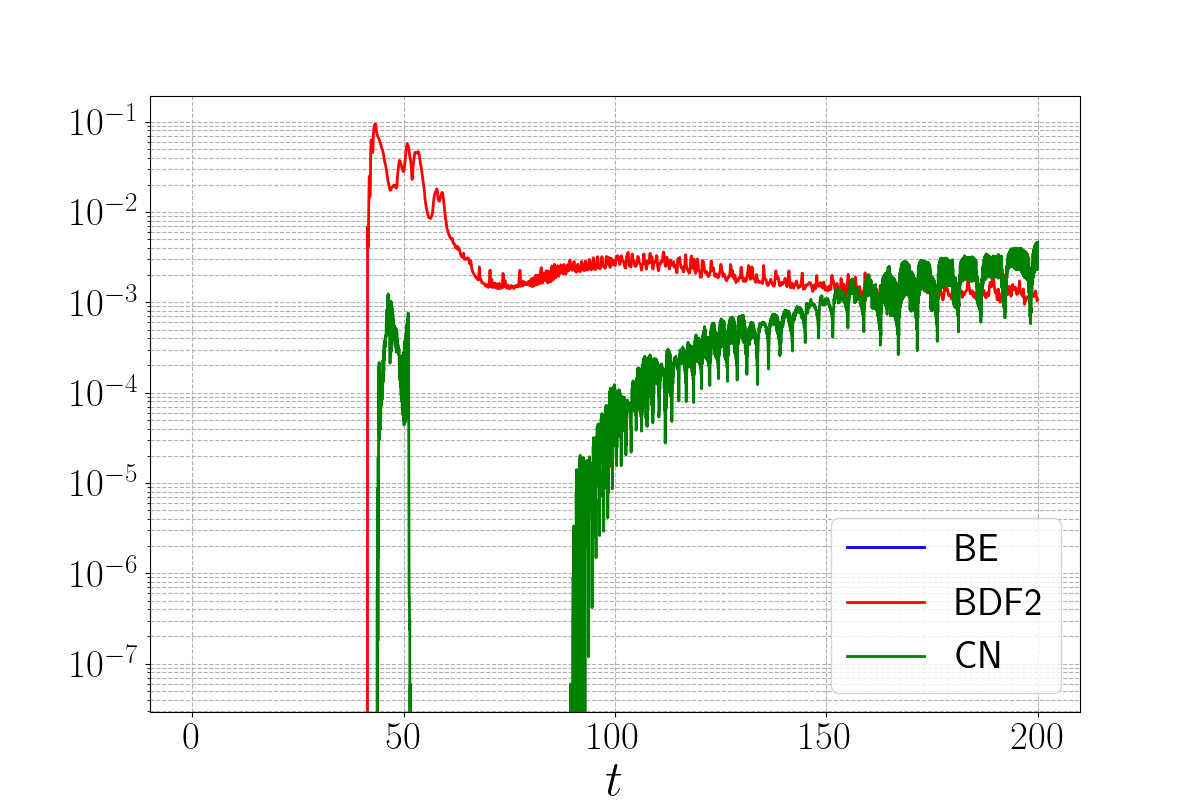}\\
\end{tabular}
\endgroup
\caption{Time integrators comparison. Energy (left panel), $-dE/dt$ (central panel), and $\frac{1}{|\Omega|}\int_\Omega \mathds{1}_{\lambda_{{\text min}}(\C) < 0}\,\dx$ (right panel) as a function of simulated time for CN (green), BDF2 (red) and BE (blue). See \Cref{sec:ode_robust} for additional details. }
\label{fig:ode_logs}
\end{figure}

\begin{figure}[htbp]
\centering
\begingroup
\setlength{\tabcolsep}{\figtabsep}
\begin{tabular}{c c c}
\bf{BE} & \bf{CN} & \bf{BDF2}\\
\includegraphics[width=0.32\textwidth,trim={90 40 30 0},clip]{figures/paper_be_4k.png} &
\includegraphics[width=0.32\textwidth,trim={90 40 30 0},clip]{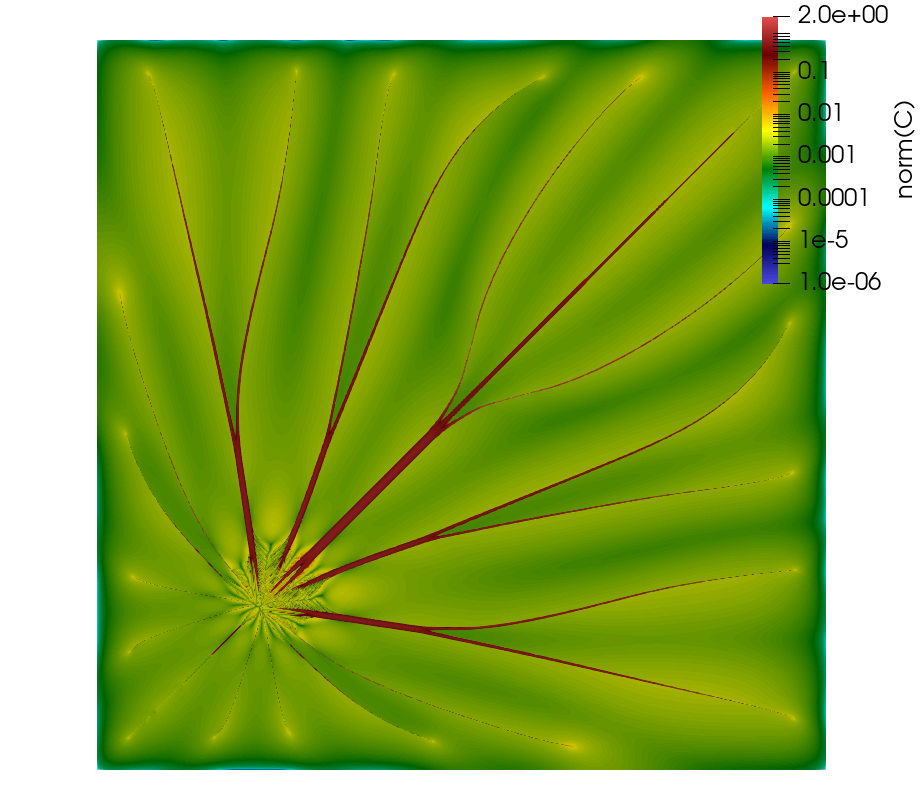} &
\includegraphics[width=0.32\textwidth,trim={90 40 30 0},clip]{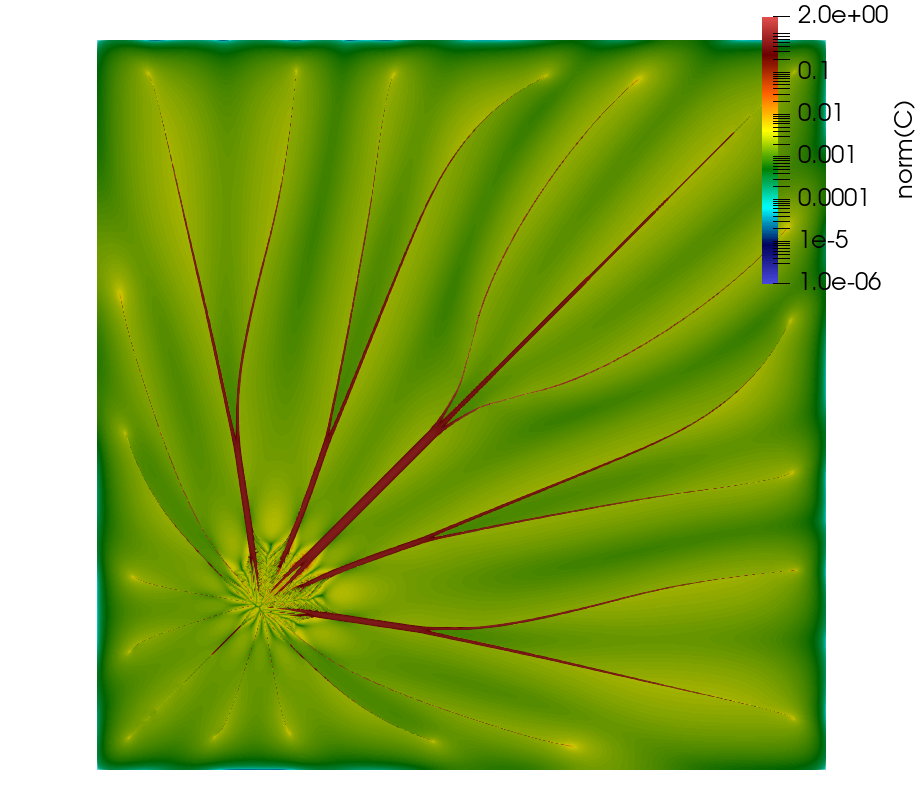} 
\end{tabular}
\endgroup
\caption{Time integrators comparison. Final state of $\Norm{\C_h}$ for BE (left panel), CN (center panel), and BDF2 (right panel). See \Cref{sec:ode_robust} for additional details.}
\label{fig:ode_ref_imgs}
\end{figure}

Next, we analyze the preservation of the structure of the PDE comparing the backward Euler solver (BE) against second-order time integrators, namely Crank--Nicolson (CN) and BDF2. We consider the same model parameters as in \Cref{sec:be_robust} and the finest $4096^2$ mesh. All solvers converge to a similar final energy value (see left panel of \Cref{fig:ode_logs}), with all being able to preserve the energy decay (central panel). While CN and BDF2 allow for larger time steps (data not shown), neither can maintain the positive semidefiniteness of $\C_h$ (see \Cref{sec:time_disc}); to quantify these errors, we compute 
\[
\frac{1}{|\Omega|}\int_\Omega \mathds{1}_{\lambda_{{\text min}}(\C_h) < 0}(x)\,\dx,
\]
where $\mathds{1}_{(\cdot)}$ is the indicator function and $\lambda_{{\text min}}(\C_h)$ is the minimum eigenvalue of $\C_h$. BE never computes solutions that lead to negative eigenvalues of $\C_h$, confirming the results of Lemma \ref{lem:be_posdef}, while BDF2 and CN fail, with BDF2 failing to preserve positive semidefiniteness even in 10\% of the domain. Nevertheless, the negative eigenvalues observed with BDF2 or CN have very small magnitudes and remain below  $r$ (data not shown).
We stress here that, at every evaluation of the residual equations \eqref{eq:semidiscrete}, we ensure the Poisson problem is always well posed by sampling $\C_h + r \mathbb{I}$ at the quadrature nodes, and by shrinking the time step if we encounter any negative eigenvalue.
In \Cref{fig:ode_ref_imgs}, we show the final state of \(\Norm{\C_h}\) for the three solvers considered, all of which preserve the symmetry of the solution. The resulting figures are essentially identical, with no discernible differences at the final time, indicating that the simulations are well resolved in both space and time.

\subsection{Robustness against Poisson regularization \texorpdfstring{$r$}{r}}\label{sec:lin_robust_r}

Here, we numerically assess the robustness of the proposed solver with respect to the regularization parameter \(r\). In semi-implicit time-stepping methods, such as those considered in \cite{astuto2023finite, astuto2023asymmetry}, one is required to solve a Poisson problem whose near-nullspace enlarges as \(r\) and \(\Norm{\C_h}\) approach zero, thereby deteriorating the numerical stability of the associated algebraic solver. In contrast, the nonlinear solver proposed here does not explicitly solve the Poisson problem within the Newton iterations. Instead, a regularized Poisson problem with a mass-dominated regularization is addressed within the Schur complement–based preconditioning step (see \Cref{sec:linear} for details). As a consequence, we expect the number of linear iterations to exhibit only a weak dependence on the value of \(r\).

\begin{figure}[htbp]
\centering
\begingroup
\setlength{\tabcolsep}{\figtabsep}
\begin{tabular}{c c c}
\includegraphics[width=0.32\textwidth,trim={15 0 80 0},clip]{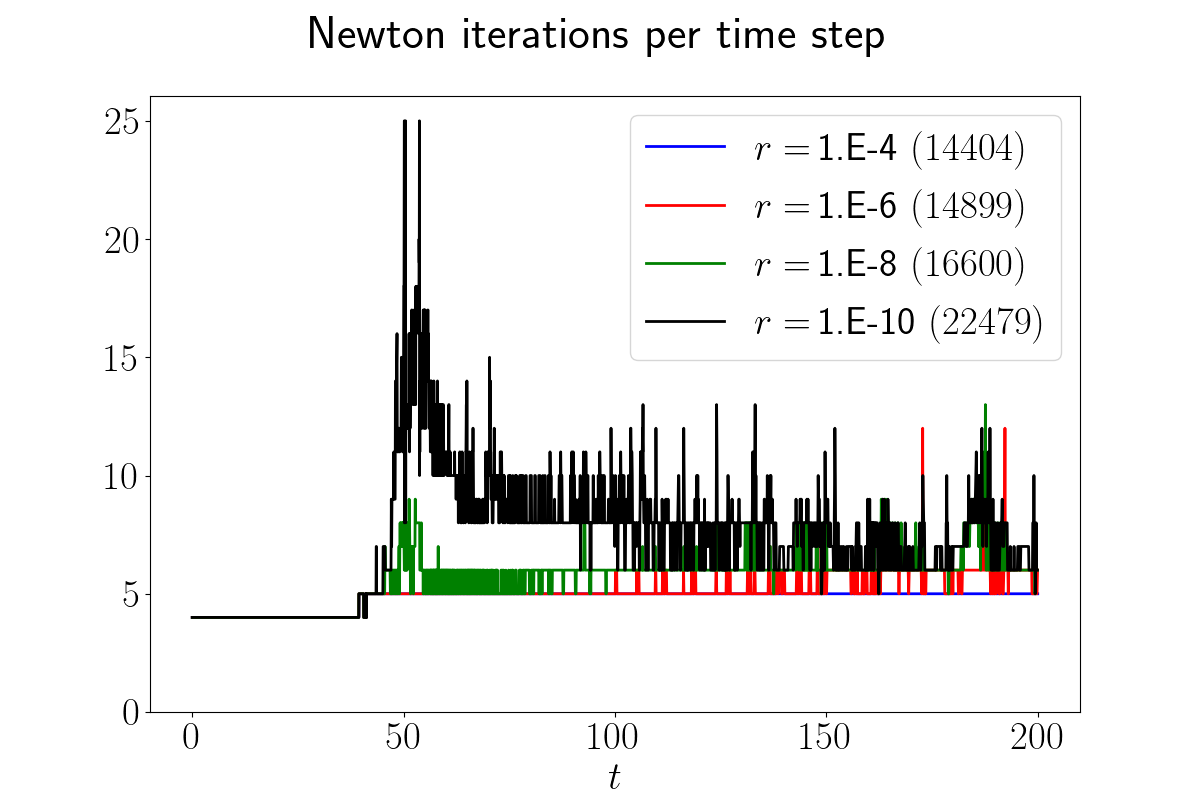} &
\includegraphics[width=0.32\textwidth,trim={15 0 80 0},clip]{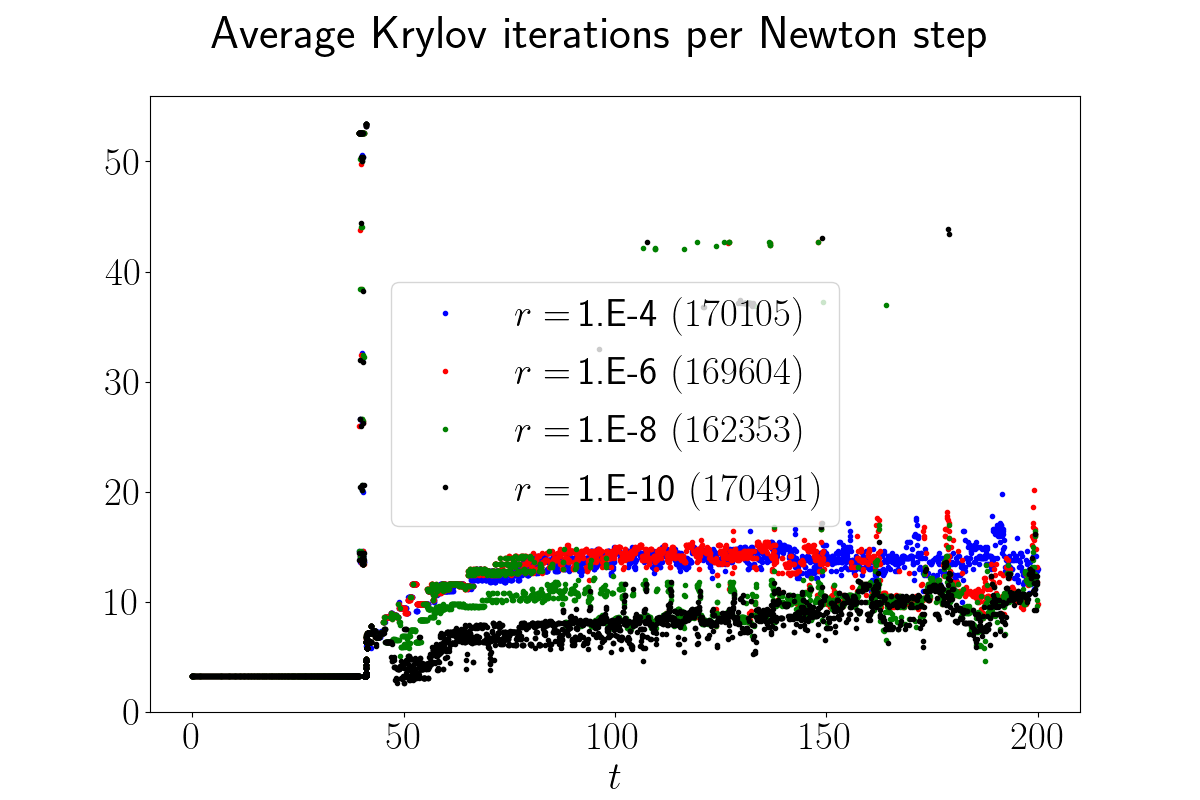} &
\includegraphics[width=0.32\textwidth,trim={15 0 80 0},clip]{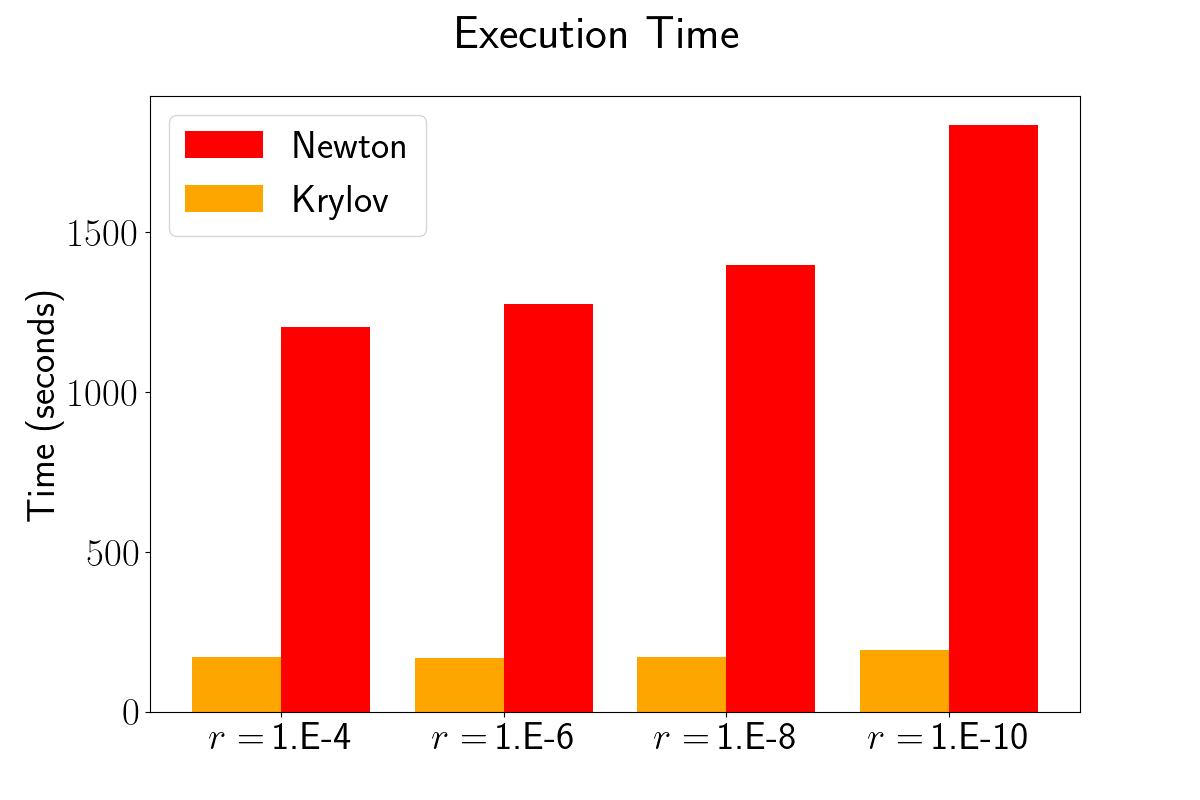}
\end{tabular}
\endgroup
\caption{Robustness with respect to the Poisson regularization parameter $r$. Newton iterations (left panel), average number of linear iterations per Newton step (central panel), and wall-clock timings (right panel) for the BE solver using the  $4096^2$ mesh and different values of  $r$ as indicated in the figures. See \Cref{sec:lin_robust_r} for additional details.}
\label{fig:box_r_its}
\end{figure}

To validate this intuition, we consider a \(1024^2\) mesh and perform a series of backward Euler simulations using the same model parameters as before, while varying the regularization parameter \(r\) from \(10^{-4}\) to \(10^{-10}\). The corresponding results are reported in \Cref{fig:box_r_its}. As shown in the central panel, the average number of linear iterations per time step decreases as \(r\) decreases, and the total number of linear iterations remains essentially independent of \(r\). Consistently, the wall-clock time spent in the Krylov solver is also insensitive to the value of \(r\) (right panel).

In contrast, the overall computational time associated with the Newton solver increases as \(r\) decreases (right panel). This behavior is explained by the larger number of Newton iterations required for convergence at smaller values of \(r\) (left panel), particularly during the onset of network formation, where an increased number of nonlinear solver failures is observed. These failures are due to the fact that the non-negativity of \(\C_h\) cannot be enforced within the Newton iterations, leading to more frequent reductions in the time step. For this configuration, we do not report the final values of the energy functional or the final state of \(\Norm{\C_h}\), as the observed differences are negligible.

\subsection{Mesh dependence}\label{sec:meshdependence}

In this section, we present numerical results obtained using the backward Euler solver and the same model parameters as in \Cref{sec:be_robust}, comparing the final state of $\Norm{\C_h}$ for different triangular and quadrilateral meshes of the unit domain. In particular, we consider the following meshes: {\bf quad structured}, a \(512\times512\) quadrilateral mesh; {\bf quad unstructured}, an unstructured mesh of 232K quadrilaterals; {\bf triangle criss-cross}, obtained by splitting each quadrilateral of the structured mesh into four triangles along its diagonals; {\bf triangle regular}, obtained by splitting each quadrilateral of the structured mesh into two triangles along the top-left to bottom-right diagonal; and {\bf triangle irregular}, a fully unstructured mesh with 300K triangles.

\begin{figure}[htbp]
\centering
\begingroup
\setlength{\tabcolsep}{\figtabsep}
\begin{tabular}{c c}
\includegraphics[width=0.32\textwidth,trim={15 0 80 35},clip]{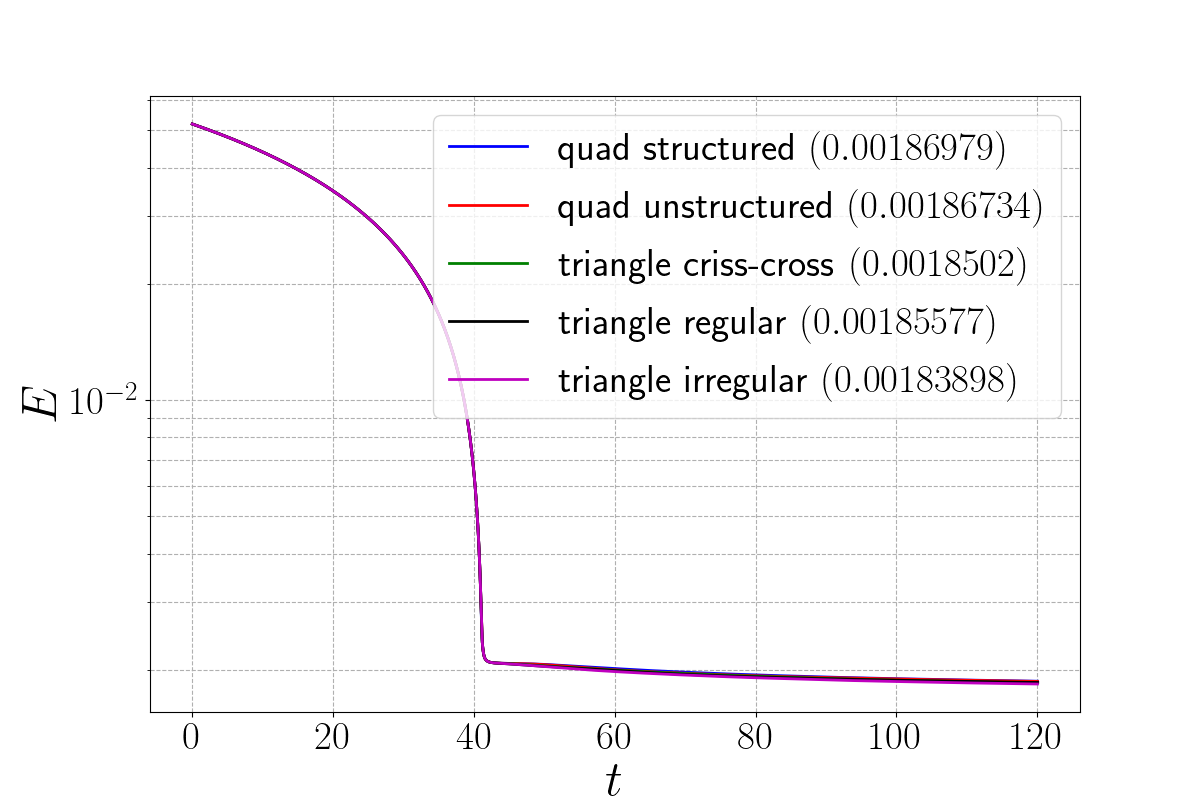} &
\includegraphics[width=0.32\textwidth,trim={15 0 80 35},clip]{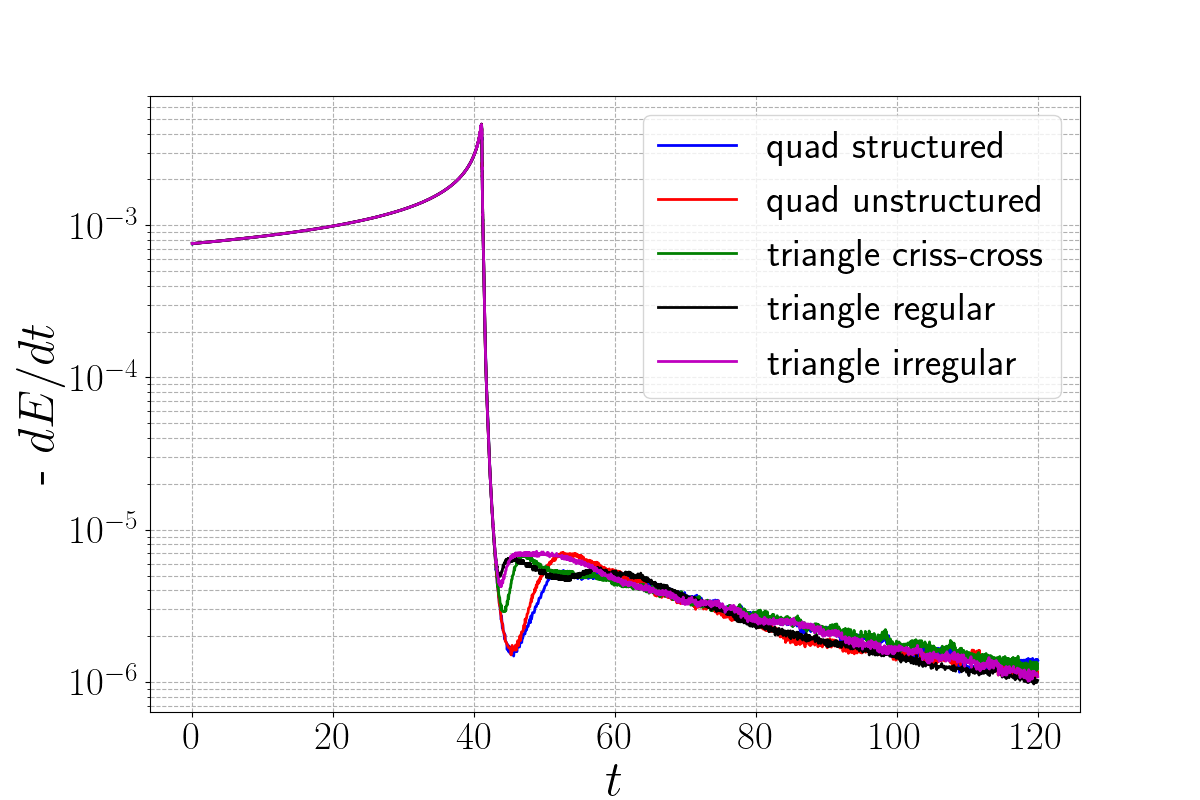}
\end{tabular}
\endgroup
\caption{Mesh dependence. Energy (left panel) and $-dE/dt$ (right panel) as a function of simulated time for different meshes of the unit square. See \Cref{sec:meshdependence} for additional details. }
\label{fig:meshdependence_logs}
\end{figure}

\begin{figure}[htbp]
\centering
\begingroup
\setlength{\tabcolsep}{\figtabsep}
\begin{tabular}{c c c c c}
quad & quad & triangle & triangle & triangle\\
structured & unstructured & criss-cross & regular & irregular\\
\includegraphics[width=0.19\textwidth,trim={200 40 140 8},clip]{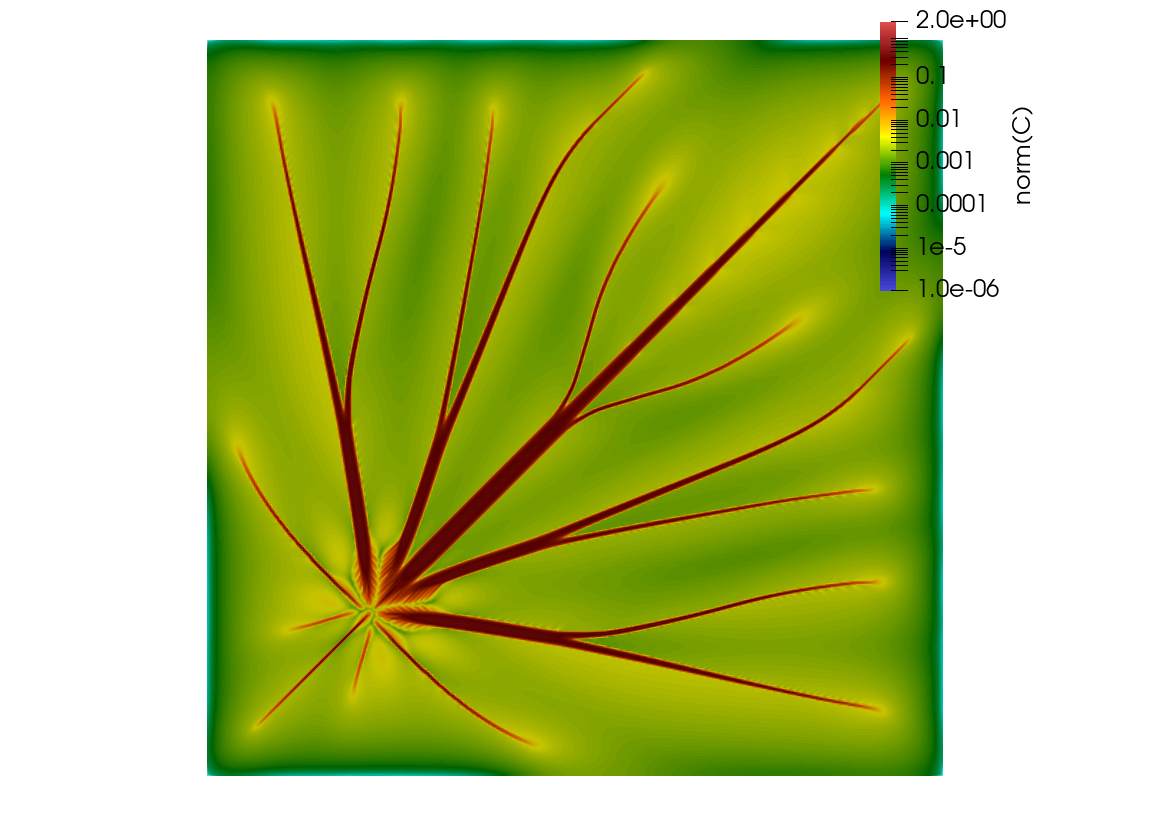} &
\includegraphics[width=0.19\textwidth,trim={200 40 140 8},clip]{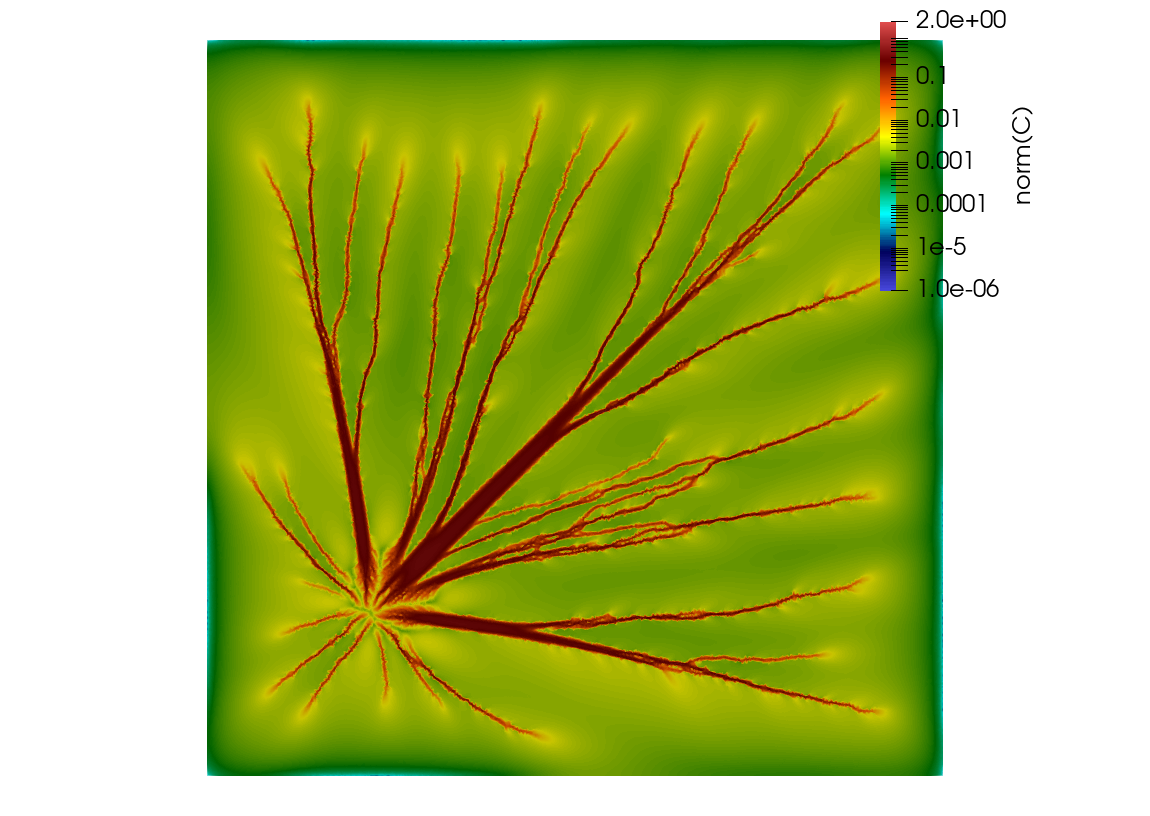} &
\includegraphics[width=0.19\textwidth,trim={200 40 140 8},clip]{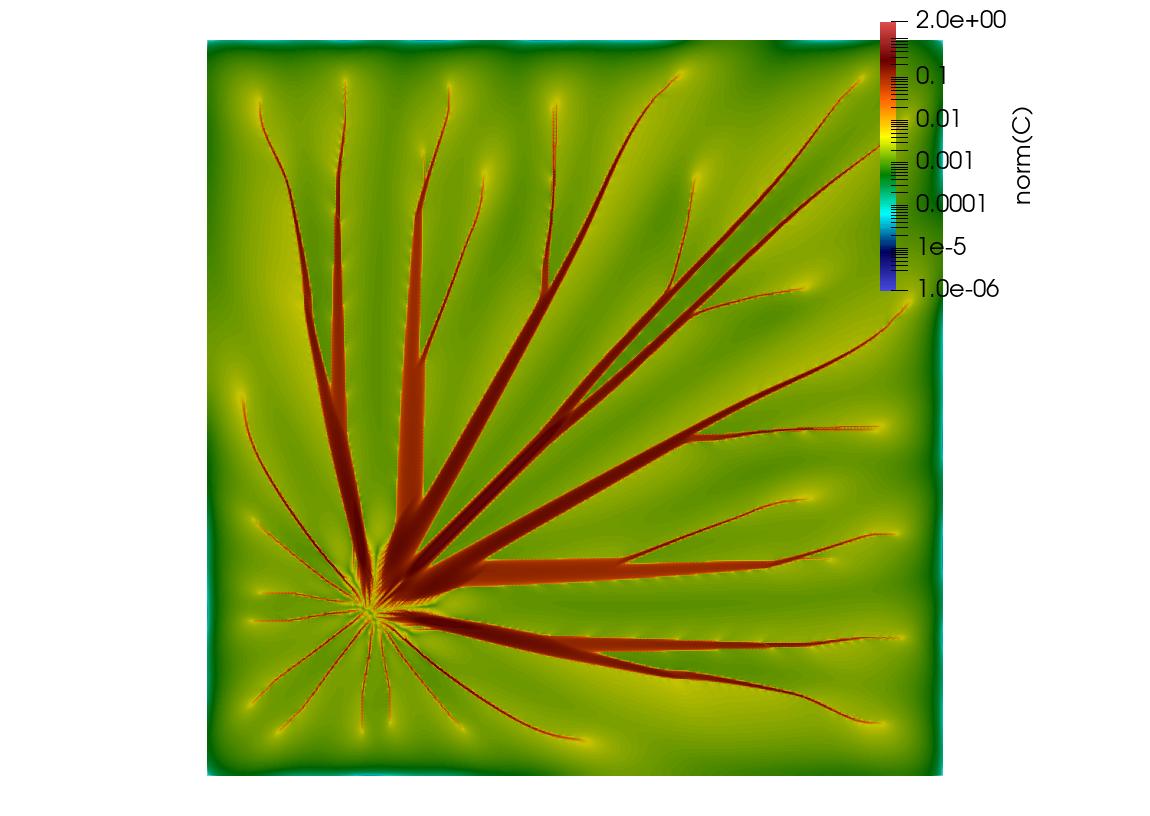} &
\includegraphics[width=0.19\textwidth,trim={200 40 140 8},clip]{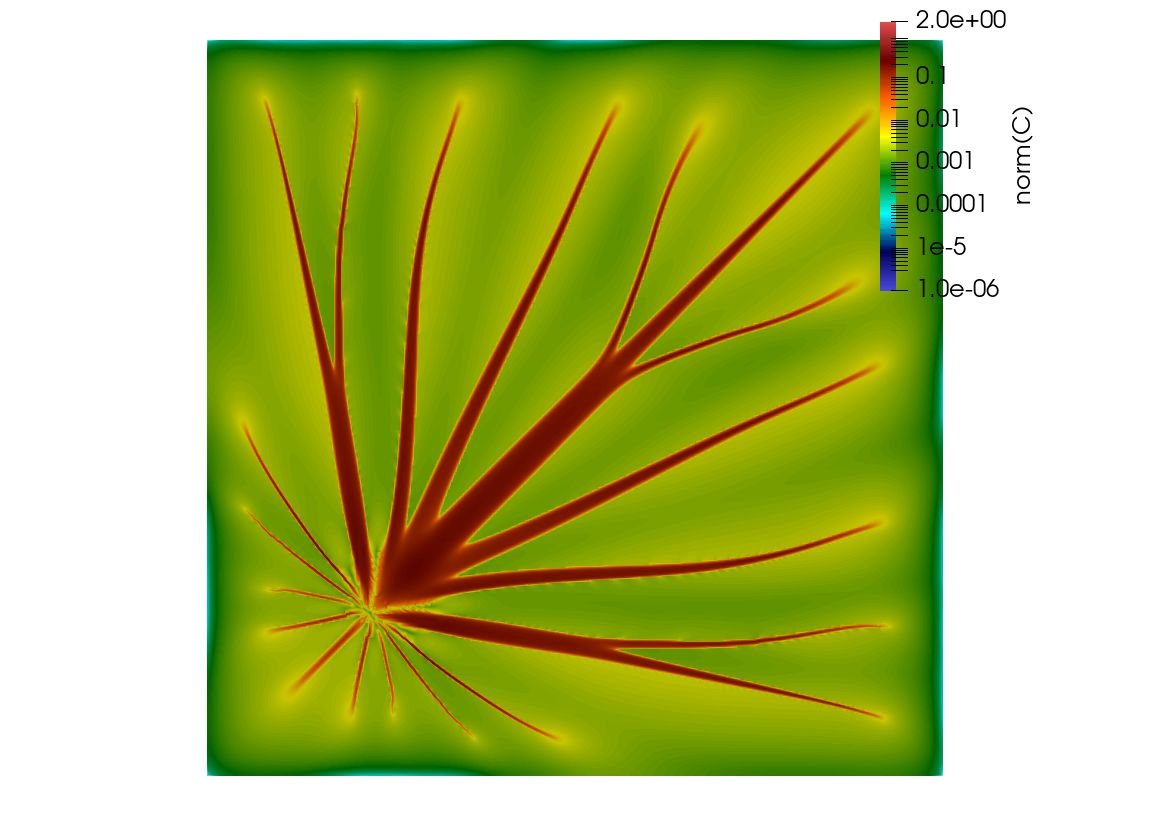} &
\includegraphics[width=0.19\textwidth,trim={200 40 140 8},clip]{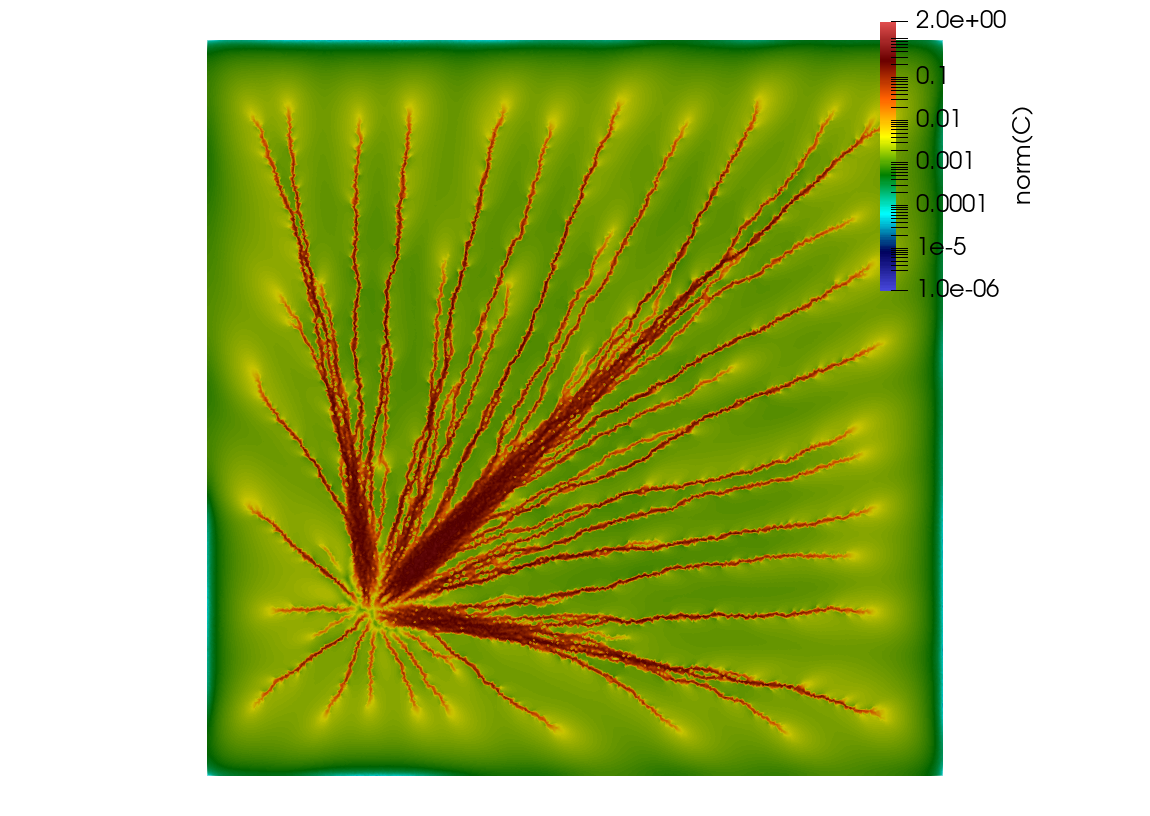}\\
\includegraphics[width=0.19\textwidth,trim={200 40 140 10},clip]{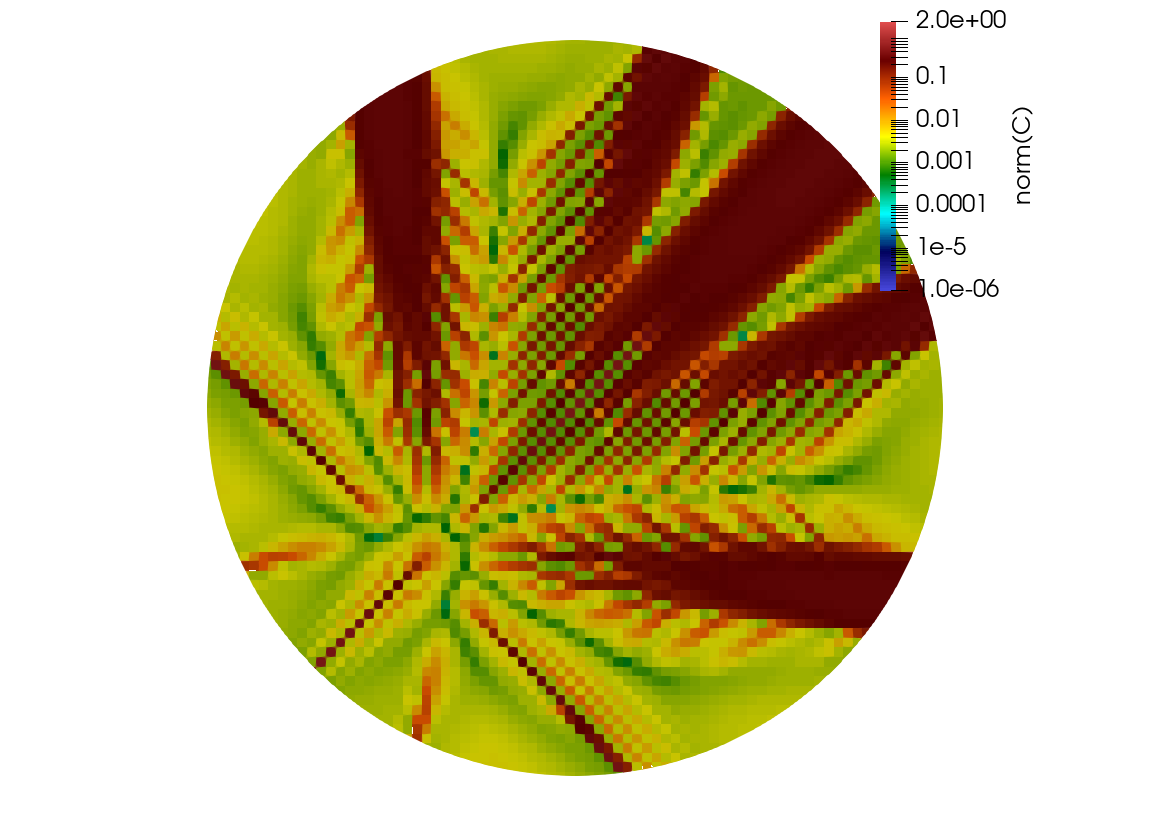} & 
\includegraphics[width=0.19\textwidth,trim={200 40 140 10},clip]{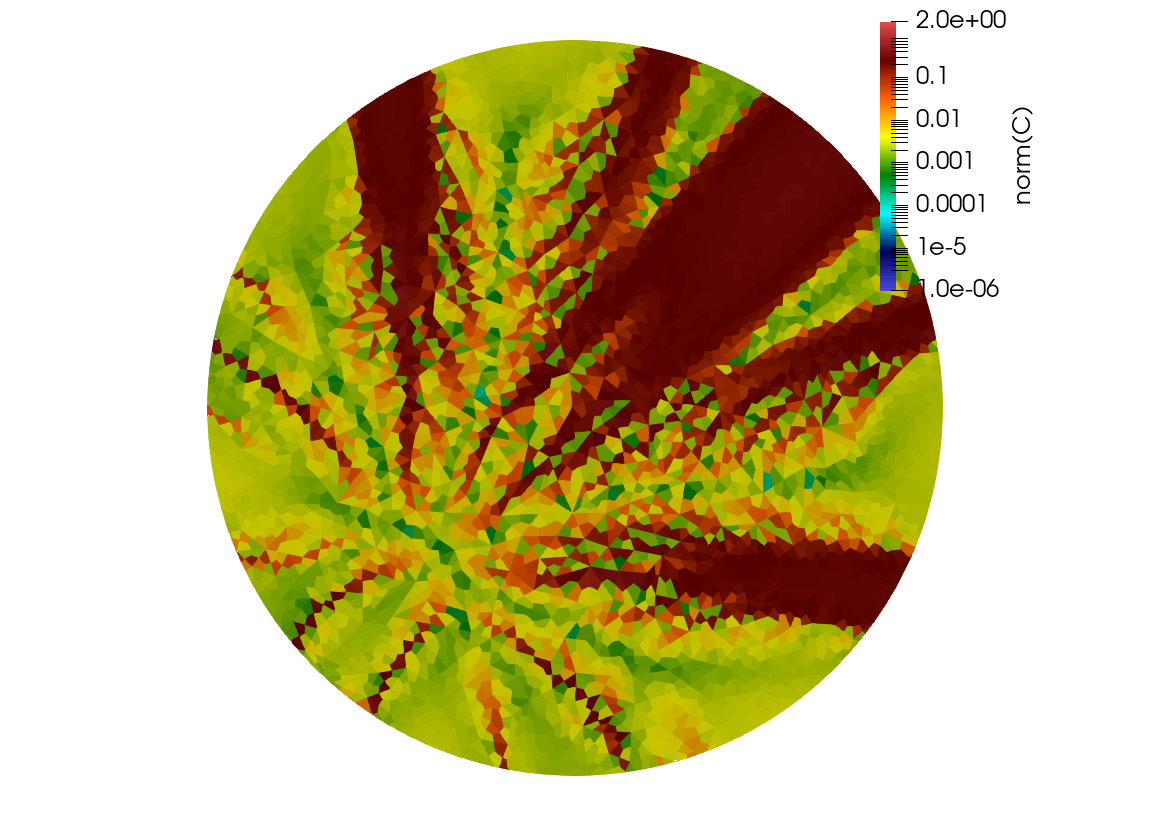} &
\includegraphics[width=0.19\textwidth,trim={200 40 140 10},clip]{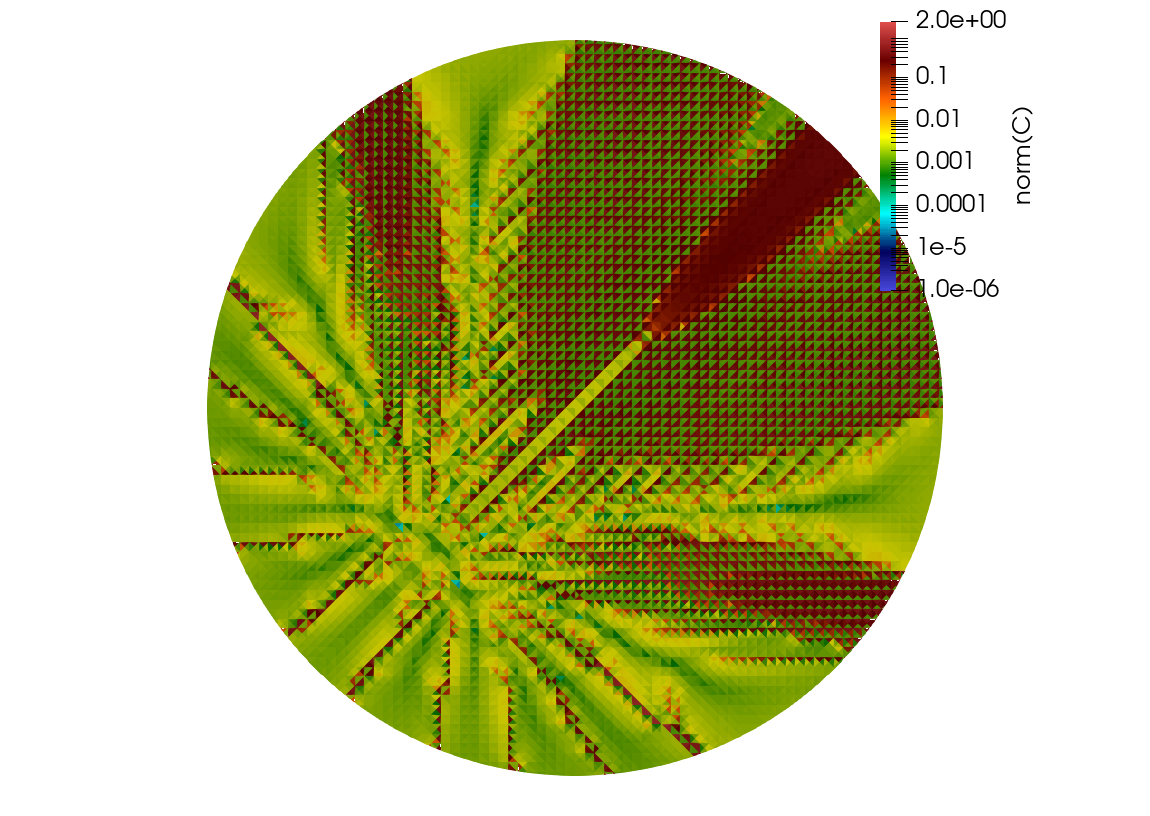} &
\includegraphics[width=0.19\textwidth,trim={200 40 140 10},clip]{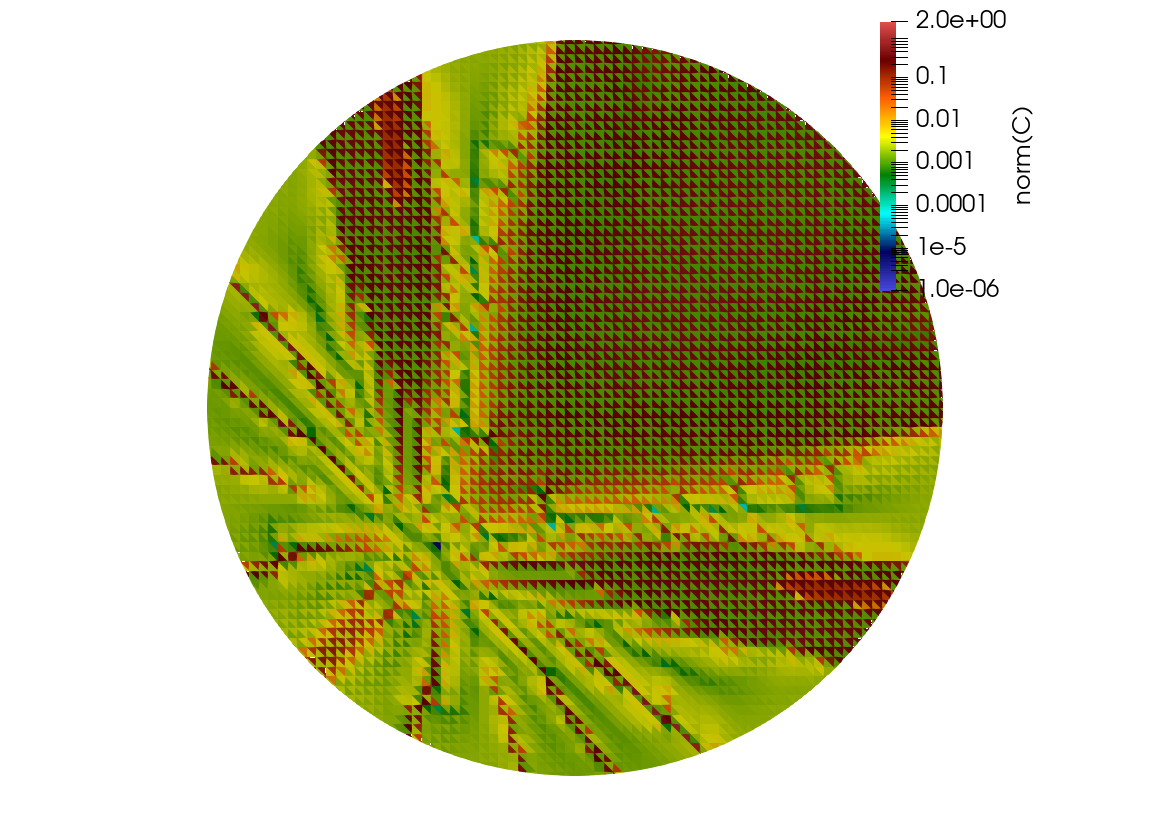} &
\includegraphics[width=0.19\textwidth,trim={200 40 140 10},clip]{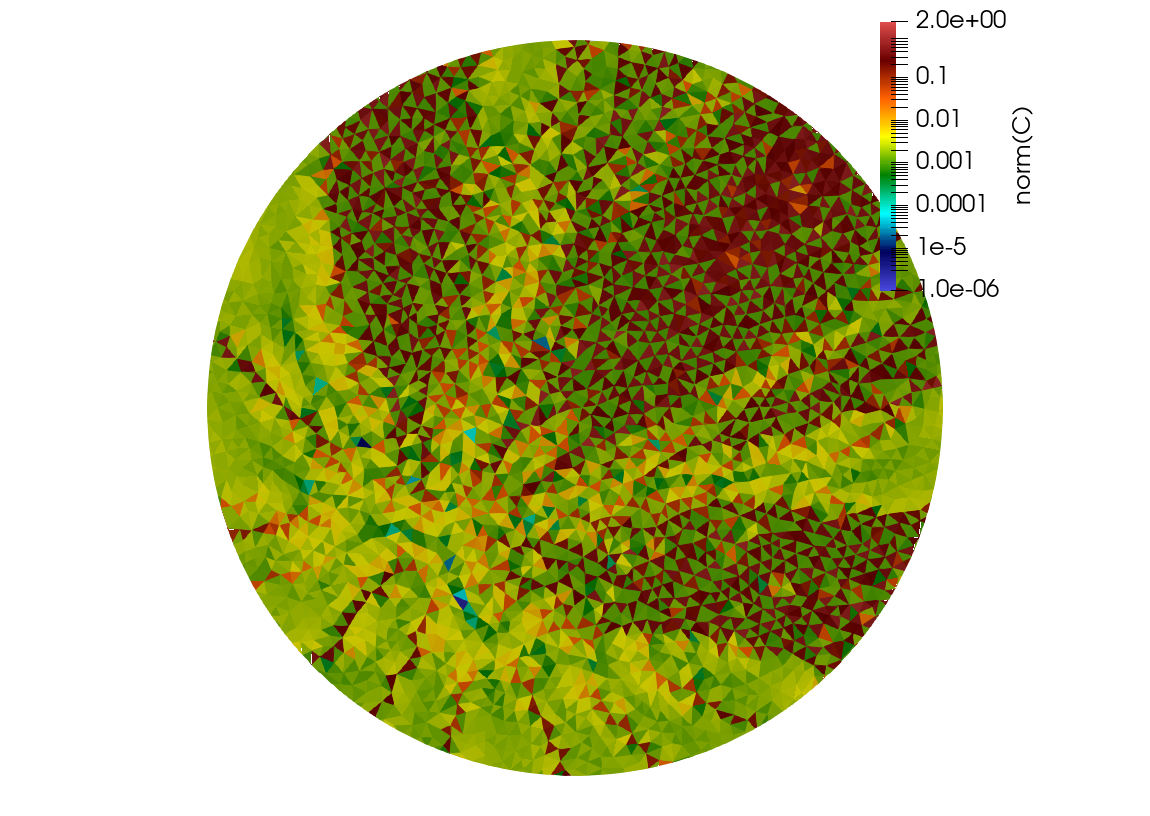}
\end{tabular}
\endgroup
\caption{Mesh dependence. Final state of $\Norm{\C_h}$ and close-ups at the source location for different meshes as labeled on top. See \Cref{sec:meshdependence} for additional details.}
\label{fig:meshdependence_imgs}
\end{figure}


All simulations achieve very similar final energy values, with the energy consistently decaying; see \Cref{fig:meshdependence_logs}. The final states of the norm of the conductivity matrix are shown in \Cref{fig:meshdependence_imgs}, where a clear dependence on the underlying mesh can be observed. The structured meshes (quad structured, triangle criss-cross, and triangle regular) yield very similar, regular, and symmetric network patterns. In contrast, the unstructured meshes yield visibly more irregular network structures. Nevertheless, the dominant branching pattern is preserved at the macroscopic level.

As illustrated in the close-up views in the bottom row of \Cref{fig:meshdependence_imgs}, quadrilateral meshes exhibit smooth inter-element transitions of \(\Norm{\C_h}\), whereas triangular meshes display a checkerboard-like pattern with strong element-to-element oscillations at small values of \(\Norm{\C_h}\). This checkerboard pattern persists under mesh refinement (data not shown), indicating that it is an intrinsic feature of the triangular discretization rather than a resolution effect.

We further note that, in the nonconvex regime considered here, adaptive mesh refinement would generally affect not only numerical accuracy but also the qualitative features of the computed network structures, since different meshes may lead to different local minima of the energy functional. While adaptivity could potentially reduce computational cost by concentrating resolution in selected regions, its interaction with mesh-dependent pattern formation and structural properties requires further investigation.

\subsection{Leaf-based simulations varying \texorpdfstring{$\gamma$}{gamma}}\label{sec:leaf_gamma}

\begin{figure}[htbp]
\centering
\begingroup
\setlength{\tabcolsep}{\figtabsep}
\begin{tabular}{c c}
$\gamma=0.75$ & $\gamma=0.5$\\
\includegraphics[width=0.45\textwidth,trim={200 40 55 10},clip]{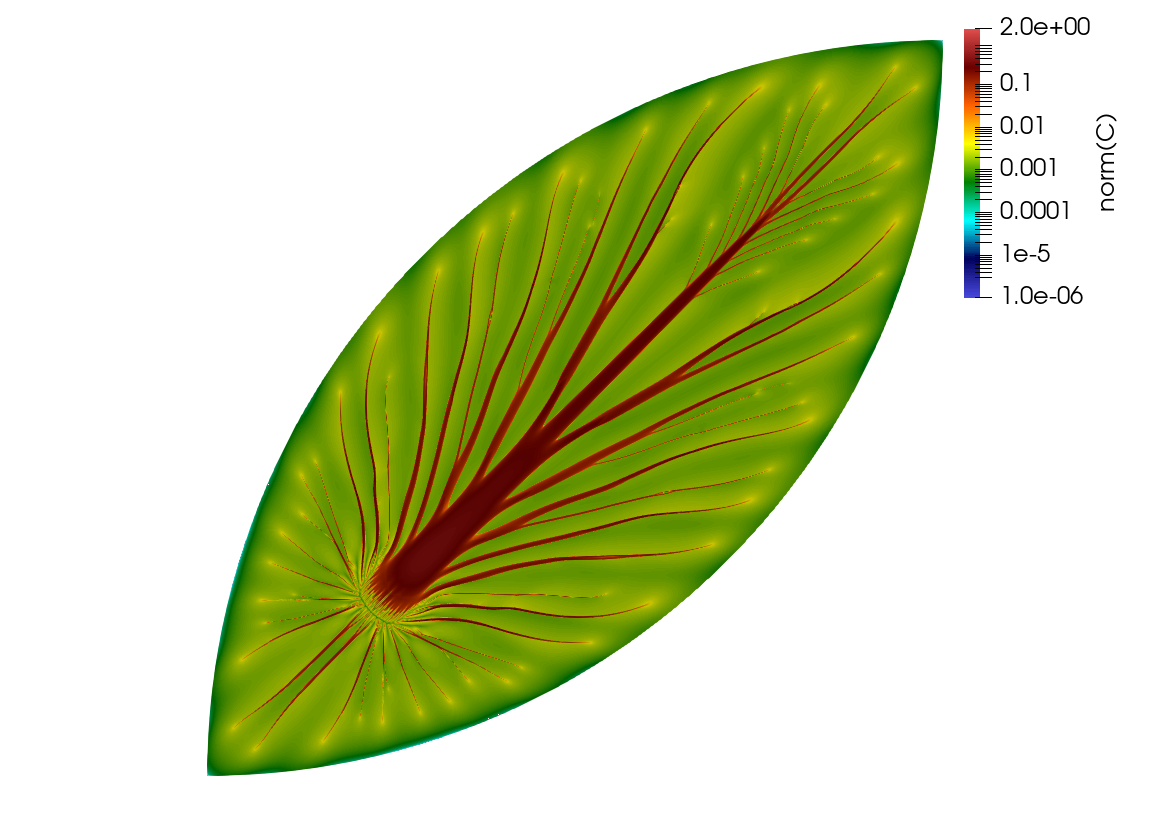} & \includegraphics[width=0.45\textwidth,trim={200 40 55 10},clip]{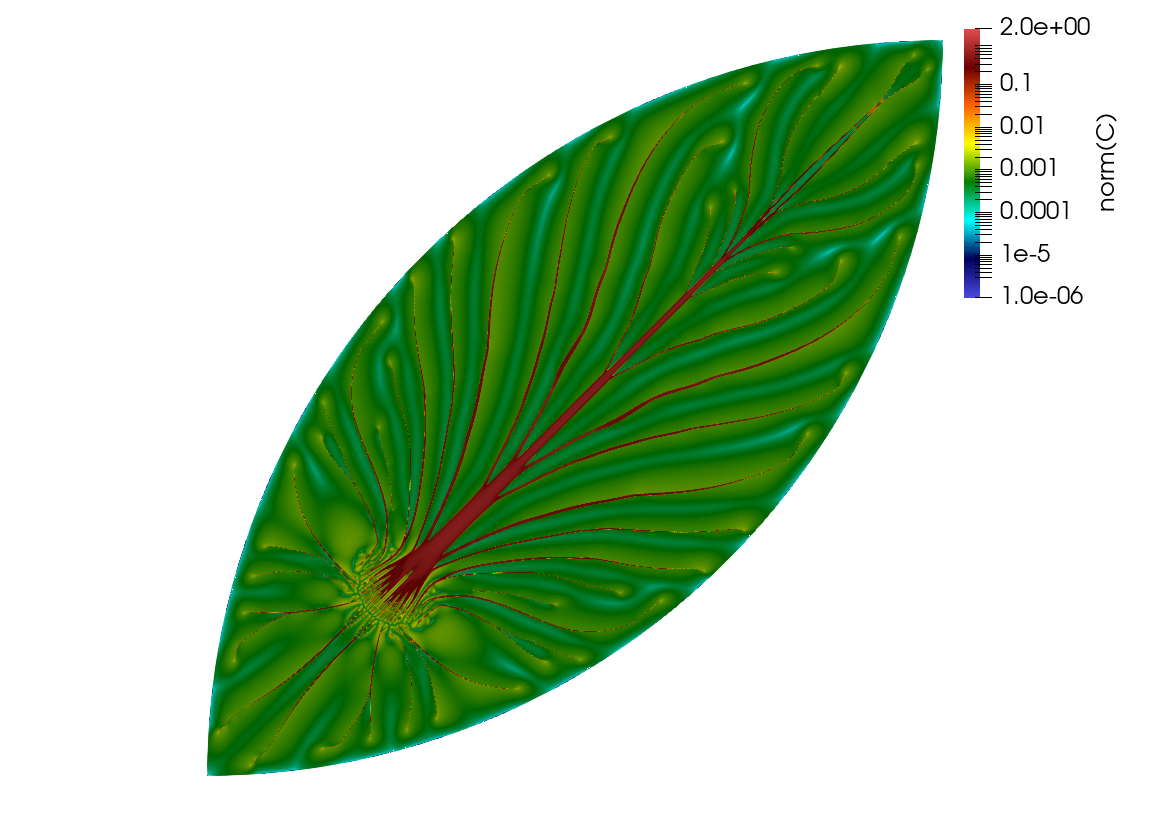}
\end{tabular}
\endgroup
\caption{Leaf-based simulations varying $\gamma$. Final state of $\Norm{\C_h}$ for different values of $\gamma$ as labeled on top. See \Cref{sec:leaf_gamma} for additional details.}
\label{fig:leaf_final}
\end{figure}

\begin{figure}[htbp]
\centering
\begingroup
\setlength{\tabcolsep}{\figtabsep}
\begin{tabular}{c c c}
\includegraphics[width=0.3\textwidth,trim={5 0 80 0},clip]{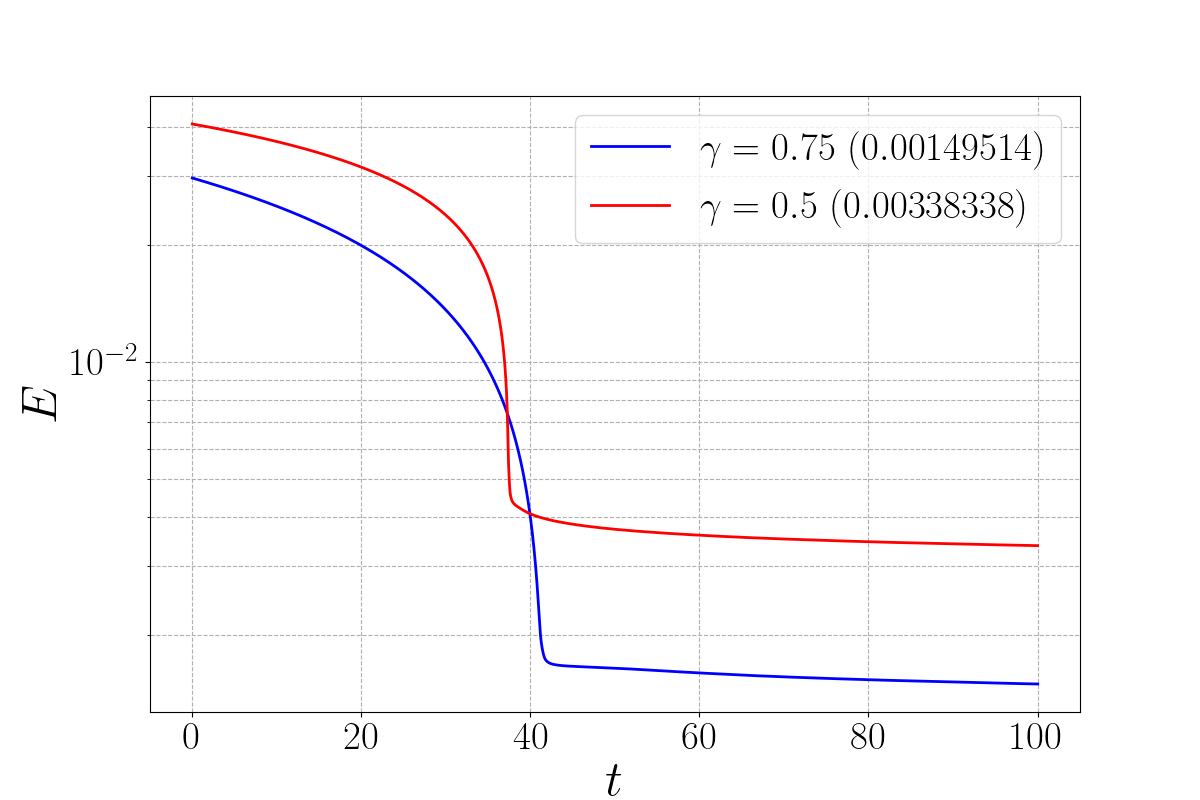} & 
\includegraphics[width=0.3\textwidth,trim={5 0 80 0},clip]{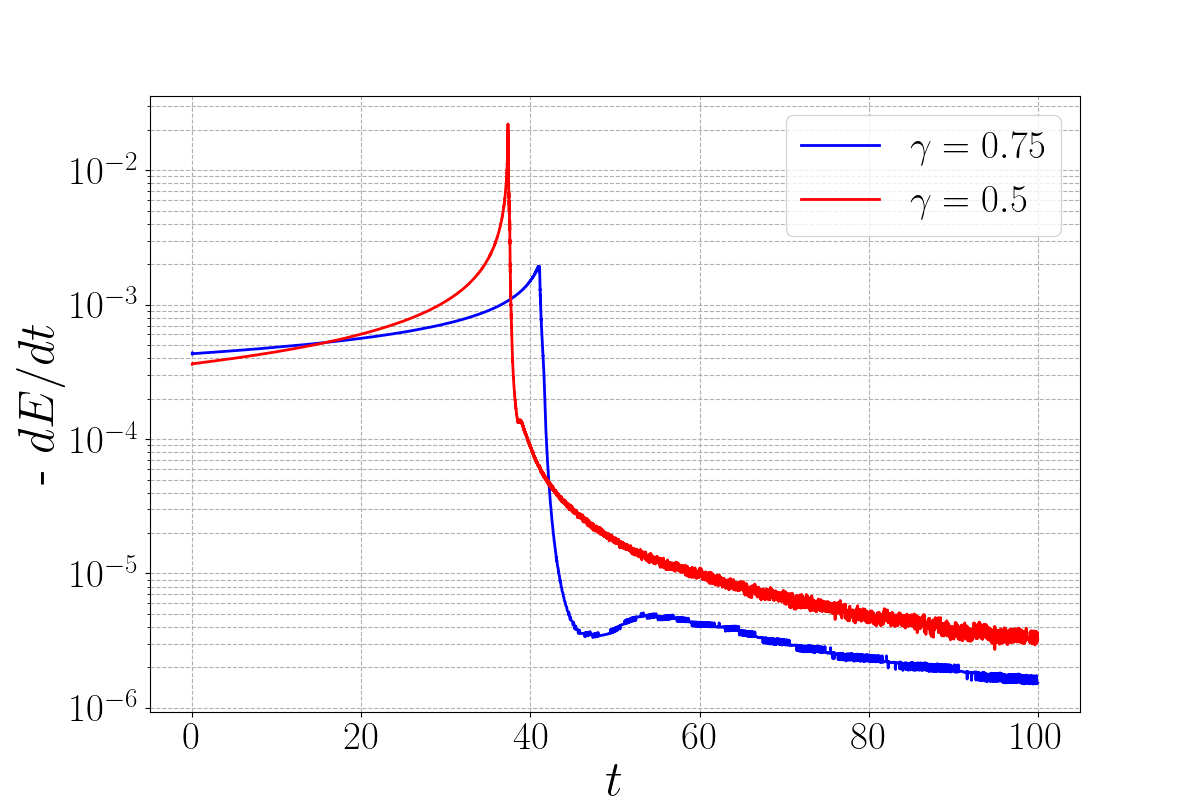} &
\includegraphics[width=0.3\textwidth,trim={5 0 80 0},clip]{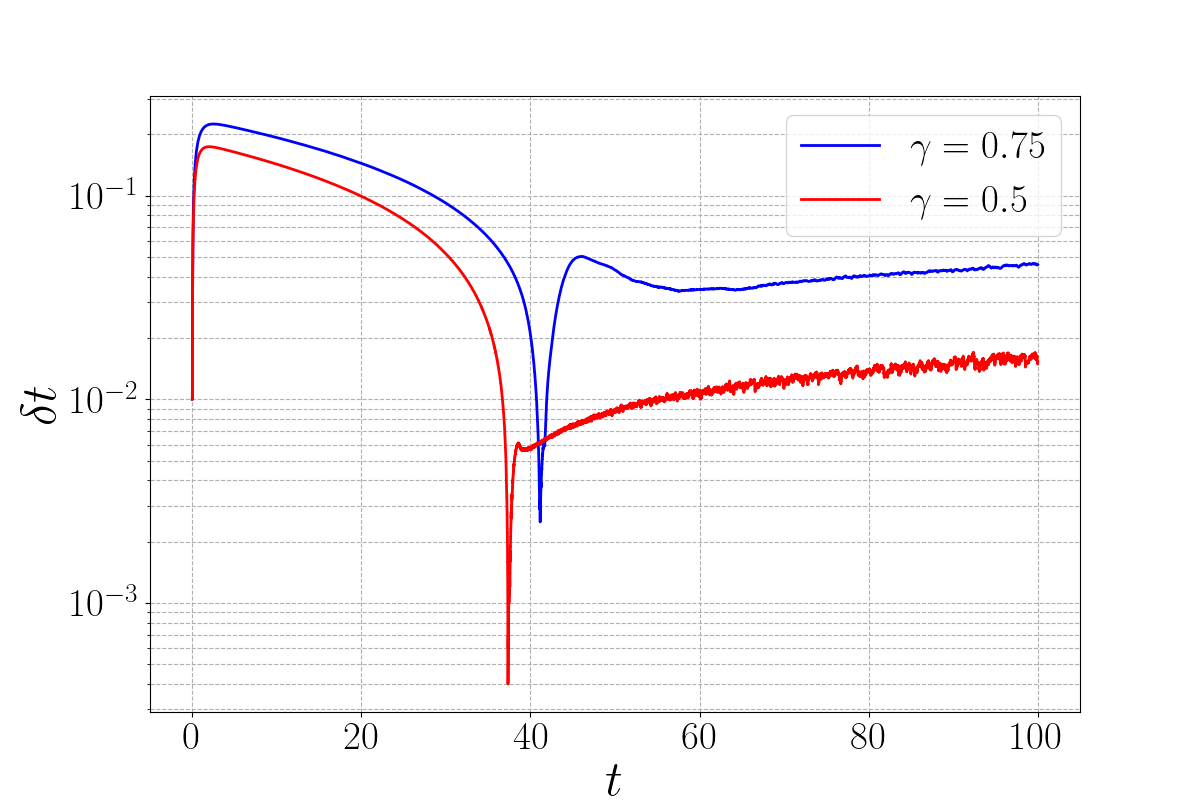} \\
\includegraphics[width=0.3\textwidth,trim={5 0 80 0},clip]{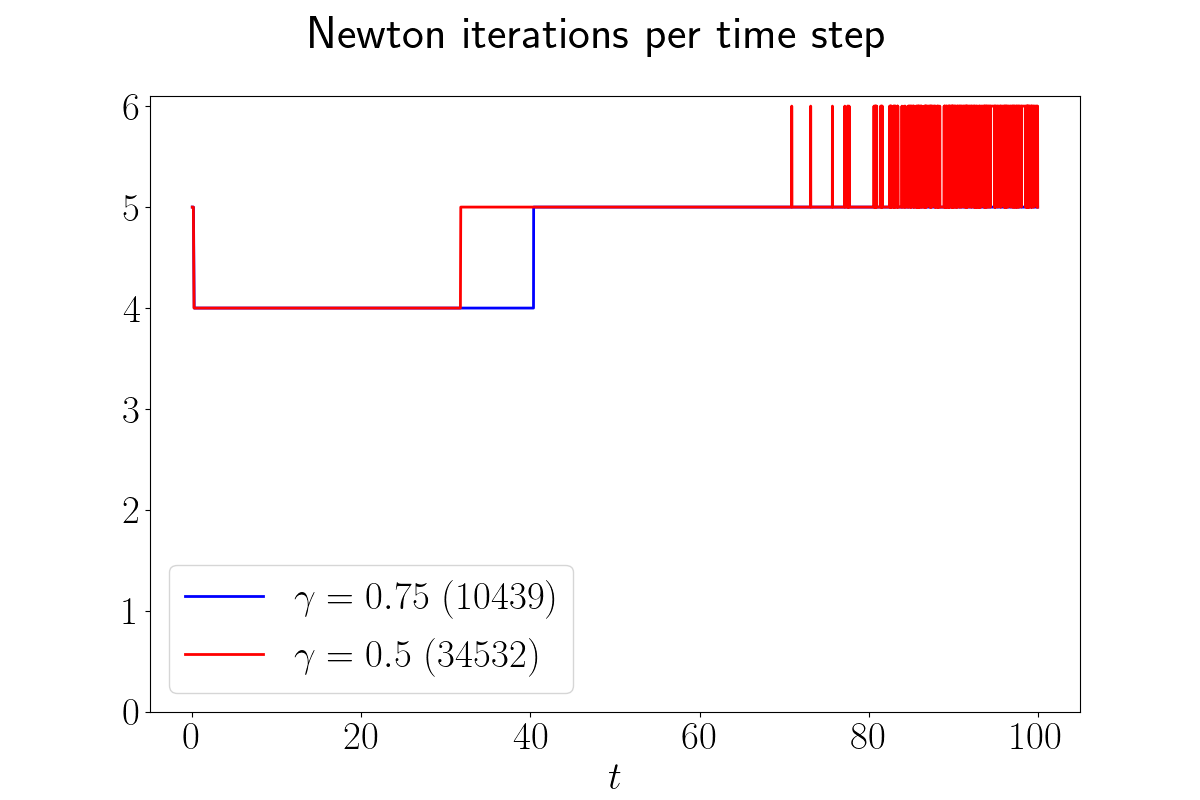} & 
\includegraphics[width=0.3\textwidth,trim={5 0 80 0},clip]{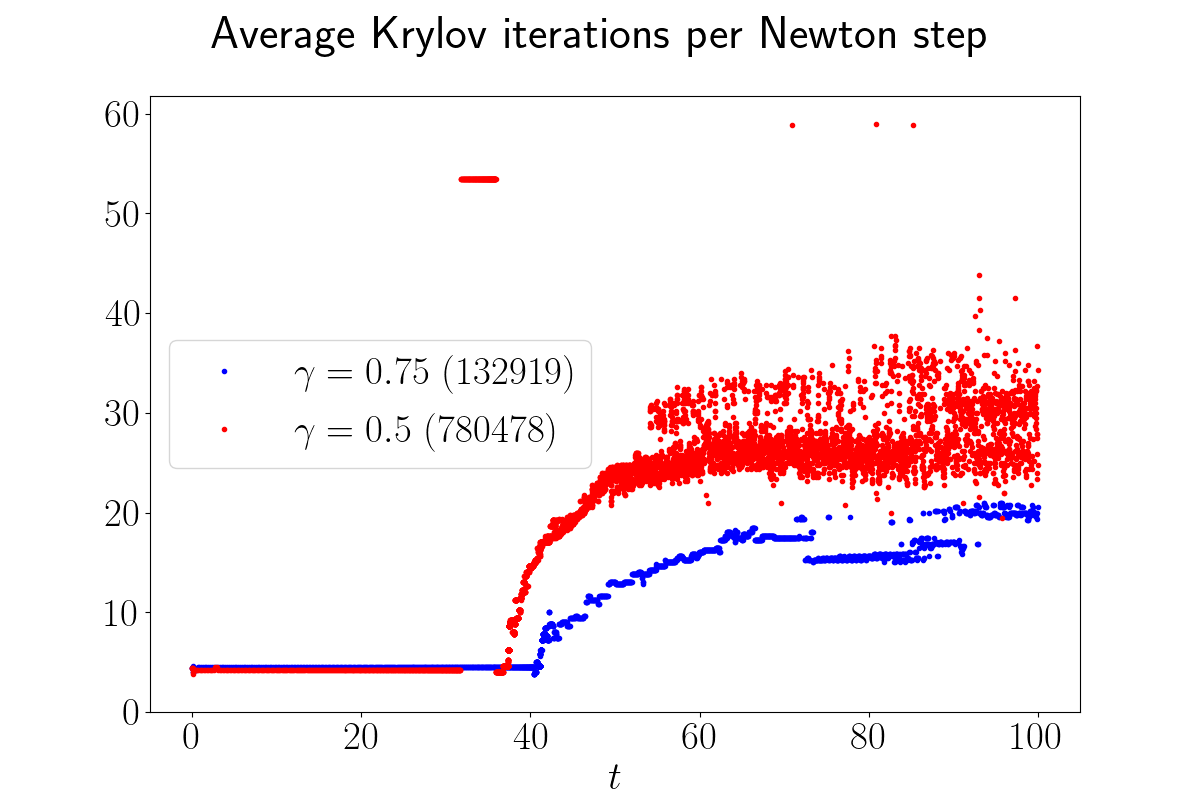} &
\includegraphics[width=0.3\textwidth,trim={5 0 80 0},clip]{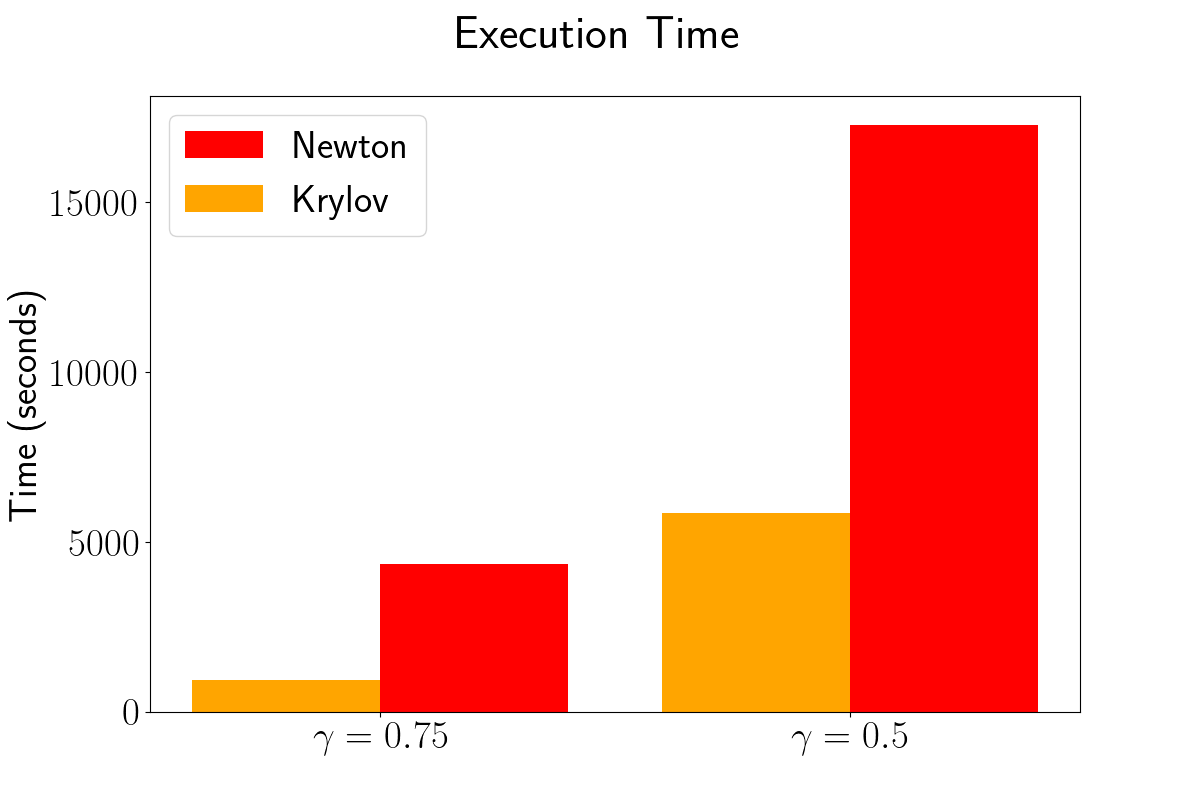} 
\end{tabular}
\endgroup
\caption{Leaf-based simulations varying $\gamma$. Top row: energy (left panel), $-dE/dt$ (central panel), and time step (right panel) in logarithmic scale as a function of simulated time for different values of $\gamma$ as indicated in the legends. Bottom row: Newton steps (left panel), average number of linear iterations per Newton step (central panel), and wall-clock timings (right panel). See \Cref{sec:leaf_gamma} for additional details. }
\label{fig:leaf_its}
\end{figure}

Here, we discuss numerical simulations using the backward Euler solver and the same model parameters as in \Cref{sec:be_robust}, while considering different values for $\gamma$; specifically, $\gamma=0.75$, and $\gamma=0.5$. The results are obtained on the triangular leaf mesh shown in \Cref{fig:leaf_mesh}, which is uniformly refined five times, yielding approximately 13 million triangles and 46 million degrees of freedom. The final state of the norm of the conductivity matrix is reported in \Cref{fig:leaf_final}, while solver performance metrics are shown in \Cref{fig:leaf_its}.

The final network structure is symmetric with respect to the domain bisector in both cases, since the mesh is designed to preserve this symmetry. In the $\gamma = 0.5$ case, the network exhibits finer ramifications and larger regions characterized by small values of $\Norm{\C_h}$, which is consistent with a stronger influence of the metabolic term $\nu \left(\Norm{\C_h}^2 + \varepsilon \right)^{\frac{\gamma-2}{2}} \C_h$. For both values of $\gamma$ considered, the energy decreases monotonically, attaining a smaller steady-state value in the $\gamma = 0.75$ case. When $\gamma = 0.5$, the initial energy is larger, and a faster transition to the network formation regime is observed. Moreover, for $\gamma = 0.5$, the solver selects smaller time steps, possibly related to the thinner network structure, resulting in a larger number of time steps and increased computational cost.

In both cases, the number of Newton iterations per time step is comparable, as the smaller time steps used for $\gamma = 0.5$ appear to mitigate the effects of the stronger nonlinearities. In contrast, the average number of linear iterations per Newton step is slightly larger for \(\gamma = 0.5\), since the regularization of the Poisson problem within the Schur complement preconditioner is less effective in this regime; nevertheless, the iteration counts remain moderate.

\subsection{Three-dimensional network formation}\label{sec:3d_results}

We close this section by presenting our three-dimensional simulation results. Here, we consider a slab domain $[0,1]\times[0,1]\times[0,0.5]$, discretized with a  $256\times256\times128$ grid of hexahedral elements. The parameters of the equations are the same as in the previous sections, with the exception of the Poisson regularization term, which is now set to $r=10^{-3}$. The source term \eqref{eq:source} is sampled at $(0.25,0.25,0.25)$, and the relative tolerance used to solve the nonlinear equations is $10^{-12}$. The total number of degrees of freedom is 58 million.

\begin{figure}[htbp]
\centering
\begingroup
\setlength{\tabcolsep}{\figtabsep}
\begin{tabular}{c c c}
\includegraphics[width=0.32\textwidth,trim={5 0 80 35},clip]{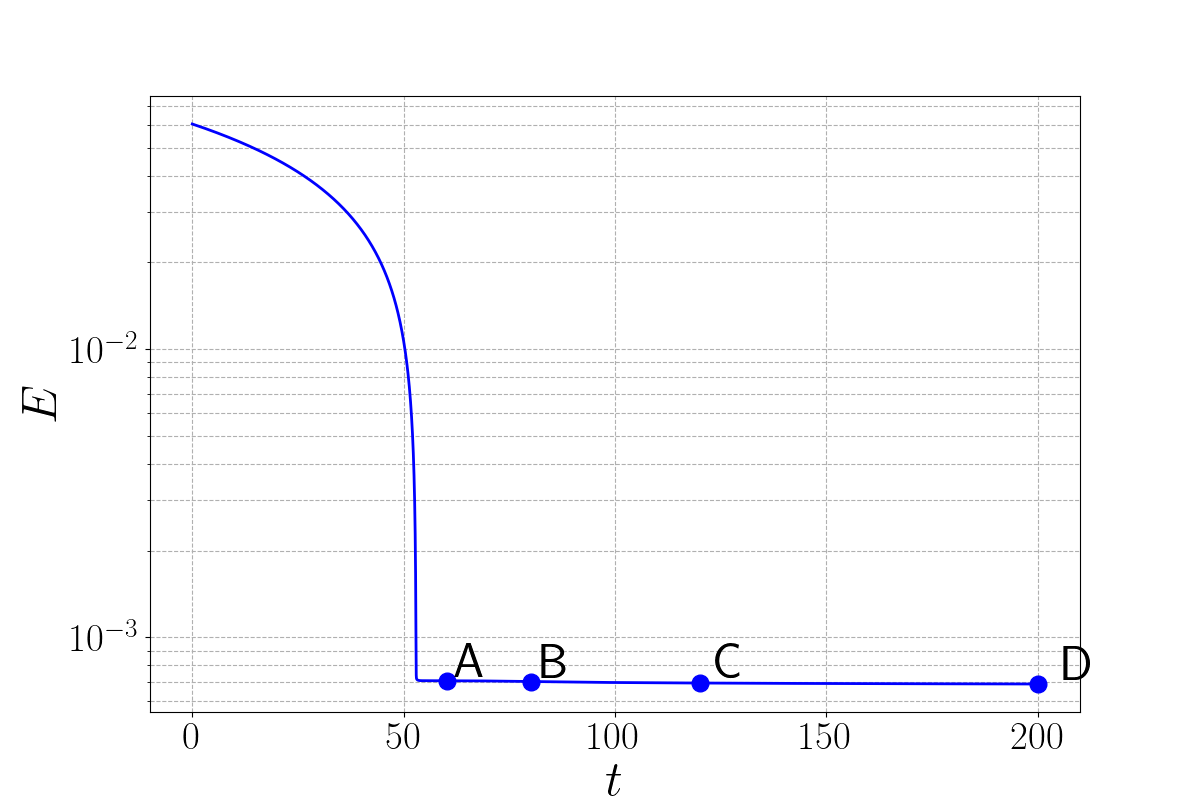} & 
\includegraphics[width=0.32\textwidth,trim={5 0 80 35},clip]{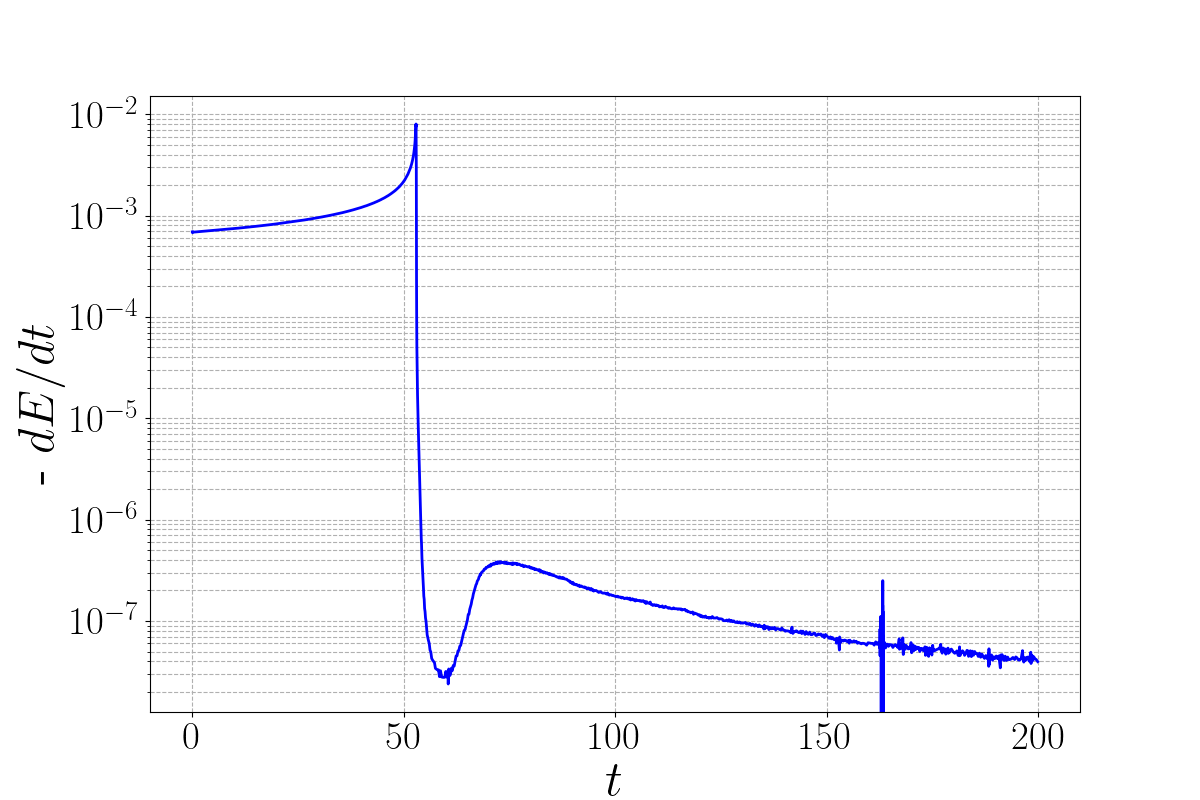} &
\includegraphics[width=0.32\textwidth,trim={5 0 80 35},clip]{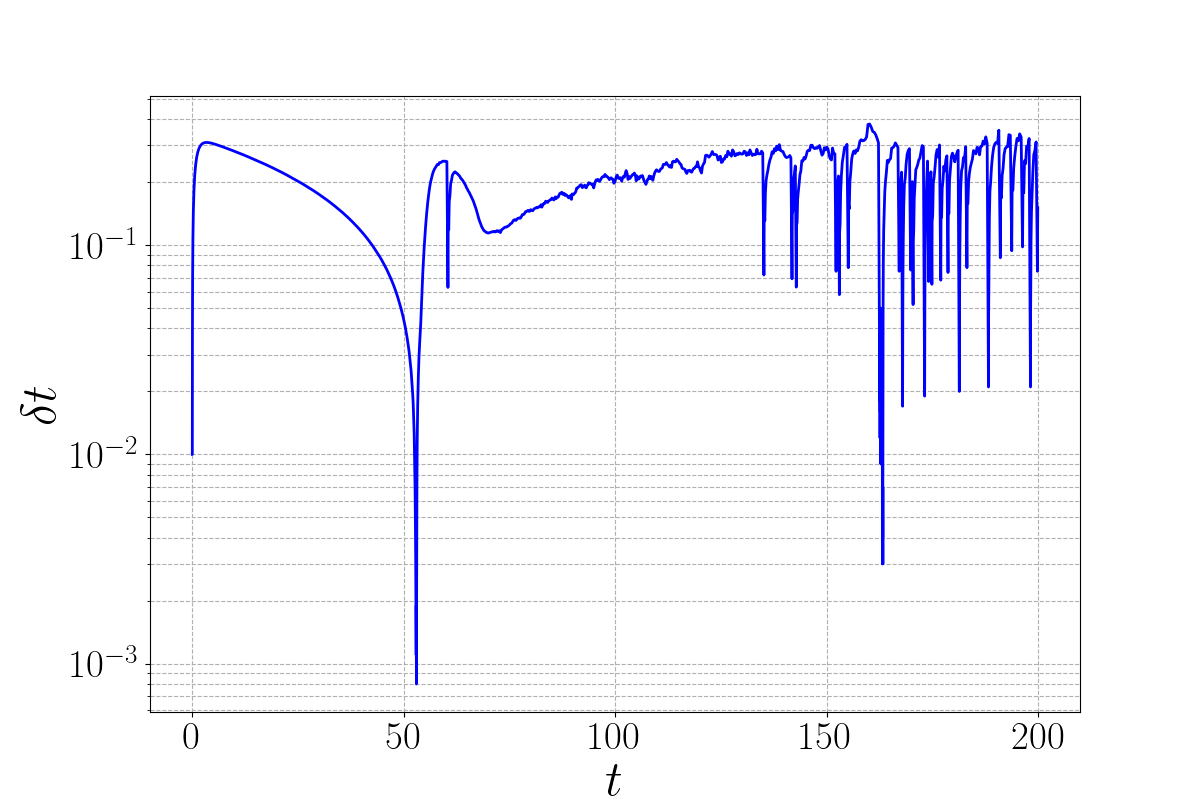}
\end{tabular}
\endgroup
\caption{Three-dimensional network formation. Energy (left panel), $-dE/dt$ (central panel), and time step (right panel), as a function of simulated time. See \Cref{sec:3d_results} for additional details. }
\label{fig:slab_sequence_logs}
\end{figure}

\begin{figure}[htbp]
\centering
\begingroup
\setlength{\tabcolsep}{\figtabsep}
\begin{tabular}{c c}
\bf{A} & \bf{B}\\
\includegraphics[width=0.45\textwidth,trim={50 20 30 0},clip]{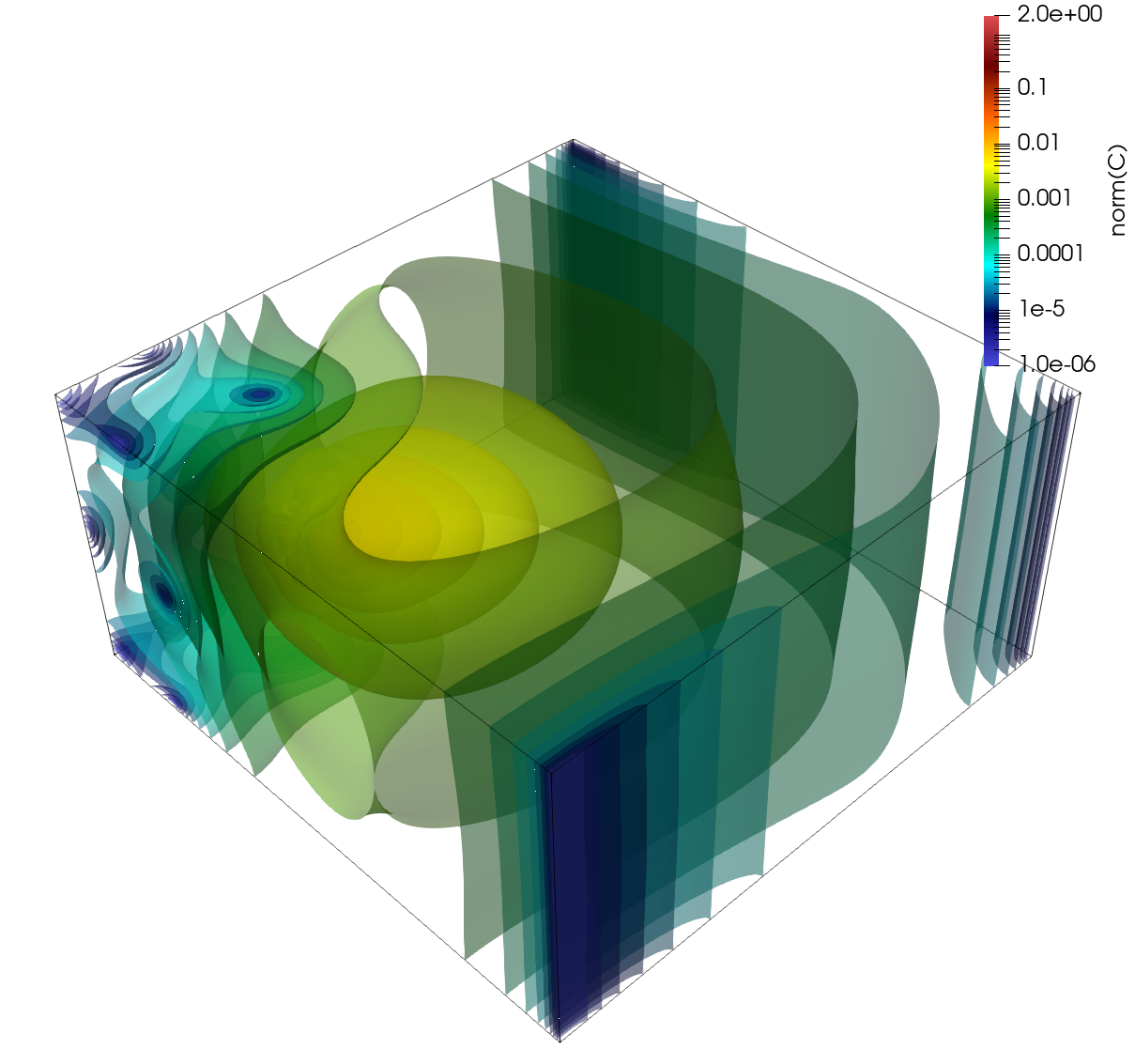} &
\includegraphics[width=0.45\textwidth,trim={50 20 30 0},clip]{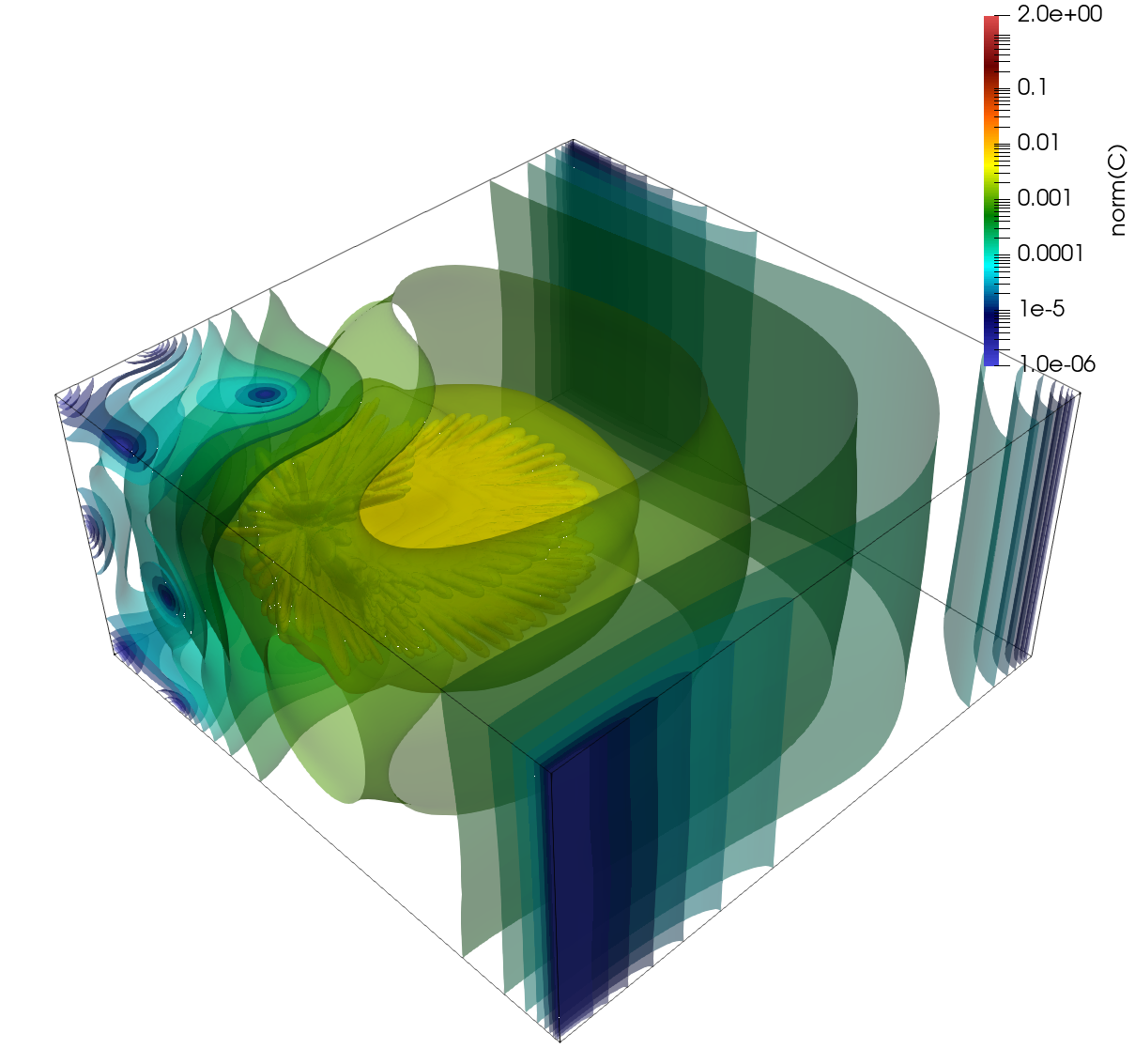}\\
\\
\bf{C} & \bf{D}\\
\includegraphics[width=0.45\textwidth,trim={50 20 30 0},clip]{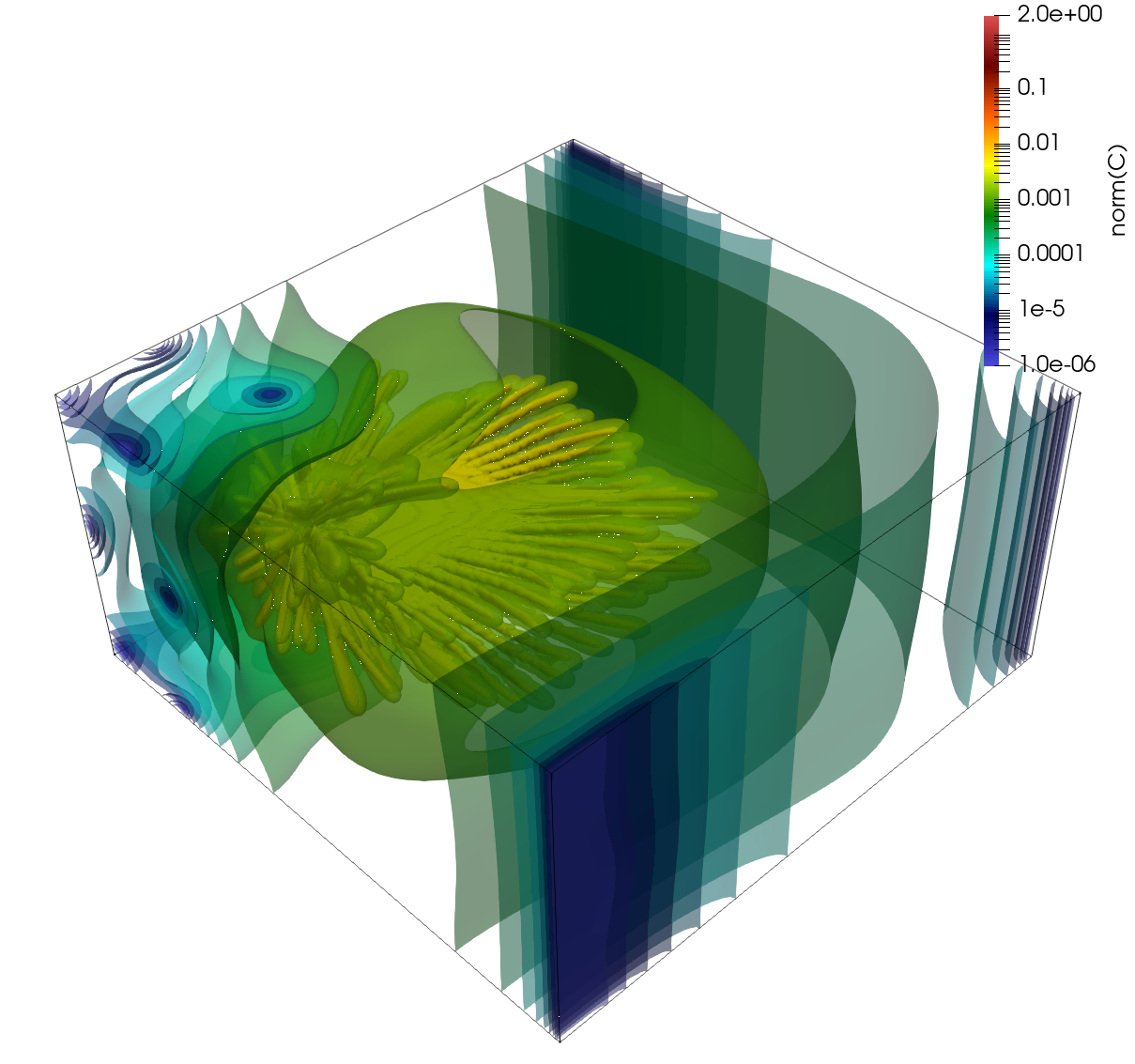} &
\includegraphics[width=0.45\textwidth,trim={50 20 30 0},clip]{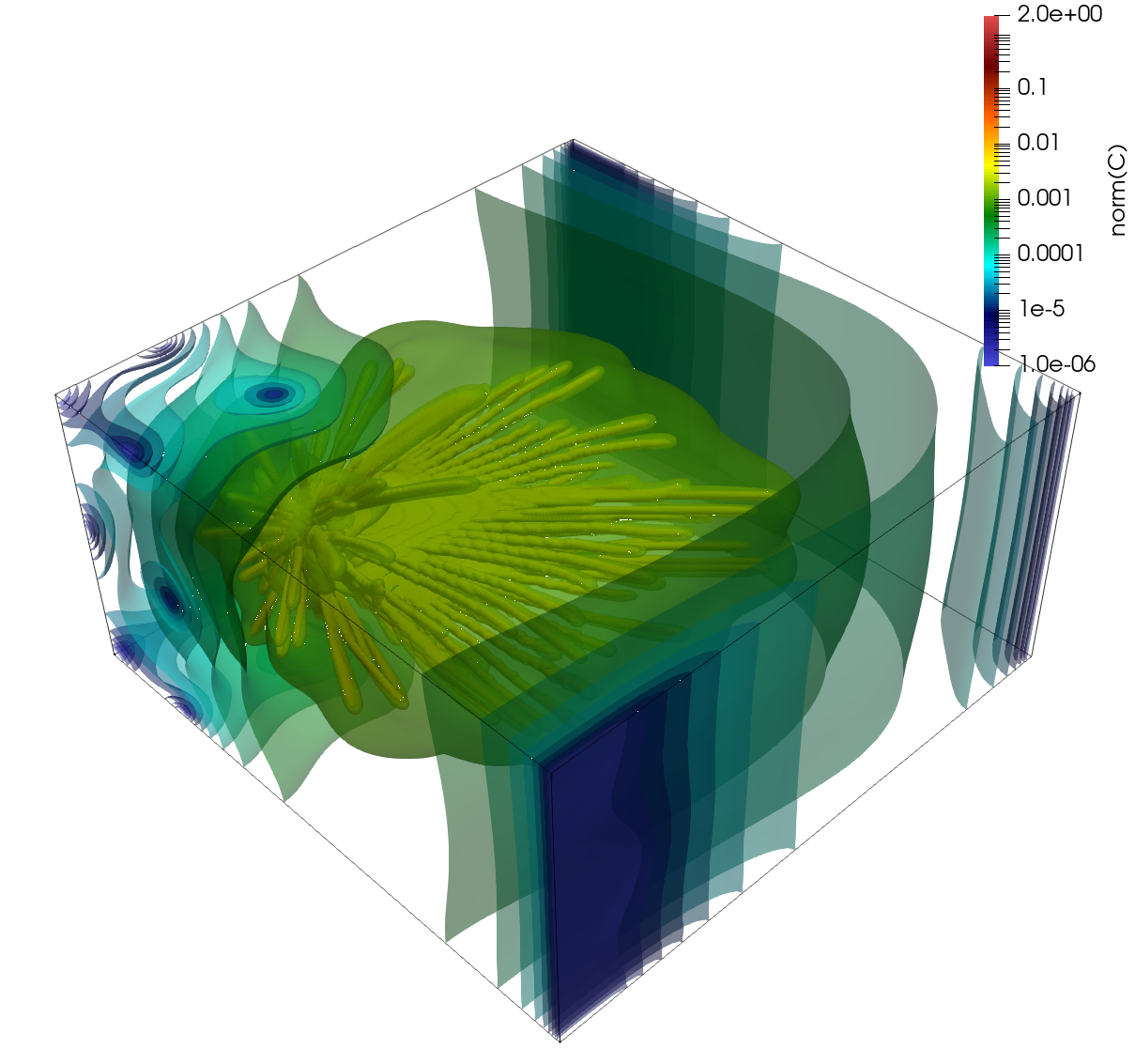}
\end{tabular}
\endgroup
\caption{Three-dimensional network formation. Contour plots of $\Norm{C_h}$ in logarithmic scale at selected time instances (see labels A--D in the left panel of \Cref{fig:slab_sequence_logs}). See \Cref{sec:3d_results} for additional details. }
\label{fig:slab_countours}
\end{figure}

\begin{figure}[htbp]
\centering
\begingroup
\setlength{\tabcolsep}{\figtabsep}
\begin{tabular}{c c}
\bf{A} & \bf{B}\\
\includegraphics[width=0.45\textwidth,trim={50 90 50 230},clip]{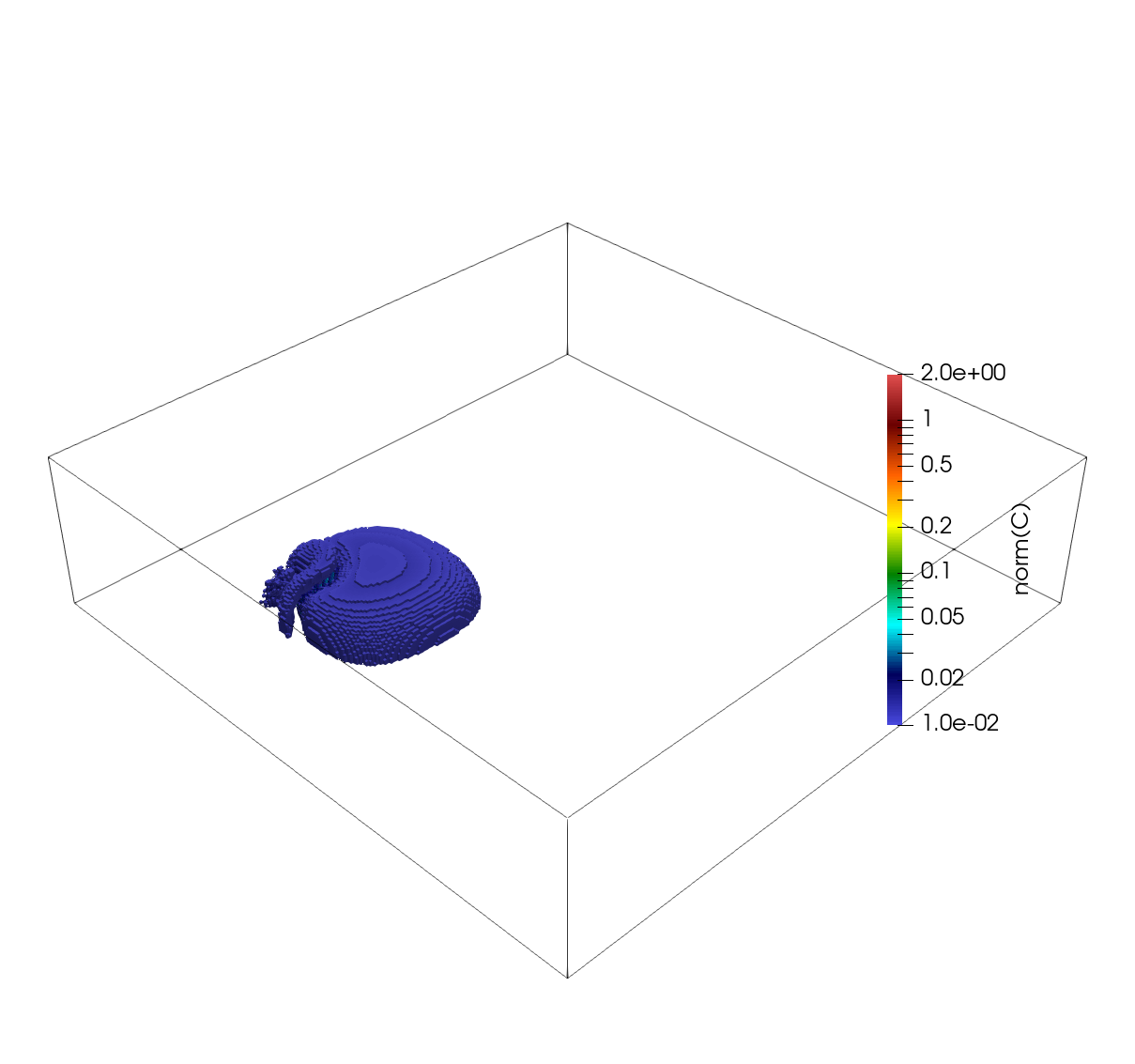} & \includegraphics[width=0.45\textwidth,trim={50 90 50 230},clip]{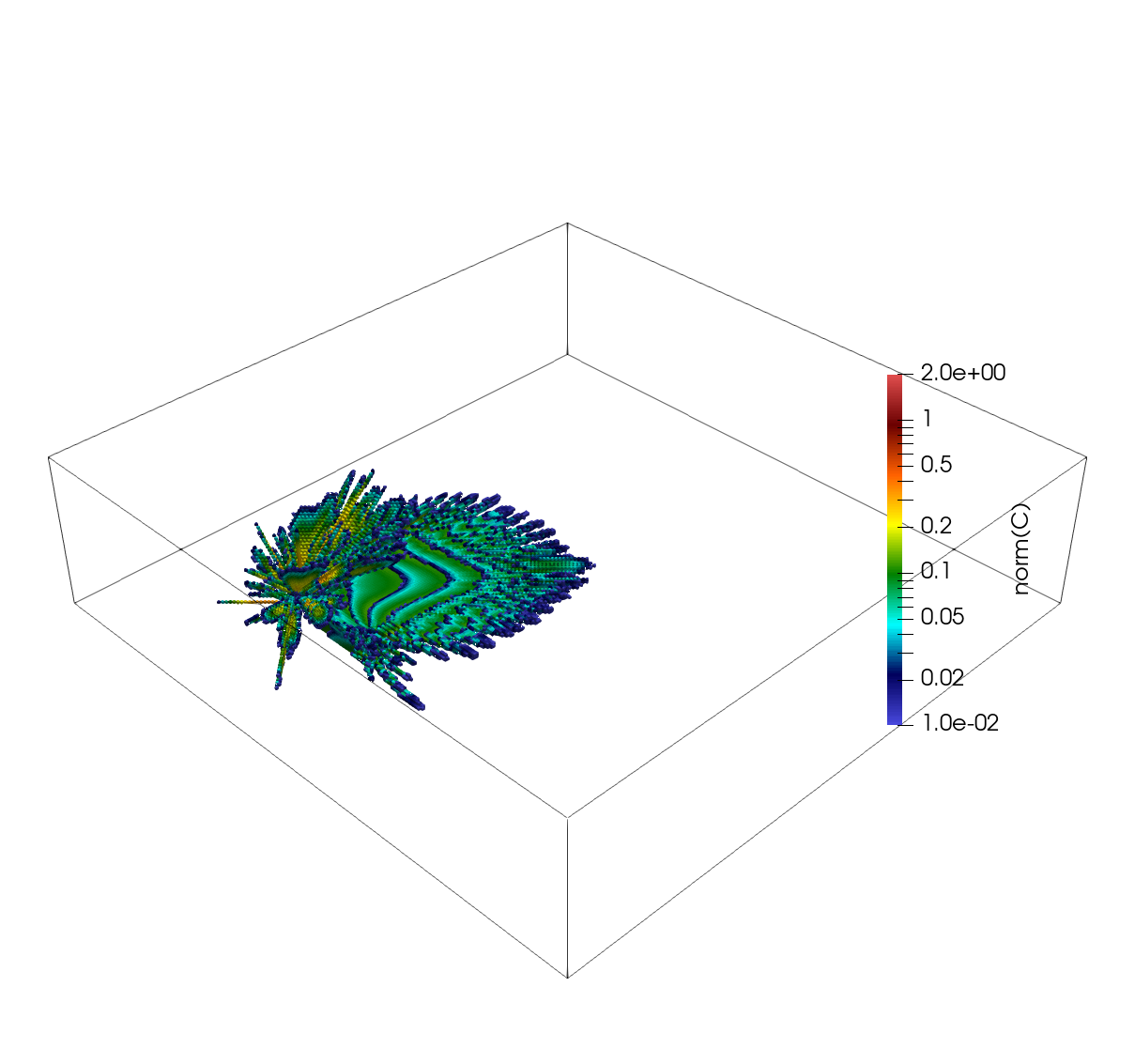}\\
\\
\bf{C} & \bf{D}\\
\includegraphics[width=0.45\textwidth,trim={50 90 50 230},clip]{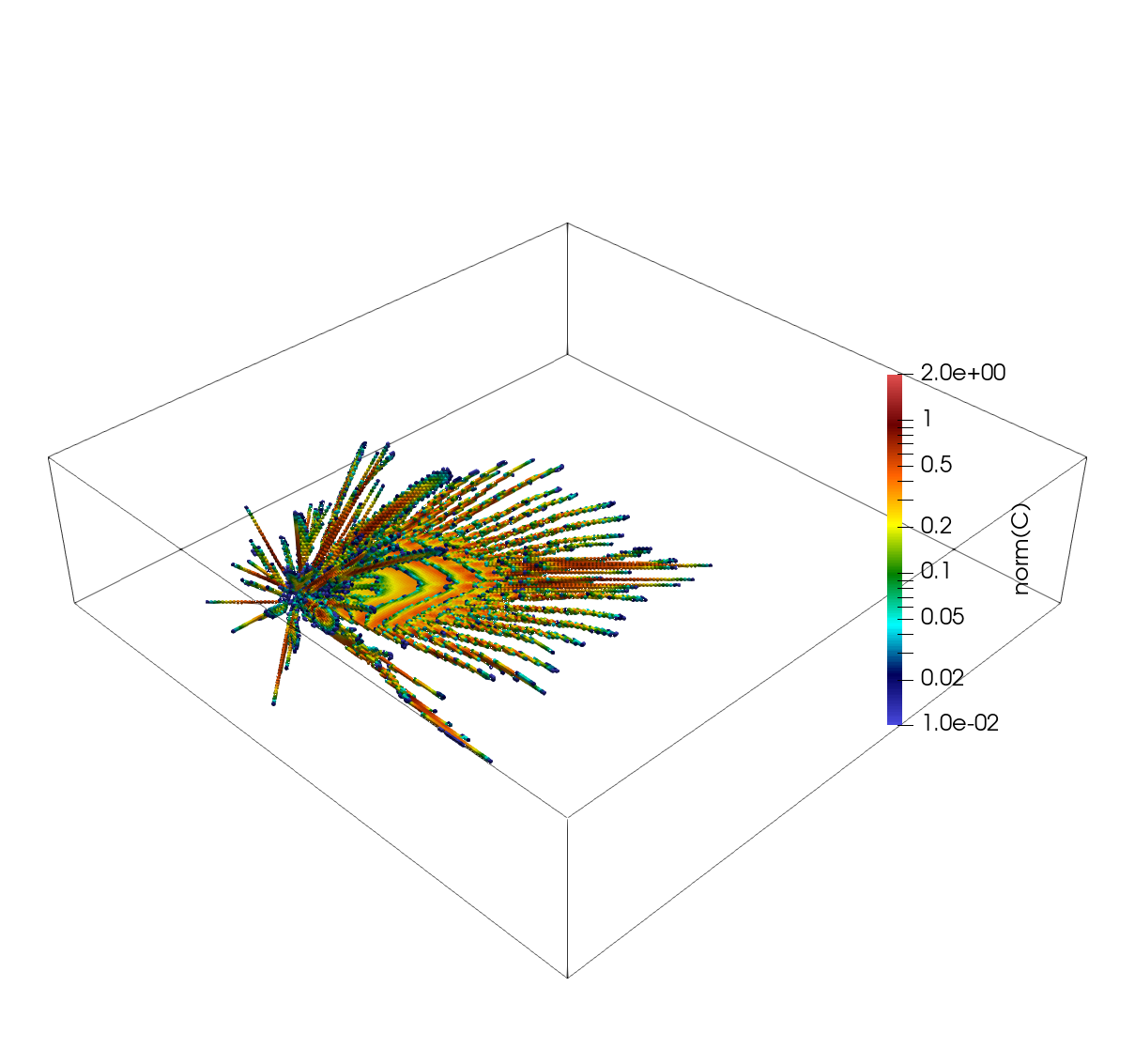} & \includegraphics[width=0.45\textwidth,trim={50 90 50 230},clip]{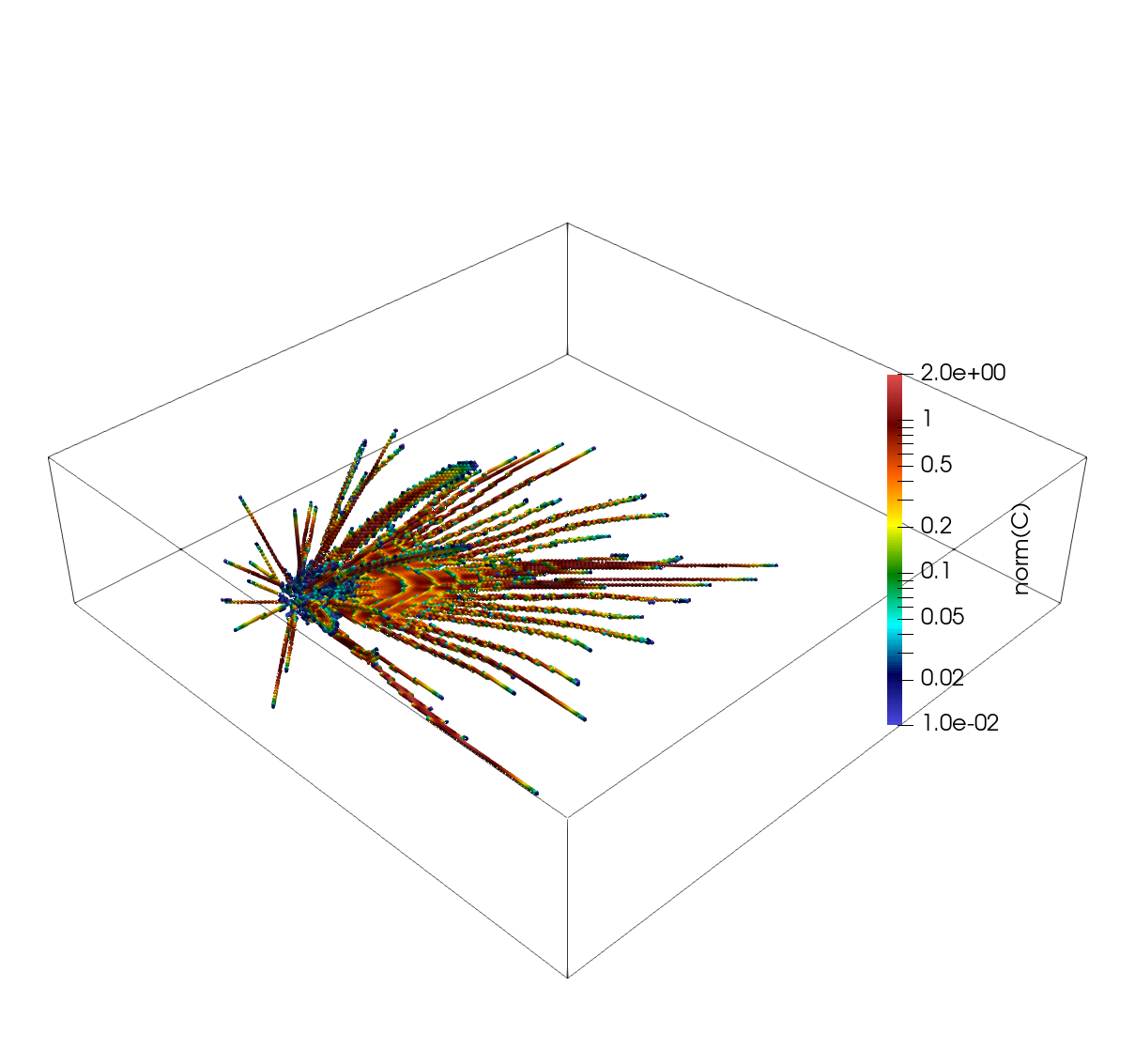}
\end{tabular}
\endgroup
\caption{Three-dimensional network formation. Threshold plot $\Norm{C_h}>10^{-2}$ on the half-space $[0,1]\times[0,1]\times[0.25,0.5]$ at selected time instances (see labels A--D in the left panel of \Cref{fig:slab_sequence_logs}). See \Cref{sec:3d_results} for additional details. }
\label{fig:slab_threshold}
\end{figure}

\Cref{fig:slab_sequence_logs} shows the energy of the system $E$ (left panel), the negative time derivative of the energy $-dE/dt$ (central panel), and the time step $\delta t$ (right panel) taken by the backward Euler solver.
In \Cref{fig:slab_countours}, we present contour plots of \(\Norm{\C_h}\) at selected time instances, as indicated by the A--D labels in the left panel of \Cref{fig:slab_sequence_logs}. The corresponding network patterns are shown in \Cref{fig:slab_threshold}, where the network pattern is obtained by retaining only those elements for which \(\Norm{\C_h} > 10^{-2}\) and by restricting the visualization to the half-space \([0,1]\times[0,1]\times[0.25,0.5]\), taking advantage of the symmetry of \(\Norm{\C_h}\) with respect to the plane \(z = 0.25\).
As in the two-dimensional case, we can again distinguish two stages of the evolution. First, the positive term $\nabla p_h\otimes\nabla p_h$ dominates, and a smooth structure is created (panel A in \Cref{fig:slab_countours} and in \Cref{fig:slab_threshold}). Then, the branching process is triggered, driven by the competition between the positive $\nabla p_h\otimes\nabla p_h$ term and the negative metabolic term. The branches then propagate through the computational domain. We observe that, similarly to the two-dimensional case, the branches in three dimensions are essentially one-dimensional structures.

\begin{figure}[htbp]
\centering
\begingroup
\setlength{\tabcolsep}{\figtabsep}
\begin{tabular}{c c}
\includegraphics[width=0.32\textwidth,trim={5 0 80 0},clip]{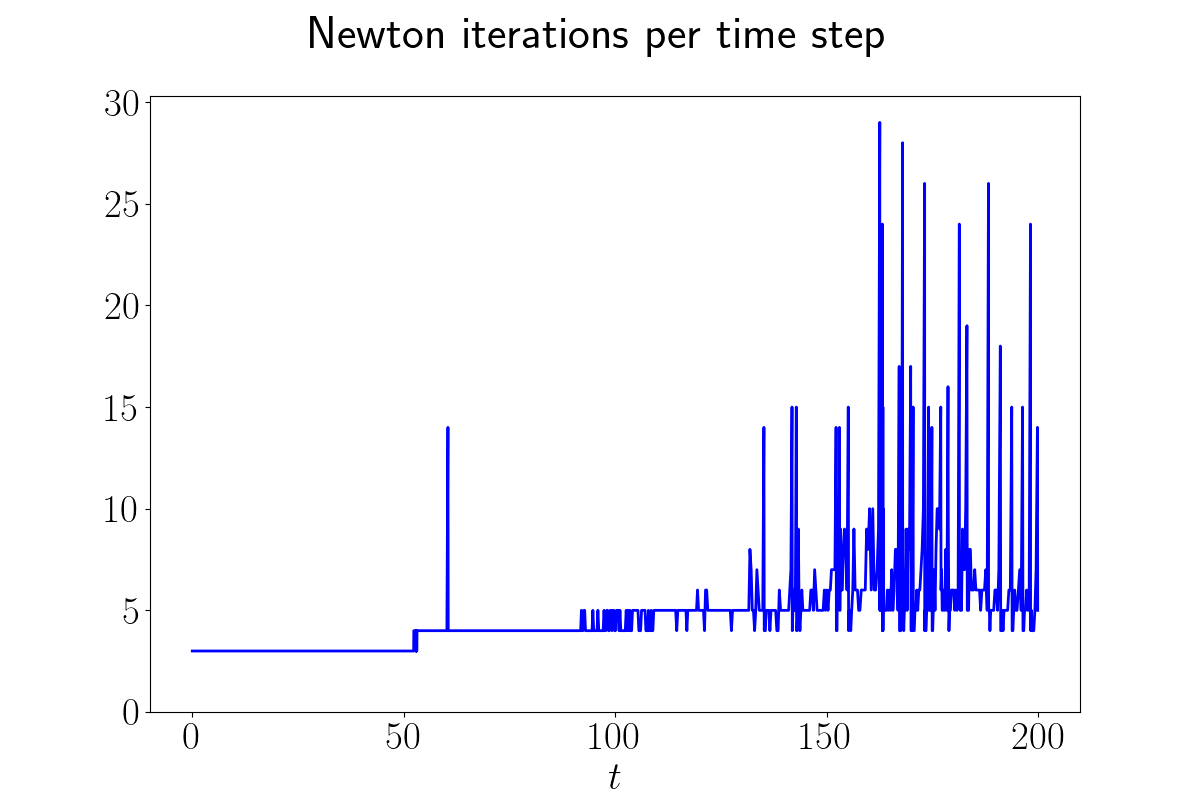} &
\includegraphics[width=0.32\textwidth,trim={5 0 80 0},clip]{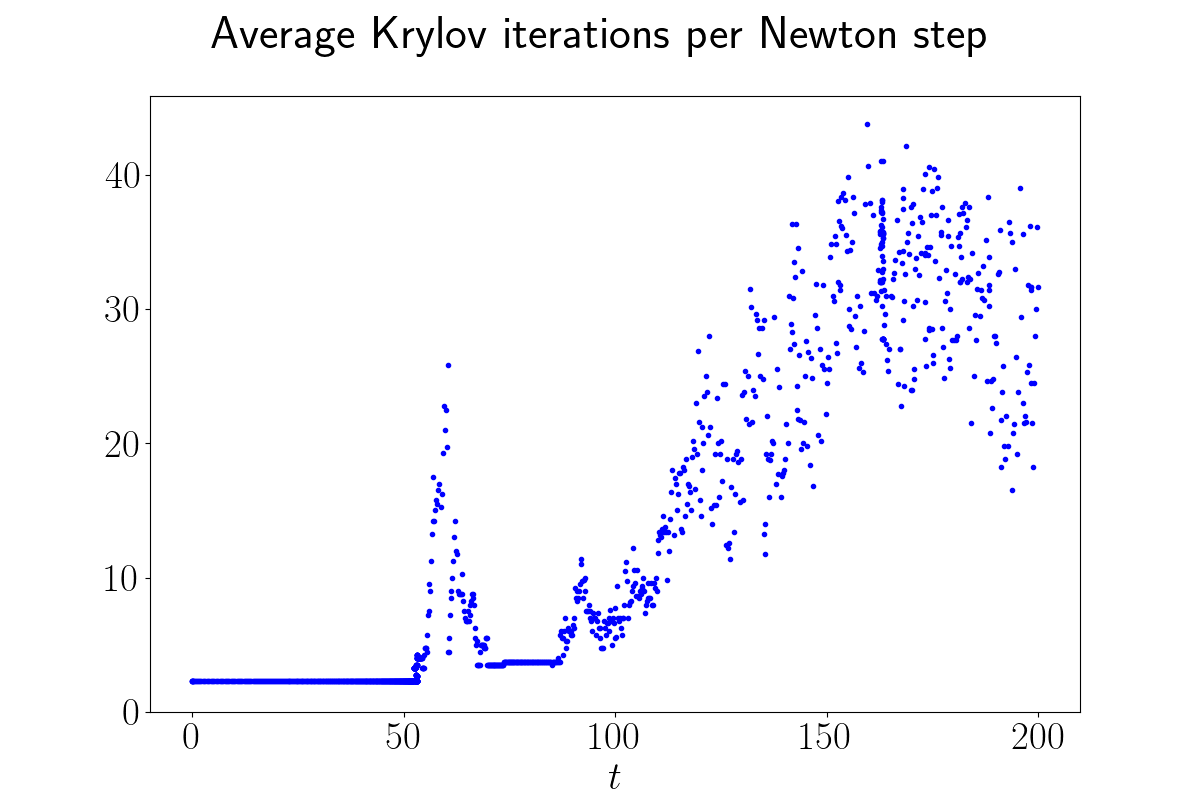} 
\end{tabular}
\endgroup
\caption{Three-dimensional network formation. Newton iterations per time step (left panel) and average number of linear iterations per Newton step (right panel) as a function of simulated time. See \Cref{sec:3d_results} for additional details. }
\label{fig:slab_sequence_alg}
\end{figure}

The number of nonlinear iterations per time step and the average number of linear iterations per Newton step are shown in \Cref{fig:slab_sequence_alg}. The nonlinearity of the problem is more severe in the late phases of network formation (left panel), while the Schur-complement preconditioner proved robust in the three-dimensional setting (right).

\section{Conclusions}\label{sec:conclusions}

We have developed robust and scalable fully implicit nonlinear finite element solvers for the simulation of biological transportation networks arising from the minimization of a nonconvex energy cost functional. Central ingredients of the proposed approach are the discretization of the conductivity matrix using a piecewise constant discontinuous finite element space and the design of effective preconditioners for the linearized systems arising within Newton’s method.

The effectiveness of the methodology is demonstrated through extensive numerical experiments in both two and three dimensions, highlighting the efficiency of the distributed-memory implementation, the robustness of the fully implicit solver, and the strong scalability of the Schur complement–based preconditioner.

We have shown that the backward Euler time integration scheme preserves the positive semidefiniteness of the conductivity matrix. Moreover, numerical evidence indicates that solution symmetries—though not positive semidefiniteness—can also be maintained when alternative time-stepping schemes such as BDF2 and Crank--Nicolson are employed.

Finally, we present, for the first time, fully three-dimensional numerical simulations of the network formation system. The results show that the main qualitative features of network formation persist in three dimensions, including the emergence of one-dimensional branching structures. The distributed-memory implementation efficiently manages the increased computational complexity, demonstrating the feasibility of large-scale three-dimensional simulations. These results pave the way for future investigations into three-dimensional network-formation models in biological systems and related physical applications.

\section*{Acknowledgements}

The research reported in this paper was funded by King Abdullah University of Science and
Technology. We are thankful for the computing resources of the Supercomputing Laboratory at
King Abdullah University of Science and Technology and, in particular, the Shaheen3 supercomputer.

\bibliographystyle{amsplain}
\bibliography{bibliography}

\end{document}